\documentclass[12pt,english]{article}
\usepackage{amsmath}
\usepackage{amsthm}

\usepackage{helvet}
\usepackage[lf]{FiraMono}
\usepackage{newtxmath}
\usepackage[T1]{fontenc}
\usepackage[utf8]{inputenc}
\usepackage{array}
\usepackage{float}
\usepackage{multirow}
\usepackage{graphicx}
\usepackage[a4paper]{geometry}
\usepackage[authoryear]{natbib}

\makeatletter

\providecommand{\tabularnewline}{\\}

\theoremstyle{plain}
\newtheorem{thm}{Theorem}
\newtheorem{prop}{Proposition}
\newtheorem{cor}{Corollary}

\ifdefined\showcaptionsetup
 \PassOptionsToPackage{caption=false}{subfig}
\fi
\usepackage{subfig}
\makeatother

\usepackage{babel}

\begin{document}
\title{Learning Rate Selection for Generalized Posteriors via Infinitesimal-Jackknife
Variance Matching}
\author{Masahiro Tanaka\thanks{Faculty of Economics, Fukuoka University, Fukuoka, Japan. Address:
8-19-1, Nanakuma, Jonan, Fukuoka, Japan 814-0180. E-mail: m.tanaka.tt@fukuoka-u.ac.jp.}}
\maketitle
\begin{abstract}
Generalized posteriors update a prior through empirical loss, with
a learning rate controlling posterior concentration. Their credible
intervals can misrepresent sampling uncertainty. We propose infinitesimal-jackknife
variance matching (IJVM), which selects the learning rate for a prespecified
scalar functional by matching its posterior variance to an IJ estimate
of the sampling variance of its posterior mean. The IJ variance is
computed from posterior covariances with observation-level losses.
A one-step update uses a pilot posterior fit and one final fit. IJVM
shares the functional-specific first-order target of rank-one weighted
score matching but requires neither loss Hessians nor gradient covariance
estimates for calibration. This facilitates applications with regular
non-smooth convex losses, including check and hinge losses, without
constructing problem-specific curvature estimators. Under suitable
regularity and Monte Carlo conditions, we establish consistency of
the one-step rate and, with additional assumptions, asymptotically
nominal credible-interval coverage. In simulations of quadratic-loss
regression, quantile regression, and hinge-loss classification, IJVM
generally achieves coverage near 95\%, although finite-sample undercoverage
remains in quantile regression. A second update can improve finite-sample
calibration in some settings, particularly at smaller sample sizes,
but the benefits of further iterations are task-dependent and not
necessarily monotone. Compared with bootstrap-based calibration, IJVM
yields comparable interval lengths with substantially fewer posterior
fits.
\end{abstract}

\section{Introduction}

\label{sec:intro} 

Generalized Bayesian inference provides a flexible framework for uncertainty
quantification when a conventional likelihood is unavailable, deliberately
misspecified, or poorly aligned with the inferential objective. Rather
than relying on a sampling density, it updates a prior through an
exponential transformation of an empirical loss function \citep{Zhang2006,Zhang2006a,Jiang2008,Bissiri2016}.
The resulting generalized posterior, often called a Gibbs posterior,
can quantify uncertainty about a population risk minimizer in settings
such as regression and classification. Its concentration is controlled
by a learning rate $\omega$, which scales the cumulative loss relative
to the prior. Although the loss determines the inferential target,
its scale does not generally ensure that posterior uncertainty has
the desired frequentist interpretation.

In particular, credible intervals from a generalized posterior need
not attain their nominal frequentist coverage, even when the posterior
concentrates around the population risk minimizer. The source of this
discrepancy is apparent from the asymptotic variances \citep{Miller2021}.
For a scalar functional $g(\theta)$ admitting a nondegenerate first-order
linearization, the generalized-posterior variance typically has the
first-order form $A_{g}/(n\omega)$, determined by the curvature of
the population risk, whereas the sampling variance of the corresponding
loss-based estimator has the form $B_{g}/n$, which also reflects
variability in the observation-level loss gradients. These expressions
agree at $\omega=\omega^{\circ}_{g}=A_{g}/B_{g}$. Selecting the learning
rate is therefore an inferential calibration problem rather than merely
a computational tuning decision. Moreover, the target is specific
to $g$: a single scalar rate need not calibrate all components or
contrasts of a multivariate parameter simultaneously.

Existing methods for choosing the learning rate pursue different objectives.
\citet{Holmes2017} propose an information-based choice for a power
likelihood, while SafeBayes selects the rate according to sequential
predictive performance \citep{Gruenwald2017}. \citet{Lyddon2019}
develop Fisher-information matching through a connection to the loss-likelihood
bootstrap, and \citet{Onizuka2026} proposes weighted score matching
(WSM), which selects the learning rate by minimizing a weighted Fisher
divergence between asymptotic distributions. Although a suitable rank-one
weight in WSM gives the same first-order target for a specified scalar
functional as the method developed here, its proposed plug-in computation
estimates the curvature and gradient covariance of the loss. These
quantities are straightforward in many smooth problems but require
additional care for piecewise-linear losses, such as check and hinge
losses, whose pointwise Hessians do not recover the population-risk
curvature. Generalized posterior calibration (GPC) \citep{Syring2019}
instead selects the learning rate by estimating credible-region coverage
through bootstrap resampling. Its effectiveness in a range of misspecified
settings is documented by \citet{Wu2023}. The price of this direct
approach is computational: estimating coverage at candidate learning
rates requires repeated fits to bootstrap-weighted data, often across
multiple rate updates. This can be expensive when posterior simulation
is itself demanding, despite attempts to improve the computational
efficiency of GPC \citep[e.g.,][]{Tanaka2026efficient}.

The infinitesimal jackknife (IJ) offers a way to estimate sampling
variability without repeatedly fitting bootstrap posteriors. It measures
the sensitivity of a posterior expectation to small changes in observation
weights. For a generalized posterior, the derivative of the posterior
mean of $g(\theta)$ with respect to an observation weight is minus
the learning rate times the posterior covariance between $g(\theta)$
and that observation's loss. Consequently, an IJ approximation to
the sampling variance of the posterior mean can be constructed from
draws obtained in a single posterior fit. \citet{Giordano2024} develop
the covariance-based IJ for Bayesian posterior expectations, and related
recent work explores robust uncertainty quantification and efficient
sensitivity analysis \citep{LuoJi2026,OhkuboIba2026}. These developments
suggest using the IJ variance not only as a standard-error estimate
but also as a benchmark for calibrating generalized-posterior dispersion.

We build on this idea to propose \emph{infinitesimal-jackknife variance
matching} (IJVM), a computationally efficient learning-rate selection
method for a prespecified scalar functional $g(\theta)$. IJVM equates
the generalized-posterior variance $U_{n}(\omega)$ of $g(\theta)$
with an IJ estimate $W_{n}(\omega)$ of the sampling variance of its
posterior mean. The IJ estimate uses posterior covariances with observation-level
losses, centered to reflect the dependence among ordinary multinomial-bootstrap
weights. The criterion $U_{n}(\omega)=W_{n}(\omega)$ leads naturally
to the multiplicative update $\omega U_{n}(\omega)/W_{n}(\omega)$.
Its Monte Carlo version requires only a pilot posterior sample and
the observation-level losses evaluated at the sampled parameter values.
A second fit at the updated learning rate then provides the posterior
used for inference. Thus, the basic one-step procedure ordinarily
requires only two posterior fits, rather than the numerous bootstrap
fits used by GPC. The calibration calculation does not require explicit
estimation of the loss Hessian, the loss-gradient covariance matrix,
or derivatives of $g$; additional updates or independent validation
fits remain available when closer finite-sample matching is desired.
This difference in the information needed for calibration, rather
than a different first-order target or a universal speed advantage,
is the main practical distinction from functional-specific WSM. In
particular, IJVM can be implemented without deriving specialized curvature
estimators for the regular non-smooth losses.

We develop a theoretical justification for this one-step procedure.
Under uniform posterior moment and local linearization conditions,
we show that $n\omega U_{n}(\omega)$ and $nW_{n}(\omega)$ converge
uniformly in probability over a compact set of learning rates to $A_{g}$
and $B_{g}$, respectively. Hence, the exact-posterior update converges
uniformly in probability to the constant $\omega^{\circ}_{g}$, so
one update is sufficient to first order even when the pilot rate is
data-dependent. We also establish existence and consistency results
for solutions of the matching equation, derive stability bounds, and
specify conditions under which Monte Carlo error does not compromise
consistency. Finally, when a sampling central limit theorem and a
suitable posterior-quantile approximation also hold, credible intervals
at a consistent selected rate attain asymptotically nominal frequentist
coverage. 

We investigate the finite-sample behavior of IJVM in homoscedastic
and heteroscedastic quadratic-loss regression, quantile regression,
and hinge-loss classification. Across these designs, one-step IJVM
gives empirical coverage close to the nominal $95\%$ level, with
interval lengths comparable to those of GPC. The principal computational
advantage of IJVM is its substantially smaller requirement for posterior
simulation. Whereas one-step IJVM requires only two posterior fits,
GPC requires repeated posterior fitting across bootstrap samples and
coverage evaluations. In our experiments, this amounts to a median
of approximately $1{,}000$--$10{,}000$ bootstrap posterior fits
for GPC, depending on the design and sample size. An ablation study
shows that the first update substantially reduces sensitivity to the
initial learning rate, while one or two additional updates can provide
further improvements in smaller samples. The extent of these improvements
depends on the inferential task, and additional iterations do not
necessarily yield better coverage. 

The remainder of the paper is organized as follows. Section~\ref{sec:method}
introduces the IJ variance-matching criterion, the one-step algorithm,
and its relationship to existing learning-rate methods. Section~\ref{sec:theory}
establishes the theoretical and Monte Carlo properties of IJVM. Section~\ref{sec:simulation}
presents the simulation study and sensitivity analysis. Section~\ref{sec:discussion}
concludes, and the Appendix provides proofs and additional computational
details.

\section{Method}

\label{sec:method}

\subsection{Generalized posterior and calibration target}

\label{subsec:posterior-calib-target}

Let $Y_{1},\ldots,Y_{n}$ be independent and identically distributed
from an unknown distribution $P$, and write $\mathcal{D}_{n}=(Y_{1},\ldots,Y_{n})$.
For $\theta\in\Theta\subset\mathbb{R}^{d}$, let $\ell_{i}(\theta)=\ell(\theta;Y_{i})$
be an observation-level loss. Define the empirical and population
risks by 
\[
R_{n}(\theta)=\frac{1}{n}\sum^{n}_{i=1}\ell_{i}(\theta),\qquad R(\theta)=E_{P}\{\ell(\theta;Y)\}.
\]
Suppose that $R$ has a unique minimizer $\theta^{\dagger}$ and let
$\widehat{\theta}_{n}$ minimize $R_{n}$. For a prior density $p$
and learning rate $\omega$ in a compact interval $\Omega=[\underline{\omega},\overline{\omega}]$
with $0<\underline{\omega}<\overline{\omega}<\infty$, the generalized
posterior is 
\begin{equation}
\pi_{\omega}(\theta\mid\mathcal{D}_{n})\propto\exp\left\{ -\omega\sum^{n}_{i=1}\ell_{i}(\theta)\right\} p(\theta).\label{eq:gen-posterior}
\end{equation}
We assume that the posterior is proper for every $\omega\in\Omega$.
All posterior expectations and covariances below are conditional on
$\mathcal{D}_{n}$.

Our inferential target is a measurable scalar functional $g:\Theta\to\mathbb{R}$,
with $g^{\dagger}=g(\theta^{\dagger})$. We associate with it a fixed
nonzero vector $\dot{g}\in\mathbb{R}^{d}$, called the functional
linearization vector. Its role is specified by the posterior asymptotic
linearity condition in \eqref{eq:ass-3}; when required for sampling
inference, the corresponding sampling-side condition is \eqref{eq:func-sampling-al}.
These conditions are imposed where invoked rather than assumed by
this definition. If $g$ is differentiable at $\theta^{\dagger}$,
the natural choice is $\dot{g}=\nabla g(\theta^{\dagger})$. Let $\psi(Y)$
be an observation-level  score with $E_{P}\{\psi(Y)\}=0$ and $V=E_{P}\{\psi(Y)\psi(Y)^{\top}\}$.
In the regular cases considered below, the population-risk curvature
$J$ is defined by 
\[
R(\theta^{\dagger}+h)-R(\theta^{\dagger})=\tfrac{1}{2}h^{\top}Jh+o(\|h\|^{2}),\qquad J\succ0,
\]
not by requiring observation-level Hessians. For smooth losses, $\psi(Y)=\nabla\ell(\theta^{\dagger};Y)$,
but differentiability of individual losses away from the population
target is not required. 
\begin{equation}
A_{g}=\dot{g}^{\top}J^{-1}\dot{g},\qquad B_{g}=\dot{g}^{\top}J^{-1}VJ^{-1}\dot{g},\qquad\omega^{\circ}_{g}=\frac{A_{g}}{B_{g}}.\label{eq:target-rate}
\end{equation}
Here $J$ is the population-risk curvature and $V$ is the covariance
of the observation-level score. For non-smooth convex losses, the
score is often the almost-sure derivative at $\theta^{\dagger}$,
although it need not exist at the empirical minimizer. Under a sampling
central limit theorem together with the sampling-side linearization
\eqref{eq:func-sampling-al}, and an appropriate generalized-posterior
normal approximation, $\sqrt{n}\{g(\widehat{\theta}_{n})-g^{\dagger}\}$
has asymptotic variance $B_{g}$, while the generalized posterior
distribution of $\sqrt{n}\{g(\theta)-g(\widehat{\theta}_{n})\}$ has
asymptotic variance $A_{g}/\omega$. Consequently, $\omega^{\circ}_{g}$
matches these two variances to first order. This target is functional-specific:
a single scalar rate need not calibrate all functionals of a multivariate
parameter simultaneously.

\subsection{Finite-sample infinitesimal-jackknife criterion}

\label{subsec:calib-finite-sample-target}

Let 
\begin{equation}
U_{n}(\omega)=\operatorname{Var}_{\pi_{\omega}}\{g(\theta)\},\qquad\xi_{i,\omega}=\operatorname{Cov}_{\pi_{\omega}}\{g(\theta),\ell_{i}(\theta)\},\qquad\overline{\xi}_{\omega}=\frac{1}{n}\sum^{n}_{i=1}\xi_{i,\omega},\label{eq:ui-xi}
\end{equation}
and define 
\begin{equation}
W_{n}(\omega)=\omega^{2}\sum^{n}_{i=1}(\xi_{i,\omega}-\overline{\xi}_{\omega})^{2}.\label{eq:ij-var}
\end{equation}
The quantity $W_{n}$ is an infinitesimal-jackknife (IJ) approximation
to the repeated-sampling variance of the generalized posterior mean
of $g(\theta)$, obtained by linearizing bootstrap reweighting.

To see this, let $M=(M_{1},\ldots,M_{n})$ denote ordinary nonparametric-bootstrap
counts, with 
\[
M\sim\operatorname{Multinomial}\left(n;\frac{1}{n},\ldots,\frac{1}{n}\right),
\]
and consider the weighted posterior 
\[
\pi_{\omega,m}(\theta\mid\mathcal{D}_{n})\propto\exp\left\{ -\omega\sum^{n}_{i=1}m_{i}\ell_{i}(\theta)\right\} p(\theta).
\]
Assuming that differentiation under the integral is valid in a neighborhood
of $m=\mathbf{1}_{n}$, the posterior-mean sensitivity identity gives
\citep{Giordano2024}
\[
\left.\frac{\partial}{\partial m_{i}}E_{\pi_{\omega,m}}\{g(\theta)\}\right|_{m=\mathbf{1}_{n}}=-\omega\xi_{i,\omega}.
\]
Since $\ensuremath{\operatorname{Cov}_{M}(M_{i},M_{j})=\mathbf{1}\{i=j\}-1/n}$,
the conditional bootstrap variance of the linearized posterior mean
is exactly 
\begin{equation}
\operatorname{Var}_{M}\left\{ -\omega\sum^{n}_{i=1}(M_{i}-1)\xi_{i,\omega}\right\} =\omega^{2}\sum^{n}_{i=1}(\xi_{i,\omega}-\overline{\xi}_{\omega})^{2}=W_{n}(\omega).\label{eq:boot-ij-identity}
\end{equation}
This is a variance of the linear approximation, not a claim that the
finite-sample bootstrap variance is exactly $W_{n}$. The centering
by $\overline{\xi}_{\omega}$ reflects the fixed bootstrap total $\sum_{i}M_{i}=n$.
Section~\ref{sec:theory} shows that, uniformly on $\Omega$, $n\omega U_{n}(\omega)$
and $nW_{n}(\omega)$ converge in probability to $A_{g}$ and $B_{g}$,
respectively. At first order, $n\omega\xi_{i,\omega}$ approximates
$\dot{g}^{\top}J^{-1}\psi(Y_{i})$ in averaged squared error; thus
$W_{n}$ has no leading-order dependence on $\omega$. We therefore
propose the finite-sample matching criterion 
\begin{equation}
U_{n}(\omega)=W_{n}(\omega).\label{eq:matching-equation}
\end{equation}

\subsection{Computation and iterative refinement}

\label{subsec:onestep-calib}

When $W_{n}(\omega)>0$, define the exact-posterior update map 
\begin{equation}
\mathcal{T}_{n}(\omega)=\omega\frac{U_{n}(\omega)}{W_{n}(\omega)}=\frac{U_{n}(\omega)}{\omega\sum^{n}_{i=1}(\xi_{i,\omega}-\overline{\xi}_{\omega})^{2}}.\label{eq:update-map}
\end{equation}
A fixed point of $\mathcal{T}_{n}$ solves \eqref{eq:matching-equation};
a boundary fixed point of the projected map need not. The multiplicative
update increases the rate if posterior variance exceeds the IJ variance,
and decreases it in the reverse case. More importantly, Theorem~\ref{thm:unif-limit-of-update-map}
shows that $\mathcal{T}_{n}$ converges uniformly to the constant
$\omega^{\circ}_{g}$, so a single update is first-order sufficient
even if the initial rate is data-dependent.

For $S\ge2$ posterior draws $\theta^{(1)},\ldots,\theta^{(S)}$ obtained
at an evaluated rate $\omega\in\Omega$ (initially $\omega^{(0)}$),
set $g_{s}=g(\theta^{(s)})$, $\ell_{si}=\ell_{i}(\theta^{(s)})$,
$\overline{g}=S^{-1}\sum_{s}g_{s}$, and $\overline{\ell}_{i}=S^{-1}\sum_{s}\ell_{si}$.
Estimate the required quantities by 
\begin{align}
\widehat{U}_{n}(\omega) & =\frac{1}{S-1}\sum^{S}_{s=1}(g_{s}-\overline{g})^{2},\nonumber \\
\widehat{\xi}_{i,\omega} & =\frac{1}{S-1}\sum^{S}_{s=1}(g_{s}-\overline{g})(\ell_{si}-\overline{\ell}_{i}),\qquad\widehat{\overline{\xi}}_{\omega}=\frac{1}{n}\sum^{n}_{i=1}\widehat{\xi}_{i,\omega},\nonumber \\
\widehat{W}_{n}(\omega) & =\omega^{2}\sum^{n}_{i=1}\left(\widehat{\xi}_{i,\omega}-\widehat{\overline{\xi}}_{\omega}\right)^{2}.\label{eq:comp-vars}
\end{align}
The \emph{infinitesimal-jackknife variance matching} (IJVM) one-step
rate is 
\begin{equation}
\widehat{\omega}^{(1)}=\Pi_{\Omega}\left\{ \omega^{(0)}\frac{\widehat{U}_{n}(\omega^{(0)})}{\widehat{W}_{n}(\omega^{(0)})}\right\} ,\label{eq:onestep-update}
\end{equation}
where $\Pi_{\Omega}$ denotes projection onto $\Omega$. The update
is computed when the estimated denominator is positive; an unstable
or degenerate denominator calls for additional simulation or a diagnostic
check rather than an unqualified update. Only draws from the pilot
posterior and the corresponding observation-level loss matrix are
required for calibration. Inference using the calibrated posterior
ordinarily entails a further posterior fit at $\widehat{\omega}^{(1)}$.
No analytical loss gradients or Hessians are needed for the calibration
calculation, although they may be used by the posterior sampler. Under
the convex-loss conditions of Section~\ref{subsec:convex-sufficient},
Corollary~\ref{cor:mc-iid} shows that, for conditionally independent
posterior draws, the Monte Carlo accuracy requirements for the computed
one-step rate hold whenever the number of draws tends to infinity.
The corresponding result for MCMC draws requires additional control
of simulation dependence. 

Although a single update is sufficient to first order, finite-sample
discrepancies between the posterior and IJ variances may remain. One
or two further updates can be useful, particularly in smaller samples,
although their benefit depends on the problem. For such refinement,
the update may be iterated with fresh posterior simulation at each
rate and optional damping:
\begin{equation}
\widehat{\omega}^{(t+1)}=\Pi_{\Omega}\left[\widehat{\omega}^{(t)}\left\{ \frac{\widehat{U}_{n}(\widehat{\omega}^{(t)})}{\widehat{W}_{n}(\widehat{\omega}^{(t)})}\right\} ^{\gamma_{t}}\right],\qquad0<\gamma_{t}\le1,\label{eq:multistep}
\end{equation}
with $\widehat{\omega}^{(0)}=\omega^{(0)}$. One can monitor the absolute
log variance ratio, 
\[
d_{t}=\left|\log\left(\frac{\widehat{U}_{n}(\widehat{\omega}^{(t)})}{\widehat{W}_{n}(\widehat{\omega}^{(t)})}\right)\right|.
\]
A reduction in this diagnostic indicates closer variance matching,
but does not necessarily imply improved finite-sample coverage. Because
selection and evaluation using the same simulation can be optimistic,
an approximate root should preferably be checked using independent
validation draws. Neither global convergence nor finite-sample improvement
of \eqref{eq:multistep} follows from first-order consistency: the
finite-sample map is nonlinear and need not be contractive.

\subsection{Relation to existing calibration methods}

Existing approaches to learning-rate selection differ primarily in
their calibration targets. Fisher-information matching (FIM) relates
generalized-posterior information to the loss-likelihood bootstrap
\citep{Lyddon2019}. SafeBayes selects the learning rate by minimizing
cumulative prequential posterior-expected loss, prioritizing predictive
performance rather than credible-region coverage \citep{Gruenwald2017}.

Generalized posterior calibration (GPC) instead tunes the rate to
match the nominal frequentist coverage of a specified credible region,
estimated through bootstrap replications and repeated posterior fitting
\citep{Syring2019}. IJVM targets $U_{n}(\omega)=W_{n}(\omega)$ for
a prespecified scalar functional $g$, rather than coverage at a particular
credibility level. Thus, IJVM calibrates variance, whereas GPC directly
calibrates coverage. Under the normal approximations used here, however,
GPC applied to an interval for the same $g$ shares the first-order
target $\omega^{\circ}_{g}$. IJVM avoids repeated bootstrap posterior
fits in its one-step update.

IJVM and weighted score matching (WSM) \citep{Onizuka2026} can share
exactly the same functional-specific first-order target. Writing $\omega^{\mathrm{WS}}(Q)=\operatorname{tr}(JQ)/\operatorname{tr}(VQ)$,
the rank-one weight $Q_{g}=J^{-1}\dot{g}\dot{g}^{\top}J^{-1}$ gives
\[
\omega^{\mathrm{WS}}(Q_{g})=\frac{\dot{g}^{\top}J^{-1}\dot{g}}{\dot{g}^{\top}J^{-1}VJ^{-1}\dot{g}}=\omega^{\circ}_{g}.
\]
Thus, the distinction is not a new asymptotic rate but how it is estimated
and which losses admit a direct implementation. The plug-in algorithm
of \citet{Onizuka2026} estimates $J$ and $V$ using empirical loss
Hessians and gradients evaluated at an empirical risk minimizer. IJVM
instead computes posterior covariances between $g(\theta)$ and the
observation-level losses; calibration requires only evaluations of
$g$ and the losses at posterior draws, not analytical or numerical
derivatives of either quantity.

This distinction is especially consequential for piecewise-linear
losses. For check-loss quantile regression and hinge-loss classification,
ordinary observation-level Hessians vanish wherever they exist, and
gradients may be undefined at fitted residuals or margins. Nevertheless,
the population risk can have positive local curvature $J$, determined
by the conditional density at a quantile or the distribution near
a classification margin; see Section~\ref{subsec:nonsmooth-examples}.
Consequently, simply averaging observation-level Hessians does not
yield the curvature needed for the plug-in WSM rate. Applying WSM
in these settings is possible in principle with separately constructed,
consistent estimators of $J$ and $V$, but requires additional problem-specific
work. By contrast, the same loss-value-based IJVM calculation applies
under the regularity conditions established in Section~\ref{sec:theory}.
This is a practical extension of the range of applications accessible
without changing the calibration algorithm.

\subsection{Extension to independent-cluster sampling}

\label{subsec:cluster-extension}

IJVM extends to i.i.d.~clusters by using entire clusters, rather
than individual observations, as reweighting units. Let $\mathcal{Y}_{1},\ldots,\mathcal{Y}_{G}$
be i.i.d.~ clusters, with $N_{c}$ observations in cluster $c$,
and let $L_{c}(\theta)=\sum^{N_{c}}_{j=1}\ell(\theta;Y_{cj})$. Writing
$\mathcal{D}^{\mathrm{CL}}_{G}=(\mathcal{Y}_{1},\ldots,\mathcal{Y}_{G})$,
the generalized posterior is $\pi_{\omega}(\theta\mid\mathcal{D}^{\mathrm{CL}}_{G})\propto\exp\{-\omega\sum^{G}_{c=1}L_{c}(\theta)\}p(\theta)$.
Define $U_{G}(\omega)=\operatorname{Var}_{\pi_{\omega}}\{g(\theta)\}$,
$\xi^{\mathrm{CL}}_{c,\omega}=\operatorname{Cov}_{\pi_{\omega}}\{g(\theta),L_{c}(\theta)\}$,
and $\bar{\xi}^{\mathrm{CL}}_{\omega}=G^{-1}\sum^{G}_{c=1}\xi^{\mathrm{CL}}_{c,\omega}$.
The cluster-level IJ variance is 
\begin{equation}
W^{\mathrm{CL}}_{G}(\omega)=\omega^{2}\sum^{G}_{c=1}\bigl(\xi^{\mathrm{CL}}_{c,\omega}-\bar{\xi}^{\mathrm{CL}}_{\omega}\bigr)^{2}.\label{eq:cluster-ij}
\end{equation}
This equals the variance of the linearized posterior mean under a
multinomial bootstrap that resamples entire clusters rather than individual
observations \citep{Tanaka2026efficient}. With hats denoting Monte
Carlo estimates, the projected one-step update is
\begin{equation}
\widehat{\omega}^{(1)}_{\mathrm{CL}}=\Pi_{\Omega}\!\left\{ \omega^{(0)}\frac{\widehat{U}_{G}(\omega^{(0)})}{\widehat{W}^{\mathrm{CL}}_{G}(\omega^{(0)})}\right\} .\label{eq:cluster-update}
\end{equation}
Implementation requires only summing the observation-level loss matrix
within each cluster before computing posterior covariances; no cluster-bootstrap
posterior fits are needed.

Under cluster-level analogues of the conditions in Section~\ref{sec:theory},
the update targets the ratio of the first-order posterior-variance
and cluster-robust sampling-variance coefficients. If this rate lies
in $\Omega$, credible intervals attain asymptotically nominal coverage
under the corresponding normal approximations. With varying cluster
sizes, $L_{c}$ targets the summed-loss risk $E\{L_{c}(\theta)\}$,
whereas using $L_{c}/N_{c}$ would define a different, equally weighted
cluster-average risk. The results concern i.i.d.~clusters, not arbitrary
complex-survey designs. Assumptions, proofs, and a small simulation
are given in Appendix~D.

\section{Theoretical properties}

\label{sec:theory}

The covariance-based IJVM construction does not require second derivatives
of the observation-level losses. We first state the assumptions directly
in terms of population-risk curvature, observation-level scores, and
posterior moments. We then give primitive conditions under which these
assumptions hold for convex losses with kinks. Proofs are given in
Appendix~A.

\subsection{Regularity conditions}

\label{subsec:reg-conds}

Throughout this section, $J$ is defined by the population-risk expansion
in Section~\ref{subsec:posterior-calib-target}, rather than by an
expectation of observation-level Hessians. Let $\psi(Y)$ be a measurable
$\mathbb{R}^{d}$-valued score, not necessarily a gradient at the
empirical minimizer, and suppose that $E_{P}\{\psi(Y)\}=0$, $V=E_{P}\{\psi(Y)\psi(Y)^{\top}\}$,
and both $J$ and $V$ are positive definite. Assume that $\sqrt{n}(\widehat{\theta}_{n}-\theta^{\dagger})=O_{P}(1)$
and 
\begin{equation}
\frac{1}{n}\sum^{n}_{i=1}\|\psi(Y_{i})\|^{2}=O_{P}(1),\qquad\frac{1}{n}\sum^{n}_{i=1}\psi(Y_{i})\psi(Y_{i})^{\top}\xrightarrow{P}V.\label{eq:ass-1}
\end{equation}
Put $X_{n}=\sqrt{n}(\theta-\widehat{\theta}_{n})$, $G_{n}=\sqrt{n}\{g(\theta)-g(\widehat{\theta}_{n})\}$,
and $D_{ni}=\sqrt{n}\{\ell_{i}(\theta)-\ell_{i}(\widehat{\theta}_{n})\}$.
We require the following uniform posterior moment conditions: 
\begin{equation}
\sup_{\omega\in\Omega}E_{\pi_{\omega}}\|X_{n}\|^{2}=O_{P}(1),\qquad\sup_{\omega\in\Omega}\left\Vert \operatorname{Var}_{\pi_{\omega}}(X_{n})-\omega^{-1}J^{-1}\right\Vert =o_{P}(1).\label{eq:ass-2}
\end{equation}
For the functional linearization vector $\dot{g}$ introduced in Section~\ref{subsec:posterior-calib-target},
require the following posterior asymptotic linearity of $g$ and averaged
observation-level loss linearization:
\begin{equation}
\sup_{\omega\in\Omega}E_{\pi_{\omega}}(G_{n}-\dot{g}^{\top}X_{n})^{2}=o_{P}(1),\qquad\sup_{\omega\in\Omega}\frac{1}{n}\sum^{n}_{i=1}E_{\pi_{\omega}}(D_{ni}-\psi(Y_{i})^{\top}X_{n})^{2}=o_{P}(1).\label{eq:ass-3}
\end{equation}
The first part of \eqref{eq:ass-3} is an assumption about the posterior-local
behavior of $g$ and does not require classical differentiability
at $\theta^{\dagger}$. If a sampling central limit theorem for $g(\widehat{\theta}_{n})$
is to be deduced from one for $\widehat{\theta}_{n}$, we additionally
require the sampling-side asymptotic linearity condition 
\begin{equation}
\sqrt{n}\{g(\widehat{\theta}_{n})-g^{\dagger}\}-\dot{g}^{\top}\sqrt{n}(\widehat{\theta}_{n}-\theta^{\dagger})=o_{P}(1).\label{eq:func-sampling-al}
\end{equation}
This additional condition is not needed for the uniform IJVM map limit
or for consistency of its learning-rate updates. In Corollary~\ref{cor:coverage},
the required sampling central limit theorem may alternatively be verified
directly. We assume the posterior moments appearing below to exist
and to be measurable; continuity of $U_{n}$ and $W_{n}$ is additionally
assumed where invoked. The score $\psi$ and the matrix $J$ enter
the analysis, not the computation of IJVM. In smooth problems $\psi(Y)$
can be the gradient at $\theta^{\dagger}$. In non-smooth problems
it is enough that the score furnish the linearization in \eqref{eq:ass-3}.
In particular, no gradient at $\widehat{\theta}_{n}$ need exist.
A total-variation Bernstein--von Mises theorem alone does not imply
\eqref{eq:ass-2}; convergence of posterior second moments is essential.

\subsection{Sufficient conditions for non-smooth convex losses}

\label{subsec:convex-sufficient}

The next result supplies verifiable conditions for \eqref{eq:ass-1}--\eqref{eq:ass-3}
and, at the same time, the estimator central limit theorem and posterior
quantile approximation needed for coverage. The global Lipschitz assumption
on the observation-level loss is convenient rather than intrinsic
to IJVM; the abstract results below also apply to non-Lipschitz losses
such as squared error. The separate sufficient condition on the functional
$g$ permits polynomial growth instead of global Lipschitz continuity. 
\begin{thm}[Convex-loss verification]
\label{thm:convex-verification} Suppose $\Theta=\mathbb{R}^{d}$
and the following conditions hold. 
\begin{enumerate}
\item For $P$-almost every $Y$, $\ell(\cdot;Y)$ is finite and convex,
and $|\ell(\theta;Y)-\ell(\theta';Y)|\leq L(Y)\|\theta-\theta'\|$
for all $\theta,\theta'$, where $E_{P}L(Y)^{2+\delta}<\infty$ for
some $\delta>0$. The population risk is finite at $\theta^{\dagger}$,
and $\ell(\cdot;Y)$ is differentiable at $\theta^{\dagger}$ almost
surely. Set $\psi(Y)=\nabla\ell(\theta^{\dagger};Y)$. 
\item $\theta^{\dagger}$ minimizes $R$ and $R(\theta^{\dagger}+h)-R(\theta^{\dagger})=\frac{1}{2}h^{\top}Jh+o(\|h\|^{2})$
as $h\to0$, where $J$ is positive definite. The matrix $V=E_{P}\{\psi(Y)\psi(Y)^{\top}\}$
is positive definite. A measurable global empirical minimizer $\widehat{\theta}_{n}$
exists with probability tending to one. 
\item The fixed prior density $p$ is bounded on $\mathbb{R}^{d}$, continuous
and strictly positive at $\theta^{\dagger}$, and the posteriors in
\eqref{eq:gen-posterior} are proper for $\omega\in\Omega$. 
\end{enumerate}
Then 
\begin{equation}
\sqrt{n}(\widehat{\theta}_{n}-\theta^{\dagger})=-J^{-1}\frac{1}{\sqrt{n}}\sum^{n}_{i=1}\psi(Y_{i})+o_{P}(1),\label{eq:convex-mest}
\end{equation}
and, with $N_{\omega}=N_{d}(0,\omega^{-1}J^{-1})$, 
\begin{equation}
\sup_{\omega\in\Omega}\left\Vert \mathcal{L}_{\pi_{\omega}}(X_{n})-N_{\omega}\right\Vert _{\mathrm{TV}}=o_{P}(1),\qquad\sup_{\omega\in\Omega}\left\Vert \operatorname{Var}_{\pi_{\omega}}(X_{n})-\omega^{-1}J^{-1}\right\Vert =o_{P}(1).\label{eq:convex-bvm}
\end{equation}
Moreover, for every fixed $q>0$, $\sup_{\omega\in\Omega}E_{\pi_{\omega}}\|X_{n}\|^{q}=O_{P}(1)$,
and the second part of \eqref{eq:ass-3} holds with this $\psi$.
Thus \eqref{eq:ass-1}--\eqref{eq:ass-2} hold. If, in addition,
$g:\mathbb{R}^{d}\to\mathbb{R}$ is measurable, differentiable at
$\theta^{\dagger}$ with $\dot{g}=\nabla g(\theta^{\dagger})\ne0$,
and there are constants $C_{g}>0$ and $m_{g}\geq1$ such that 
\begin{equation}
|g(\theta)|\leq C_{g}(1+\|\theta\|^{m_{g}}),\qquad\theta\in\mathbb{R}^{d},\label{eq:func-growth-poly}
\end{equation}
then the functional posterior linearization in \eqref{eq:ass-3} and
the sampling-side linearization in \eqref{eq:func-sampling-al} hold.
In addition, for every fixed $q>0$, $\sup_{\omega\in\Omega}E_{\pi_{\omega}}|G_{n}|^{q}=O_{P}(1)$.
Furthermore, 
\begin{equation}
\sup_{\omega\in\Omega}\sqrt{n}\left|E_{\pi_{\omega}}\{g(\theta)\}-g(\widehat{\theta}_{n})\right|=o_{P}(1).\label{eq:convex-postmean}
\end{equation}
Consequently, for any possibly random $\omega_{n}\in\Omega$, including
a data-dependent learning rate, 
\begin{equation}
\sqrt{n}\left[E_{\pi_{\omega_{n}}}\{g(\theta)\}-g^{\dagger}\right]\rightsquigarrow N(0,B_{g}).\label{eq:convex-postmean-clt}
\end{equation}
The uniform posterior-quantile expansion \eqref{eq:quant-expansion}
also holds for every fixed $0<\beta<1$. 
\end{thm}

The theorem establishes the empirical local quadratic expansion and
uniform integrability of posterior moments jointly, even though individual
losses need not be twice differentiable. In the proof, convexity propagates
local quadratic curvature to a linear lower bound on the empirical
risk outside a root-$n$ neighborhood. This controls posterior tails
uniformly over the compact interval $\Omega$. These assumptions exclude
nonregular cases such as $\ell(\theta;Y)=|\theta|$ when $Y=0$ almost
surely: the generalized posterior then concentrates at rate $n^{-1}$,
and its centered IJ variance is zero. The theorem permits non-differentiability
of individual losses at data-dependent parameter values, including
empirical minimizers, but requires almost-sure differentiability at
the population minimizer and a locally quadratic population risk.
It therefore covers regular non-smooth convex estimation problems,
rather than nonregular problems with nonquadratic local risk or non-Gaussian
limiting posteriors.

The uniform equivalence in \eqref{eq:convex-postmean} shows that,
to first order, the generalized posterior mean and the empirical risk
minimizer have the same sampling distribution, even for non-smooth
convex losses. Consequently, the asymptotic sampling variance targeted
by the IJ estimator is $B_{g}/n$, matching the limit of $W_{n}(\omega)$
in Theorem~\ref{thm:unif-limit-of-update-map}. This provides a direct
connection between the covariance-based IJ criterion and frequentist
uncertainty about the generalized posterior mean.

\subsection{Exact-posterior update and matching roots}

\label{subsec:exact-map-match-root} 
\begin{thm}[Uniform limit of the IJVM map]
\label{thm:unif-limit-of-update-map} Under \eqref{eq:ass-1}--\eqref{eq:ass-3},
\begin{equation}
\sup_{\omega\in\Omega}|n\omega U_{n}(\omega)-A_{g}|=o_{P}(1),\qquad\sup_{\omega\in\Omega}|nW_{n}(\omega)-B_{g}|=o_{P}(1).\label{eq:limits-ab}
\end{equation}
Consequently $W_{n}$ is positive throughout $\Omega$ with probability
tending to one, and on that event 
\begin{equation}
\sup_{\omega\in\Omega}|\mathcal{T}_{n}(\omega)-\omega^{\circ}_{g}|=o_{P}(1).\label{limit-map}
\end{equation}
For any fixed or data-dependent $\omega^{(0)}\in\Omega$, if $\omega^{\circ}_{g}\in\Omega$,
then $\Pi_{\Omega}\{\mathcal{T}_{n}(\omega^{(0)})\}\xrightarrow{P}\omega^{\circ}_{g}$. 
\end{thm}

The proof uses $\psi(Y_{i})$ in place of a gradient at $\widehat{\theta}_{n}$.
The leading IJ score satisfies 
\begin{equation}
\sup_{\omega\in\Omega}\frac{1}{n}\sum^{n}_{i=1}\left\{ n\omega\xi_{i,\omega}-\dot{g}^{\top}J^{-1}\psi(Y_{i})\right\} ^{2}=o_{P}(1).\label{eq:ij-score-limit}
\end{equation}
Thus a single exact-posterior update is first-order sufficient irrespective
of the pilot rate. 
\begin{prop}[Finite-sample stability]
\label{prop:stab-of-onestep-calib} Suppose $\omega^{\circ}_{g}\in\Omega$,
and on a given realization $U_{n},W_{n}$ are well defined and 
\[
\sup_{\omega\in\Omega}|n\omega U_{n}(\omega)-A_{g}|\leq\delta_{U},\qquad\sup_{\omega\in\Omega}|nW_{n}(\omega)-B_{g}|\leq\delta_{W}<B_{g}
\]
for $\delta_{U},\delta_{W}\geq0$. Set 
\[
\varepsilon_{n}=\frac{B_{g}\delta_{U}+A_{g}\delta_{W}}{B_{g}(B_{g}-\delta_{W})},\qquad\omega^{(1)}_{n}=\Pi_{\Omega}\{\mathcal{T}_{n}(\omega^{(0)})\},\quad\omega^{(0)}\in\Omega.
\]
Then $W_{n}>0$ on $\Omega$ and 
\begin{equation}
\sup_{\omega\in\Omega}|\mathcal{T}_{n}(\omega)-\omega^{\circ}_{g}|\leq\varepsilon_{n}.\label{eq:map-bound}
\end{equation}
For every root $\omega^{*}_{n}\in\Omega$ of $U_{n}=W_{n}$, 
\begin{equation}
|\omega^{(1)}_{n}-\omega^{*}_{n}|\leq2\varepsilon_{n}.\label{eq:omega1-bound}
\end{equation}
Whether or not a root exists, the one-step residual satisfies 
\begin{equation}
n|U_{n}(\omega^{(1)}_{n})-W_{n}(\omega^{(1)}_{n})|\leq\frac{2(B_{g}+\delta_{W})}{\underline{\omega}}\varepsilon_{n}.\label{eq:onestep-calib-bound}
\end{equation}
If, in addition, $\omega^{\circ}_{g}\in\operatorname{int}(\Omega)$,
$\mathcal{T}_{n}$ is continuous, and $\varepsilon_{n}\leq\operatorname{dist}(\omega^{\circ}_{g},\partial\Omega)$,
at least one matching root exists in $\Omega$. 
\end{prop}

The bounds concern the IJ variance criterion, not finite-sample coverage.
Uniform asymptotic convergence does not provide explicit nonasymptotic
constants. 
\begin{thm}[Matching and approximate roots]
\label{thm:exit-consistent-appr-root} Under Theorem~\ref{thm:unif-limit-of-update-map},
suppose $\omega^{\circ}_{g}\in\operatorname{int}(\Omega)$ and $U_{n}-W_{n}$
is continuous on $\Omega$ with probability tending to one. Then $U_{n}=W_{n}$
has at least one solution in $\Omega$ with probability tending to
one, and all such solutions are uniformly close to $\omega^{\circ}_{g}$
in probability. Moreover, any possibly random $\widetilde{\omega}_{n}\in\Omega$
satisfying 
\begin{equation}
n|U_{n}(\widetilde{\omega}_{n})-W_{n}(\widetilde{\omega}_{n})|=o_{P}(1)\label{eq:approx-root}
\end{equation}
satisfies $\widetilde{\omega}_{n}\xrightarrow{P}\omega^{\circ}_{g}$. 
\end{thm}

These conclusions do not establish uniqueness of a finite-sample root
or convergence of the unmodified multistep iteration.

\subsection{Monte Carlo implementation}

\label{subsec:mc-impl-coverage}

Let $P_{\mathrm{joint}}$ denote the joint law of the data and all
simulation used by the procedure. The following conditions specify
Monte Carlo accuracy at the scale required for asymptotic calibration;
a fixed number of posterior draws need not satisfy them as $n$ grows. 
\begin{thm}[Computed one-step rate and verified refinement]
\label{thm:computed} Assume Theorem~\ref{thm:unif-limit-of-update-map}
and $\omega^{\circ}_{g}\in\Omega$.

\noindent\textup{(a) One-step rate.} Let $\omega^{(0)}\in\Omega$
be fixed or random, possibly selected using pilot simulation. At this
rate, suppose the estimates in \eqref{eq:comp-vars} satisfy 
\begin{align}
n|\widehat{U}_{n}(\omega^{(0)})-U_{n}(\omega^{(0)})| & =o_{P_{\mathrm{joint}}}(1),\label{eq:mc-bound-u}\\
\frac{1}{n}\sum^{n}_{i=1}\left[n\omega^{(0)}\{\widehat{\xi}_{i,\omega^{(0)}}-\xi_{i,\omega^{(0)}}\}\right]^{2} & =o_{P_{\mathrm{joint}}}(1).\label{eq:mc-bound-xi}
\end{align}
Then the denominator of \eqref{eq:onestep-update} is positive with
probability tending to one, and $\widehat{\omega}^{(1)}\xrightarrow{P_{\mathrm{joint}}}\omega^{\circ}_{g}$.

\smallskip{}
 \noindent\textup{(b) Independently verified refinement.} Suppose
also that the continuity condition of Theorem~\ref{thm:exit-consistent-appr-root}
holds. Let $\widetilde{\omega}_{n}\in\Omega$ be any candidate constructed
from the data and selection-stage simulation. Conditional on these,
freeze the candidate and generate independent validation draws from
$\pi_{\widetilde{\omega}_{n}}$. Use superscript $\mathrm{val}$ for
validation estimates and suppose 
\begin{align}
n|\widehat{U}^{\mathrm{val}}_{n}-U_{n}| & =o_{P_{\mathrm{joint}}}(1),\label{eq:val-bound-u}\\
\frac{1}{n}\sum^{n}_{i=1}\left[n\widetilde{\omega}_{n}\{\widehat{\xi}^{\mathrm{val}}_{i,\widetilde{\omega}_{n}}-\xi_{i,\widetilde{\omega}_{n}}\}\right]^{2} & =o_{P_{\mathrm{joint}}}(1),\label{eq:val-bound-xi}\\
n|\widehat{U}^{\mathrm{val}}_{n}-\widehat{W}^{\mathrm{val}}_{n}| & =o_{P_{\mathrm{joint}}}(1),\label{eq:verif-requir}
\end{align}
where arguments $\widetilde{\omega}_{n}$ are suppressed in the first
and third lines. Then $\widetilde{\omega}_{n}\xrightarrow{P_{\mathrm{joint}}}\omega^{\circ}_{g}$. 
\end{thm}

Part (b) is a statement about a candidate checked with independent
draws, not a convergence theorem for an arbitrary iterative algorithm.
\begin{cor}[Monte Carlo accuracy under independent posterior draws]
\label{cor:mc-iid} Suppose the conditions of Theorem~\ref{thm:convex-verification}
hold, and let $g$ be differentiable at $\theta^{\dagger}$ with $\dot{g}=\nabla g(\theta^{\dagger})\ne0$
and satisfy \eqref{eq:func-growth-poly}.

Let $\mathcal{F}_{n}$ be a sigma-field containing the data and any
simulation used to select an evaluated rate $\omega_{n}\in\Omega$,
where $\omega_{n}$ is $\mathcal{F}_{n}$-measurable. Conditional
on $\mathcal{F}_{n}$, suppose that $\theta^{(1)},\ldots,\theta^{(S_{n})}$
are independent draws from $\pi_{\omega_{n}}$, where $S_{n}\geq2$
is deterministic and $S_{n}\to\infty$.

Then the estimators in \eqref{eq:comp-vars} satisfy 
\begin{align}
n\{\widehat{U}_{n}(\omega_{n})-U_{n}(\omega_{n})\} & =O_{P_{\mathrm{joint}}}(S^{-1/2}_{n}),\label{eq:mc-iid-u}\\
\frac{1}{n}\sum^{n}_{i=1}\left[n\omega_{n}\{\widehat{\xi}_{i,\omega_{n}}-\xi_{i,\omega_{n}}\}\right]^{2} & =O_{P_{\mathrm{joint}}}(S^{-1}_{n}).\label{eq:mc-iid-xi}
\end{align}
Consequently, the Monte Carlo accuracy conditions \eqref{eq:mc-bound-u}
and \eqref{eq:mc-bound-xi} hold when $\omega_{n}=\omega^{(0)}$.

The same conclusions apply to independent validation draws generated
after $\widetilde{\omega}_{n}$ has been fixed. In particular, if
the number of validation draws tends to infinity, conditions \eqref{eq:val-bound-u}
and \eqref{eq:val-bound-xi} hold. 
\end{cor}

Under this corollary, the independent posterior draws must be generated
after the evaluated rate has been selected. Reusing the draws that
determine a simulation-based pilot rate requires additional justification.
For dependent MCMC draws, analogous conclusions require suitable control
of Monte Carlo error for the relevant posterior moments. 

\subsection{Frequentist coverage}

\label{subsec:coverage}

Variance consistency alone does not imply credible-interval coverage.
A sampling central limit theorem and posterior quantile approximation
are additionally needed. The convex-loss verification theorem gives
both for functionals satisfying \eqref{eq:func-growth-poly} and differentiability
at the population target, but the following statement also applies
under the more general abstract conditions. 
\begin{cor}[Coverage at a consistent selected rate]
\label{cor:coverage} Let $\widehat{\omega}_{n}\in\Omega$ be any
jointly random rate satisfying $\widehat{\omega}_{n}\xrightarrow{P_{\mathrm{joint}}}\omega^{\circ}_{g}\in\Omega$.
For $0<\alpha<1$ and $0<\beta_{L}<\beta_{U}<1$ with $\beta_{U}-\beta_{L}=1-\alpha$,
let $q_{\beta,n}(\omega)$ denote the exact $\beta$-quantile of $g(\theta)$
under $\pi_{\omega}$, and let $z_{\beta}$ be the standard normal
$\beta$-quantile. Suppose that, for $\beta\in\{\beta_{L},\beta_{U}\}$,
\begin{equation}
\sup_{\omega\in\Omega}\left|\sqrt{n}\{q_{\beta,n}(\omega)-g(\widehat{\theta}_{n})\}-z_{\beta}\sqrt{A_{g}/\omega}\right|=o_{P}(1),\label{eq:quant-expansion}
\end{equation}
and that $\sqrt{n}\{g(\widehat{\theta}_{n})-g^{\dagger}\}\rightsquigarrow N(0,B_{g})$.
Then 
\[
P_{\mathrm{joint}}\left\{ g^{\dagger}\in[q_{\beta_{L},n}(\widehat{\omega}_{n}),q_{\beta_{U},n}(\widehat{\omega}_{n})]\right\} \longrightarrow1-\alpha.
\]
The same conclusion holds for simulated endpoints $\widehat{q}_{\beta,n}$
if 
\begin{equation}
\sqrt{n}\left|\widehat{q}_{\beta,n}(\widehat{\omega}_{n})-q_{\beta,n}(\widehat{\omega}_{n})\right|=o_{P_{\mathrm{joint}}}(1),\qquad\beta\in\{\beta_{L},\beta_{U}\}.\label{eq:cond-endpoint}
\end{equation}
\end{cor}

\subsection{Two non-smooth examples}

\label{subsec:nonsmooth-examples}

\paragraph{Quantile regression with check loss.}

Let $(X,Y)$ denote a covariate-response pair, write $x=x(X)\in\mathbb{R}^{d}$
for a fixed feature map, and suppose that the conditional $\tau$-quantile
of $Y$ given $X$ is $x^{\top}\theta^{\dagger}$, where $0<\tau<1$.
The observation-level check loss is 
\[
\ell(\theta;Y,X)=\rho_{\tau}(Y-x^{\top}\theta),\qquad\rho_{\tau}(u)=u\{\tau-\mathbf{1}(u<0)\}.
\]
Suppose $E|Y|<\infty$ and $E\|x\|^{2+\delta}<\infty$ for some $\delta>0$.
Write $q(X)=x^{\top}\theta^{\dagger}$. Assume that $F_{Y\mid X}(q(X)\mid X)=\tau$
almost surely, and that the conditional density $f_{Y\mid X}(y\mid X)$
exists and is continuous at $q(X)$ almost surely. Suppose further
that there exist a constant $\eta>0$ and a nonnegative random variable
$M(X)$ such that, almost surely, 
\[
\sup_{|s|\leq\eta\|x\|}f_{Y\mid X}(q(X)+s\mid X)\leq M(X),\qquad E\{M(X)\|x\|^{2}\}<\infty.
\]
Assume that the matrices $J$ and $V$ defined below are positive
definite. The check loss is convex and Lipschitz in $\theta$ with
envelope $\|x\|$, and its almost-sure score at the risk minimizer
is 
\[
\psi(Y,X)=x\{\mathbf{1}(Y<x^{\top}\theta^{\dagger})-\tau\}.
\]
Under the correctly specified conditional-quantile restriction, the
population-risk curvature and score covariance are 
\[
J=E\!\left[f_{Y\mid X}(x^{\top}\theta^{\dagger}\mid X)xx^{\top}\right],\qquad V=\tau(1-\tau)E(xx^{\top}).
\]
For $g(\theta)=e^{\top}_{j}\theta$, where $e_{j}$ is a coordinate
vector, 
\begin{equation}
A_{g}=e^{\top}_{j}J^{-1}e_{j},\qquad B_{g}=e^{\top}_{j}J^{-1}VJ^{-1}e_{j},\qquad\omega^{\circ}_{g}=\frac{A_{g}}{B_{g}}.\label{eq:quantile-target}
\end{equation}
With a bounded prior density positive and continuous at $\theta^{\dagger}$
and an existing empirical minimizer, the convex-loss verification
theorem applies. Thus the non-differentiability of the check loss
does not prevent first-order IJVM calibration or asymptotically valid
intervals, subject to the stated Monte Carlo conditions. 

\paragraph{Support vector machine classifier with hinge loss.}

Let $x=(1,Z)^{\top}$, $Y\in\{-1,1\}$, and consider the hinge loss
\[
\ell(\theta;Y,Z)=2(1-Yx^{\top}\theta)_{+},\qquad p_{y}(z)=P(Y=y\mid Z=z).
\]
Assume $Z$ has a continuous positive density $f_{Z}$, the functions
$p_{y}$ are continuous and positive at the relevant margins, and
the population minimizer $\theta^{\dagger}$ has $\theta^{\dagger}_{2}\ne0$.
The loss is convex and Lipschitz with envelope $2\|x\|$. Provided
$E\|x\|^{2+\delta}<\infty$ and there is no point mass at either margin,
the almost-sure score is 
\[
\psi(Y,Z)=-2Yx\,\mathbf{1}\{Yx^{\top}\theta^{\dagger}<1\}.
\]
Put $z_{y}=(y-\theta^{\dagger}_{1})/\theta^{\dagger}_{2}$ and $x(z)=(1,z)^{\top}$.
Differentiating the \emph{population} hinge risk gives 
\begin{align}
J & =\frac{2}{|\theta^{\dagger}_{2}|}\sum_{y\in\{-1,1\}}p_{y}(z_{y})f_{Z}(z_{y})x(z_{y})x(z_{y})^{\top},\label{eq:hinge-j}\\
V & =4E\!\left[xx^{\top}\mathbf{1}\{Yx^{\top}\theta^{\dagger}<1\}\right].\label{eq:hinge-v}
\end{align}
The two distinct margin points and their positive weights make $J$
positive definite; under positive label probabilities on intervals,
$V$ is also positive definite. Consequently, for a bounded proper
prior that is positive and continuous at the minimizer, the convex-loss
verification theorem covers the hinge-loss classifier. The slope functional
$g(\theta)=\theta_{2}$ has the calibration target $\omega^{\circ}_{g}=A_{g}/B_{g}$
with $J$ and $V$ in \eqref{eq:hinge-j}--\eqref{eq:hinge-v}. 

\section{Simulation}

\label{sec:simulation}

\subsection{Simulation design}

\label{subsec:sim-design}

We consider four designs: homoscedastic quadratic-loss regression
(LinReg-homo), heteroscedastic quadratic-loss regression (LinReg-hetero),
median regression with check loss (QuantReg), and hinge-loss classification
(SVM). QuantReg and SVM assess the non-smooth convex-loss theory of
Section~\ref{sec:theory}. For each design and $n\in\{100,200,400,800\}$,
we construct equal-tailed $95\%$ generalized-posterior credible intervals.
Population calibration targets and implementation details are collected
in Appendix~B.

\paragraph{LinReg-homo and LinReg-hetero: linear regression.}

Following the independent-observation regression design adapted from
\citet{Wu2023}, generate 
\[
Y_{i}=X^{\top}_{i}\theta_{0}+\varepsilon_{i},\qquad X_{i}\stackrel{\mathrm{i.i.d.}}{\sim}N_{4}(0,\Sigma),\qquad\Sigma_{jk}=0.2^{|j-k|},\qquad\theta_{0}=(1,1,2,-1)^{\top},
\]
with conditionally independent errors $\varepsilon_{i}\mid X_{i}\sim N(0,\sigma^{2}_{i})$.
LinReg-homo is homoscedastic, with $\sigma_{i}=1$. In LinReg-hetero,
let $a=\Phi^{-1}(0.05)$, $b=\Phi^{-1}(0.95)$, and $\varsigma^{2}=0.05(0.01)^{2}+0.90(0.1)^{2}+0.05$,
where $\Phi(\cdot)$ denotes the standard normal distribution function.
Set 
\[
\sigma_{i}=\varsigma^{-1}\begin{cases}
0.01, & X_{i1}<a,\\
0.1, & a\leq X_{i1}\leq b,\\
1, & X_{i1}>b.
\end{cases}
\]
Thus $E(\varepsilon^{2}_{i})=1$ in both designs, but in LinReg-hetero
the conditional variance depends strongly on the first covariate.
We use $\ell_{i}(\theta)=\tfrac{1}{2}(Y_{i}-X^{\top}_{i}\theta)^{2}$
and the fixed prior $\theta\sim N_{4}(0,10^{2}I_{4})$, with inferential
target $g(\theta)=\theta_{1}$ and truth $g^{\dagger}=1$. The first-order
calibration target in \eqref{eq:target-rate} is $\omega^{\circ}_{g}=1$
for LinReg-homo and approximately $0.270$ for LinReg-hetero.

\paragraph{QuantReg: median regression with check loss.}

Following the design in Section~5.1 of \citet{Martin2022}, generate
independent pairs $(X_{i},Y_{i})$ from 
\[
X_{i}+2\sim\chi^{2}_{4},\qquad Y_{i}\mid X_{i}\sim N(2+X_{i},2^{2}),\qquad x_{i}=(1,X_{i})^{\top}.
\]
Here $\tau=1/2$, the conditional median is $x^{\top}_{i}\theta^{\dagger}$
with $\theta^{\dagger}=(2,1)^{\top}$, and the observation-level loss
is 
\[
\ell_{i}(\theta)=\rho_{\tau}(Y_{i}-x^{\top}_{i}\theta),\qquad\rho_{\tau}(u)=u\{\tau-\mathbf{1}(u<0)\}.
\]
We use the fixed proper prior $\theta\sim N_{2}(0,10^{2}I_{2})$ and
target the slope $g(\theta)=\theta_{2}$, with $g^{\dagger}=1$. The
exact first-order rate for this scalar functional is calculated in
Appendix~B; it differs from the joint-credible-region learning rate
discussed by \citet{Martin2022}.

\paragraph{SVM: hinge-loss classification.}

Following the design in Section~5.2 of \citet{Martin2022}, generate
independent observations with 
\[
Z_{i}\sim N(1,1),\quad x_{i}=(1,Z_{i})^{\top},\quad\Pr(Y_{i}=1\mid Z_{i}=z)=F_{3}(1-z),\quad Y_{i}\in\{-1,1\},
\]
where $F_{3}$ is the Student-$t$ distribution function with three
degrees of freedom and $x_{i}$ is a feature vector. The observation-level
loss is $\ell_{i}(\theta)=2\max\{1-Y_{i}x^{\top}_{i}\theta,0\}$,
and the fixed Laplace prior is 
\[
p(\theta)\propto\exp\!\left\{ -\sum^{2}_{j=1}\frac{|\theta_{j}|}{\nu\kappa_{j}}\right\} ,\qquad\nu=10,\quad\kappa_{1}=\kappa_{2}=1.
\]
These prior scales $\kappa_{j}$ are fixed across datasets. The population
minimizer has the form $\theta^{\dagger}=c^{\dagger}(1,-1)^{\top}$.
We target the slope $g(\theta)=\theta_{2}$, so $g^{\dagger}=-c^{\dagger}$.
The hinge loss is non-smooth but satisfies the convex-loss conditions
in Section~\ref{subsec:nonsmooth-examples}. The value of $c^{\dagger}$
and the population calibration target are computed in Appendix~B.

\subsection{Calibration methods and computational protocol}

\label{subsec:sim-protocol}

All methods use the same loss and prior, varying only the learning
rate $\omega\in\Omega=[0.01,10]$. IJVM starts at $\omega^{(0)}=0.5$;
IJVM($k$), $k\in\{1,2,5\}$, performs $k$ undamped updates with
fresh posterior draws at each rate, followed by a final posterior
fit. For the linear-regression designs, we compare IJVM with SafeBayes
\citep{Gruenwald2017}, FIM \citep{Lyddon2019}, WSM-I, WSM-M, and
WSM-g \citep{Onizuka2026}, and GPC \citep{Syring2019}. For QuantReg
and SVM, we compare IJVM with GPC. We do not apply the original Hessian-based
WSM plug-in computation to these two designs because, for check and
hinge losses, the observation-level Hessians do not provide a consistent
estimate of the nonzero population-risk curvature. A comparison with
a non-smooth adaptation of WSM would require a separately specified
curvature estimator and is not undertaken here. GPC uses $B=500$
multinomial bootstrap samples to calibrate $95\%$ equal-tailed credible
intervals, with at most $50$ coverage evaluations and tolerance $0.01$.
The algorithms are detailed in Appendix~B.

For LinReg-homo and LinReg-hetero, coverage and interval-length evaluations
use $5{,}000$ independent draws per fit from the exact Gaussian generalized
posterior. For QuantReg and SVM, we use data-augmentation Gibbs samplers
\citep{Kozumi2011,Polson2011}, with $1{,}000$ burn-in iterations
and $5{,}000$ retained draws per fit. For QuantReg, the asymmetric-Laplace
scale is $1/\omega$ in an unweighted fit, and the Gibbs sampler is
implemented in C++. For statistical evaluations, the stated draw budgets
apply to original-data, bootstrap-weighted, and final fits; see Appendix~B.

IJVM($k$) requires $k$ calibration posterior fits and one final
fit, whereas each GPC coverage evaluation requires $500$ bootstrap
fits. For each completed design and sample size, coverage, mean interval
length, and wall-clock times are estimated from $N_{\mathrm{rep}}=1{,}000$
independent datasets. Timing includes calibration and final fitting,
but excludes data generation and shared preprocessing. Simulations
are run in Python, with up to $63$ parallel workers on an AMD Ryzen
Threadripper 9980X CPU. 

\subsection{Simulation results}

\label{subsec:sim-result}

Table~\ref{tab:coverage} reports empirical coverage of nominal $95\%$
credible intervals. With $1{,}000$ simulated datasets, the Monte
Carlo standard error is approximately $0.007$ when the coverage probability
is $0.95$. In the homoscedastic regression design (LinReg-homo),
coverage for the three IJVM variants ranged from $0.941$ to $0.962$,
comparable to that of FIM, WSM, and GPC; SafeBayes, by contrast, substantially
undercovered. Under heteroscedasticity (LinReg-hetero), IJVM coverage
ranged from $0.926$ to $0.956$, compared with $0.914$--$0.947$
for GPC. WSM-g also achieved coverage close to the nominal level,
whereas SafeBayes, FIM, WSM-I, and WSM-M showed pronounced undercoverage.
In the non-smooth median regression design (QuantReg), IJVM coverage
improved with sample size but remained below nominal, ranging from
$0.904$ to $0.943$, while GPC maintained coverage between $0.951$
and $0.955$. In the non-smooth hinge-loss classification design (SVM),
IJVM coverage ranged from $0.937$ to $0.954$, with differences from
GPC diminishing at larger sample sizes. The interval lengths reported
in Figures~\ref{sm:fig:length-dgp1}--\ref{sm:fig:length-dgp4}
in the Appendix decreased with sample size and were generally comparable
between IJVM and GPC. Additional IJVM updates sometimes improved finite-sample
coverage. For example, at $n=100$, coverage increased from $0.904$
to $0.917$ in QuantReg and from $0.941$ to $0.950$ in LinReg-homo
when moving from one to two updates. In other settings, the differences
were smaller or mixed, and five updates did not consistently outperform
two. Thus, the benefits of further iteration are task-dependent rather
than systematic.

\begin{table}
\caption{Empirical coverage of nominal $95\%$ intervals}

\label{tab:coverage}
\centering{}%
\begin{tabular}{llr@{\extracolsep{0pt}.}lr@{\extracolsep{0pt}.}lr@{\extracolsep{0pt}.}lr@{\extracolsep{0pt}.}l}
\hline 
DGP & Method & \multicolumn{8}{c}{$n$}\tabularnewline
\cline{3-10}
 &  & \multicolumn{2}{c}{100} & \multicolumn{2}{c}{200} & \multicolumn{2}{c}{400} & \multicolumn{2}{c}{800}\tabularnewline
\hline 
\multirow{9}{*}{LinReg-homo} & SafeBayes & 0&514 & 0&516 & 0&537 & 0&529\tabularnewline
 & FIM & 0&938 & 0&940 & 0&952 & 0&962\tabularnewline
 & WSM-I & 0&943 & 0&945 & 0&954 & 0&960\tabularnewline
 & WSM-M & 0&944 & 0&946 & 0&952 & 0&962\tabularnewline
 & WSM-g & 0&944 & 0&946 & 0&955 & 0&959\tabularnewline
 & GPC & 0&956 & 0&954 & 0&962 & 0&965\tabularnewline
 & IJVM(1) & 0&941 & 0&945 & 0&953 & 0&962\tabularnewline
 & IJVM(2) & 0&950 & 0&945 & 0&955 & 0&959\tabularnewline
 & IJVM(5) & 0&945 & 0&944 & 0&953 & 0&960\tabularnewline
\hline 
\multirow{9}{*}{LinReg-hetero} & SafeBayes & 0&307 & 0&296 & 0&251 & 0&264\tabularnewline
 & FIM & 0&539 & 0&592 & 0&617 & 0&679\tabularnewline
 & WSM-I & 0&791 & 0&805 & 0&796 & 0&798\tabularnewline
 & WSM-M & 0&788 & 0&793 & 0&796 & 0&791\tabularnewline
 & WSM-g & 0&927 & 0&947 & 0&935 & 0&955\tabularnewline
 & GPC & 0&914 & 0&935 & 0&929 & 0&947\tabularnewline
 & IJVM(1) & 0&929 & 0&944 & 0&937 & 0&956\tabularnewline
 & IJVM(2) & 0&926 & 0&942 & 0&940 & 0&953\tabularnewline
 & IJVM(5) & 0&928 & 0&947 & 0&934 & 0&954\tabularnewline
\hline 
\multirow{4}{*}{QuantReg} & GPC & 0&951 & 0&953 & 0&954 & 0&955\tabularnewline
 & IJVM(1) & 0&904 & 0&926 & 0&934 & 0&937\tabularnewline
 & IJVM(2) & 0&917 & 0&929 & 0&934 & 0&943\tabularnewline
 & IJVM(5) & 0&909 & 0&924 & 0&935 & 0&938\tabularnewline
\hline 
\multirow{4}{*}{SVM} & GPC & 0&964 & 0&960 & 0&949 & 0&951\tabularnewline
 & IJVM(1) & 0&945 & 0&952 & 0&940 & 0&949\tabularnewline
 & IJVM(2) & 0&945 & 0&952 & 0&941 & 0&945\tabularnewline
 & IJVM(5) & 0&946 & 0&954 & 0&937 & 0&939\tabularnewline
\hline 
\end{tabular}
\end{table}

Table~\ref{tab:time} and Figures~\ref{sm:fig:time-dgp1}--\ref{sm:fig-time-dgp4}
in the Appendix summarize wall-clock time. IJVM(1) was faster than
GPC at all sample sizes in LinReg-homo, QuantReg, and SVM, with GPC's
median time approximately $1.7$--$30$, $14$--$37$, and $2.2$--$12$
times that of IJVM(1), respectively. In LinReg-hetero, IJVM(1) was
faster for $n\leq400$, but GPC was faster at $n=800$. The relative
computational advantage of IJVM(1) therefore depended on the design
and sample size. SafeBayes, FIM, and WSM were generally faster than
IJVM(1) in the linear-regression designs, although SafeBayes, FIM,
WSM-I, and WSM-M substantially undercovered in LinReg-hetero, whereas
WSM-g maintained coverage close to the nominal level. Additional IJVM
updates increased computation time without consistently improving
coverage or interval length.

\begin{table}
\caption{Median wall-clock time (s)}

\label{tab:time}
\centering{}%
\begin{tabular}{llr@{\extracolsep{0pt}.}lr@{\extracolsep{0pt}.}lr@{\extracolsep{0pt}.}lr@{\extracolsep{0pt}.}l}
\hline 
DGP & Method & \multicolumn{8}{c}{$n$}\tabularnewline
\cline{3-10}
 &  & \multicolumn{2}{c}{100} & \multicolumn{2}{c}{200} & \multicolumn{2}{c}{400} & \multicolumn{2}{c}{800}\tabularnewline
\hline 
\multirow{9}{*}{LinReg-homo} & SafeBayes & 0&029 & 0&063 & 0&110 & 0&200\tabularnewline
 & FIM & 0&001 & 0&002 & 0&003 & 0&003\tabularnewline
 & WSM-I & 0&001 & 0&001 & 0&002 & 0&002\tabularnewline
 & WSM-M & 0&001 & 0&001 & 0&001 & 0&001\tabularnewline
 & WSM-g & 0&001 & 0&001 & 0&001 & 0&001\tabularnewline
 & GPC & 1&214 & 1&152 & 1&173 & 1&233\tabularnewline
 & IJVM(1) & 0&041 & 0&151 & 0&349 & 0&710\tabularnewline
 & IJVM(2) & 0&083 & 0&291 & 0&686 & 1&394\tabularnewline
 & IJVM(5) & 0&200 & 0&720 & 1&666 & 3&465\tabularnewline
\hline 
\multirow{9}{*}{LinReg-hetero} & SafeBayes & 0&035 & 0&070 & 0&133 & 0&205\tabularnewline
 & FIM & 0&002 & 0&003 & 0&004 & 0&003\tabularnewline
 & WSM-I & 0&001 & 0&002 & 0&002 & 0&002\tabularnewline
 & WSM-M & 0&001 & 0&002 & 0&002 & 0&001\tabularnewline
 & WSM-g & 0&001 & 0&002 & 0&002 & 0&001\tabularnewline
 & GPC & 0&349 & 0&339 & 0&361 & 0&384\tabularnewline
 & IJVM(1) & 0&023 & 0&068 & 0&136 & 0&692\tabularnewline
 & IJVM(2) & 0&044 & 0&127 & 0&271 & 1&390\tabularnewline
 & IJVM(5) & 0&103 & 0&308 & 0&653 & 3&369\tabularnewline
\hline 
\multirow{4}{*}{QuantReg} & GPC & 2&067 & 3&659 & 6&163 & 11&983\tabularnewline
 & IJVM(1) & 0&056 & 0&109 & 0&216 & 0&872\tabularnewline
 & IJVM(2) & 0&084 & 0&165 & 0&331 & 1&449\tabularnewline
 & IJVM(5) & 0&174 & 0&339 & 0&705 & 3&200\tabularnewline
\hline 
\multirow{4}{*}{SVM} & GPC & 0&664 & 0&904 & 1&128 & 2&129\tabularnewline
 & IJVM(1) & 0&057 & 0&113 & 0&252 & 0&974\tabularnewline
 & IJVM(2) & 0&087 & 0&171 & 0&409 & 1&318\tabularnewline
 & IJVM(5) & 0&176 & 0&357 & 0&885 & 3&177\tabularnewline
\hline 
\end{tabular}
\end{table}

The number of posterior fits provides a complementary measure of computational
cost. IJVM(1) requires one pilot posterior fit for calibration and
one additional fit for final inference. By contrast, each GPC bootstrap-coverage
evaluation requires $500$ bootstrap posterior fits under our implementation.
Table~\ref{sm:tab:gpc-niter} reports median numbers of GPC coverage
evaluations of 17--20, 5, 12--13, and 2--4 for LinReg-homo, LinReg-hetero,
QuantReg, and SVM, respectively. With 500 bootstrap samples per evaluation,
these correspond to approximately 1,000--10,000 bootstrap posterior
fits across the four designs, excluding additional original-data and
final fits. Thus, one-step IJVM substantially reduces the number of
posterior fits while achieving coverage and interval lengths comparable
to GPC in these experiments. The size of the wall-clock advantage
remains dependent on the sampler and computational implementation.
The relatively short GPC runtimes in SVM reflect the small number
of coverage evaluations, Gibbs sampling restricted to observations
with nonzero bootstrap weights, and parallel computation across bootstrap
samples. Each IJVM update instead uses the full dataset.

\subsection{Ablation study}

\label{subsec:ablation-study}

\begin{figure}[H]
\caption{Empirical coverage of $95\%$ intervals for different pilot rates:
LinReg-homo}

\label{fig:ablation-pilot-dgp-1}
\begin{centering}
\subfloat[$n=100$]{
\centering{}\includegraphics[width=0.4\columnwidth]{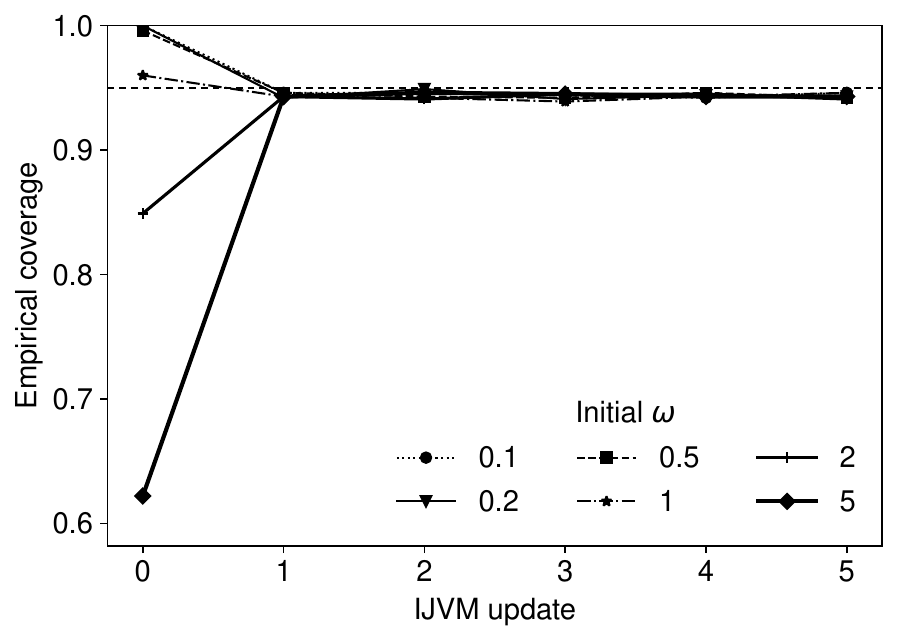}} \subfloat[$n=200$]{
\centering{}\includegraphics[width=0.4\columnwidth]{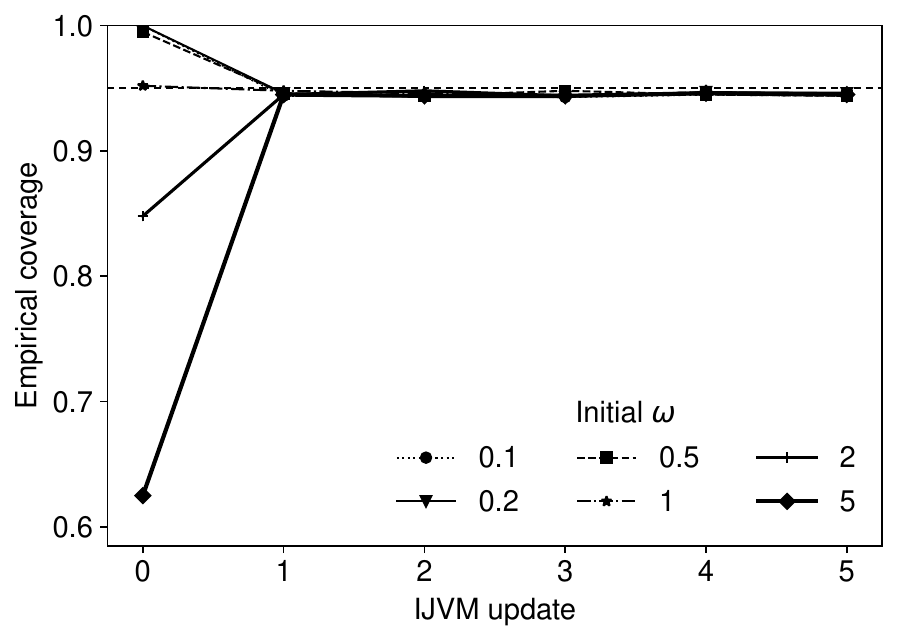}}
\par\end{centering}
\centering{}\subfloat[$n=400$]{
\centering{}\includegraphics[width=0.4\columnwidth]{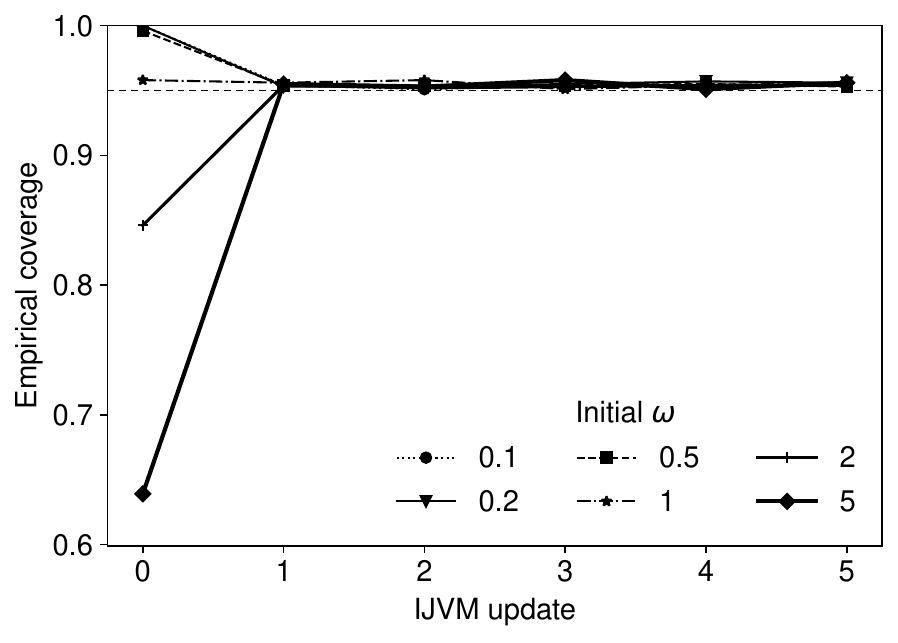}} \subfloat[$n=800$]{
\centering{}\includegraphics[width=0.4\columnwidth]{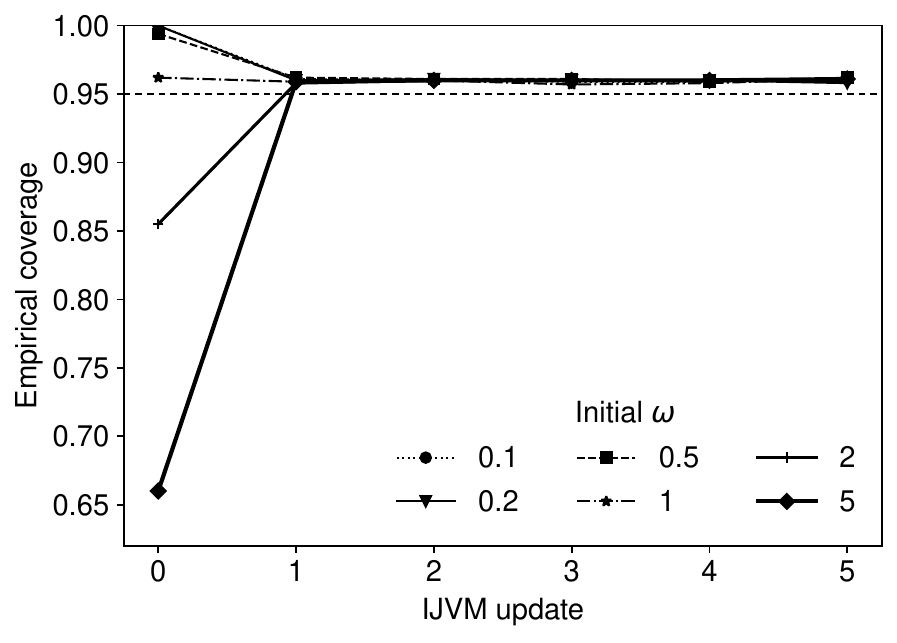}}
\end{figure}

\begin{figure}[H]

\caption{Empirical coverage of $95\%$ intervals for different pilot rates:
LinReg-hetero}

\label{fig:ablation-pilot-dgp-2}
\begin{centering}
\subfloat[$n=100$]{
\centering{}\includegraphics[width=0.4\columnwidth]{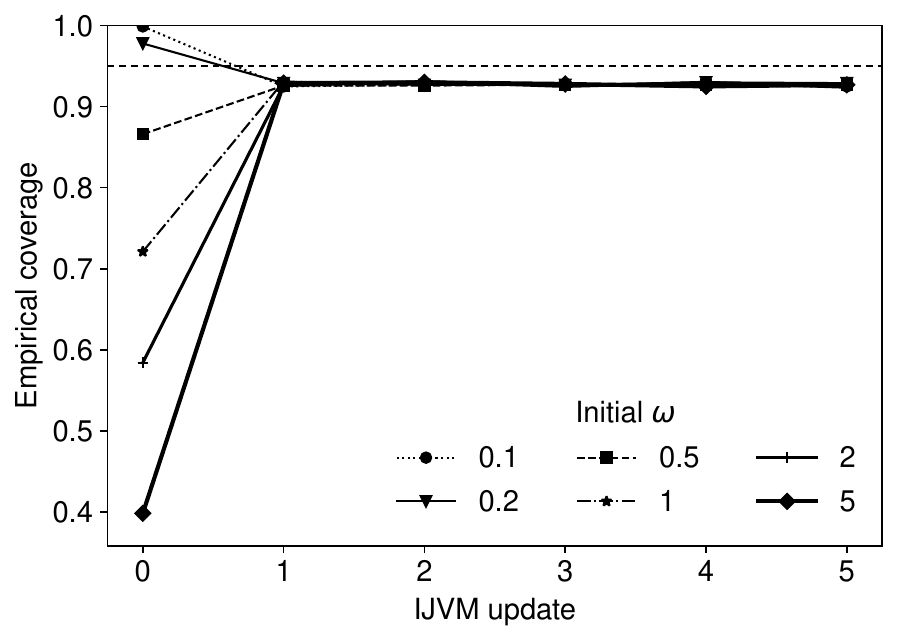}} \subfloat[$n=200$]{
\centering{}\includegraphics[width=0.4\columnwidth]{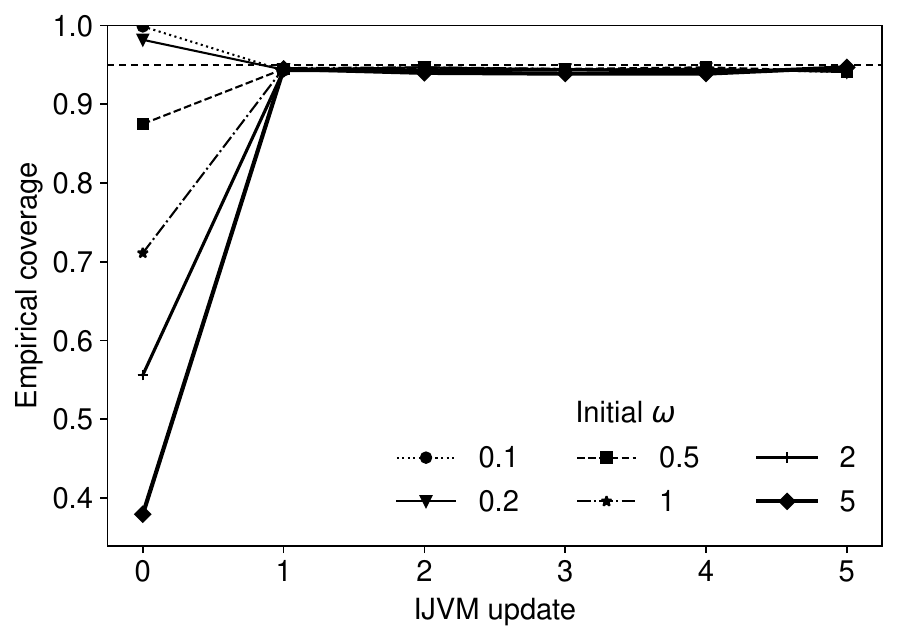}}
\par\end{centering}
\centering{}\subfloat[$n=400$]{
\centering{}\includegraphics[width=0.4\columnwidth]{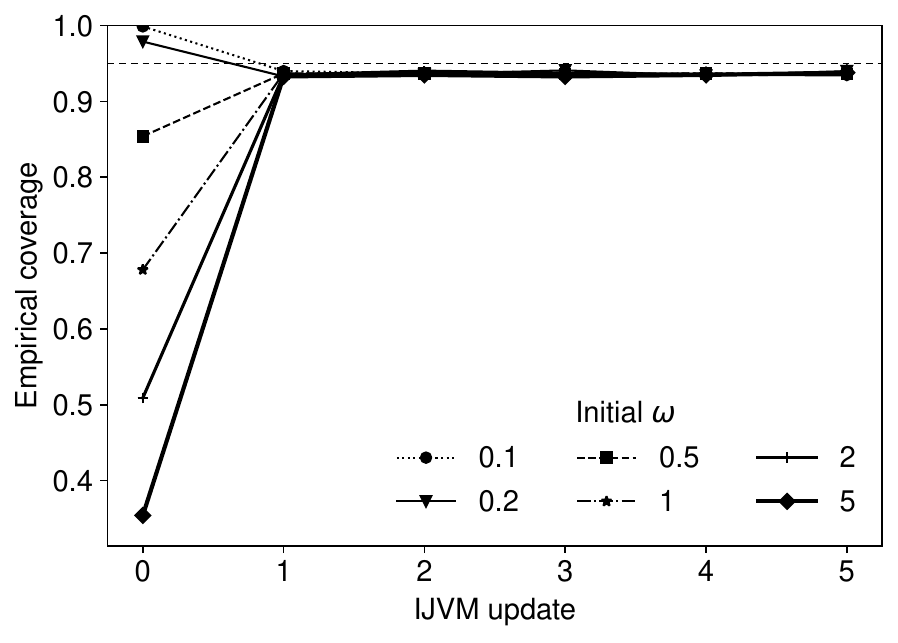}} \subfloat[$n=800$]{
\centering{}\includegraphics[width=0.4\columnwidth]{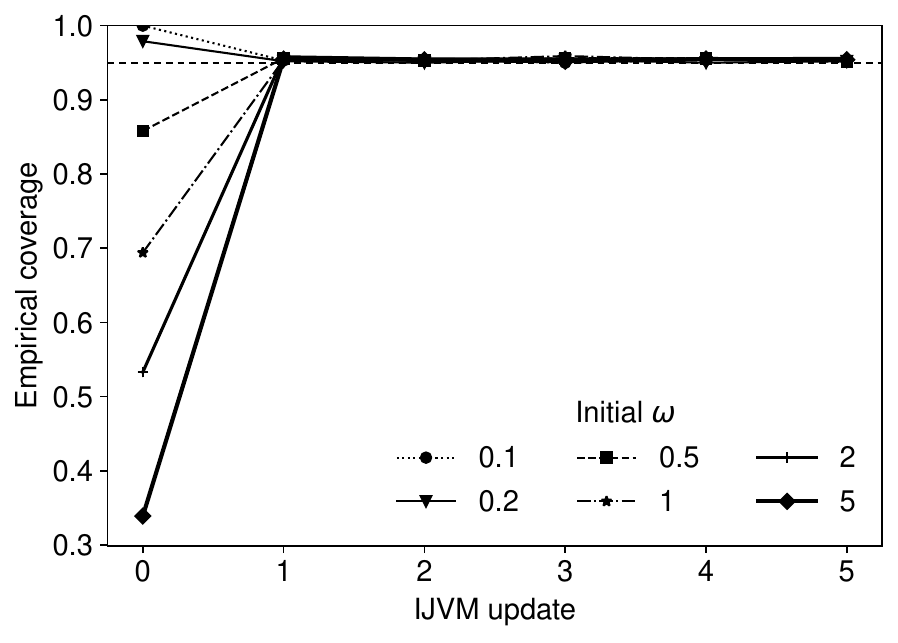}}
\end{figure}

\begin{figure}[H]

\caption{Empirical coverage of $95\%$ intervals for different pilot rates:
QuantReg}

\label{fig:ablation-pilot-dgp-3}
\begin{centering}
\subfloat[$n=100$]{
\centering{}\includegraphics[width=0.4\columnwidth]{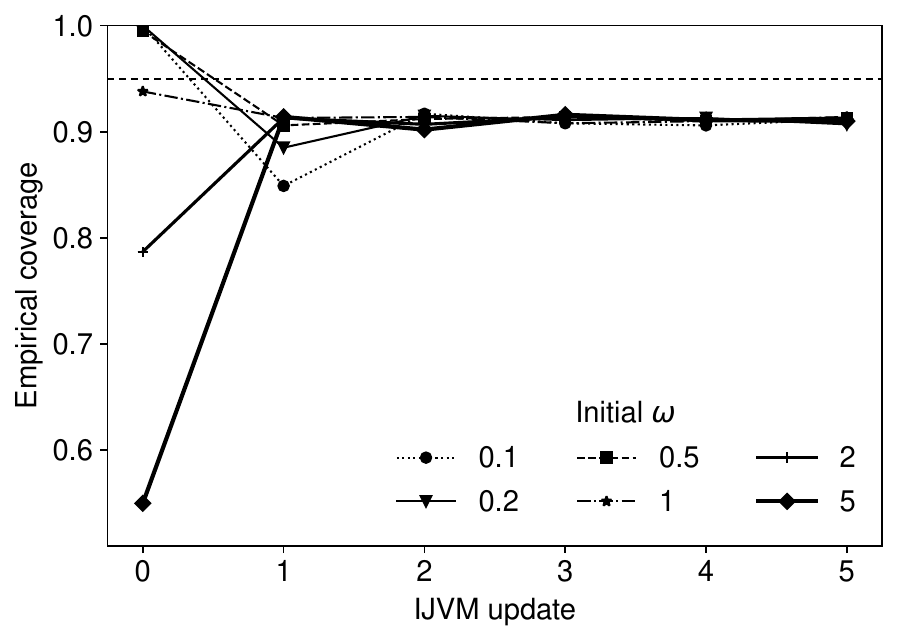}} \subfloat[$n=200$]{
\centering{}\includegraphics[width=0.4\columnwidth]{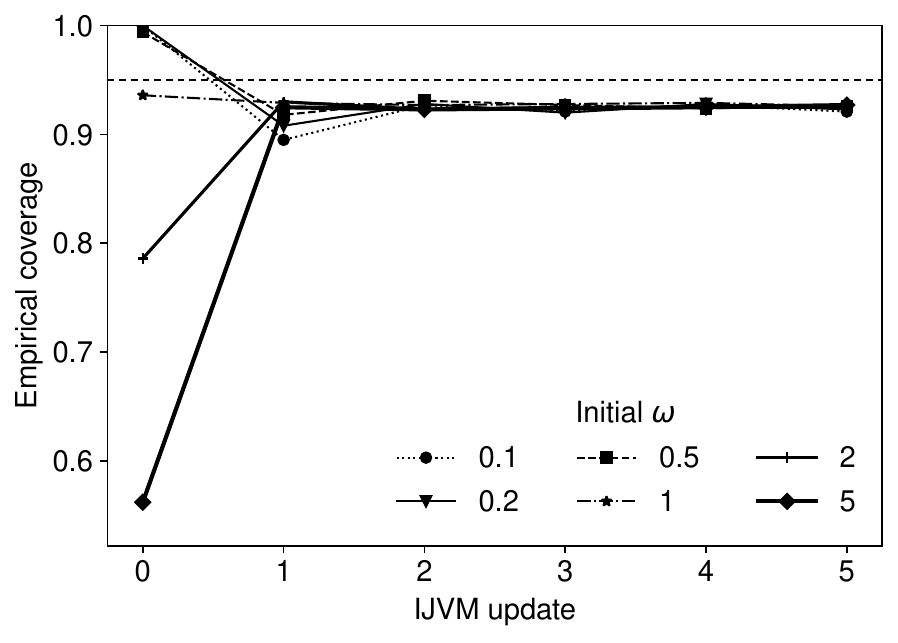}}
\par\end{centering}
\begin{centering}
\subfloat[$n=400$]{
\centering{}\includegraphics[width=0.4\columnwidth]{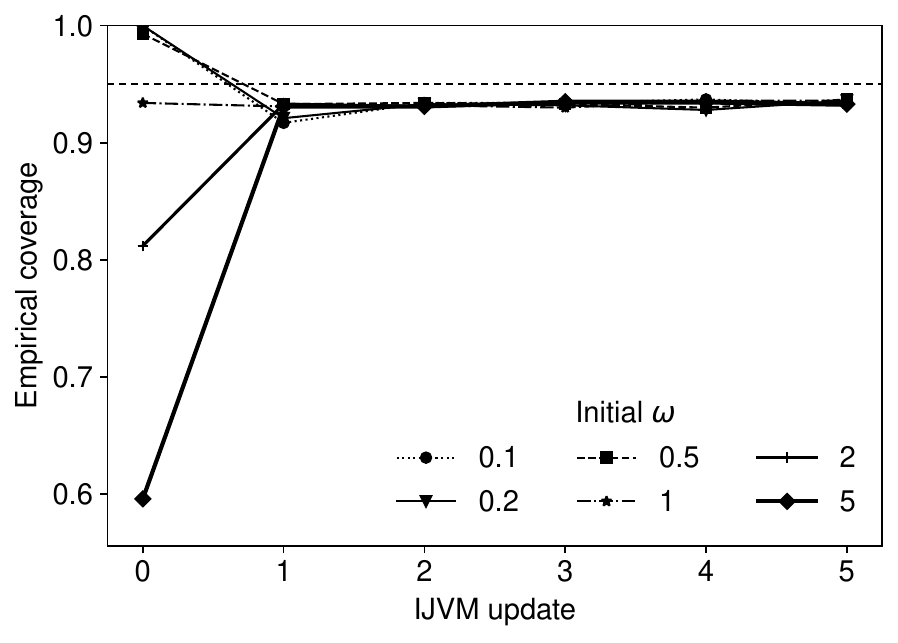}} \subfloat[$n=800$]{
\centering{}\includegraphics[width=0.4\columnwidth]{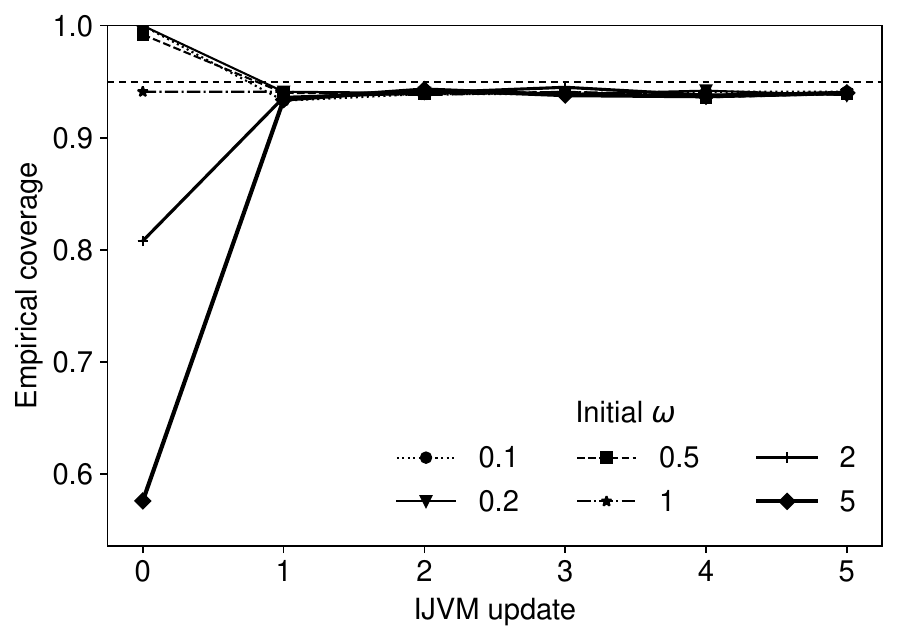}}
\par\end{centering}
\end{figure}

\begin{figure}

\caption{Empirical coverage of $95\%$ intervals for different pilot rates:
SVM}

\label{fig:ablation-pilot-dgp-4}
\begin{centering}
\subfloat[$n=100$]{
\centering{}\includegraphics[width=0.4\columnwidth]{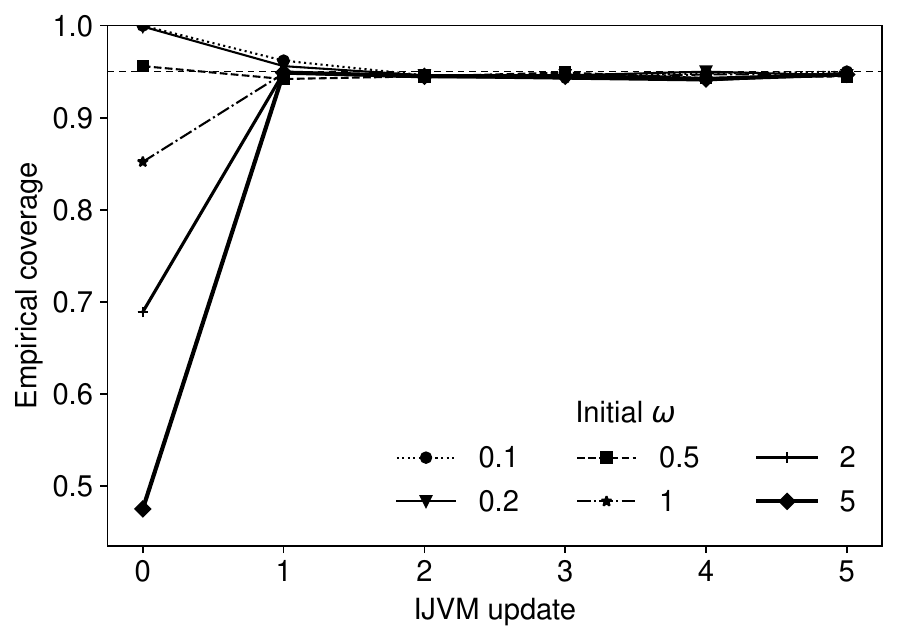}} \subfloat[$n=200$]{
\centering{}\includegraphics[width=0.4\columnwidth]{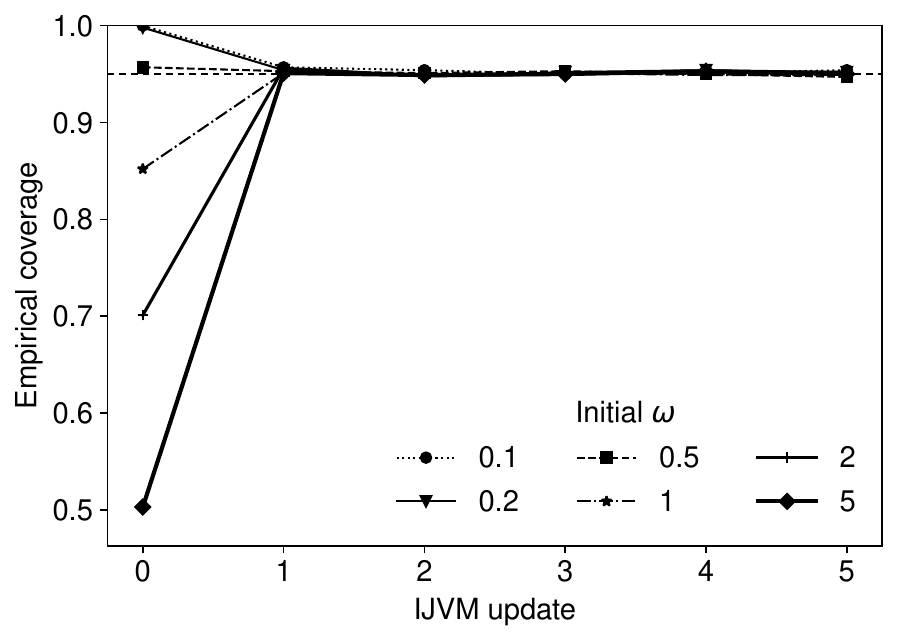}}
\par\end{centering}
\centering{}\subfloat[$n=400$]{
\centering{}\includegraphics[width=0.4\columnwidth]{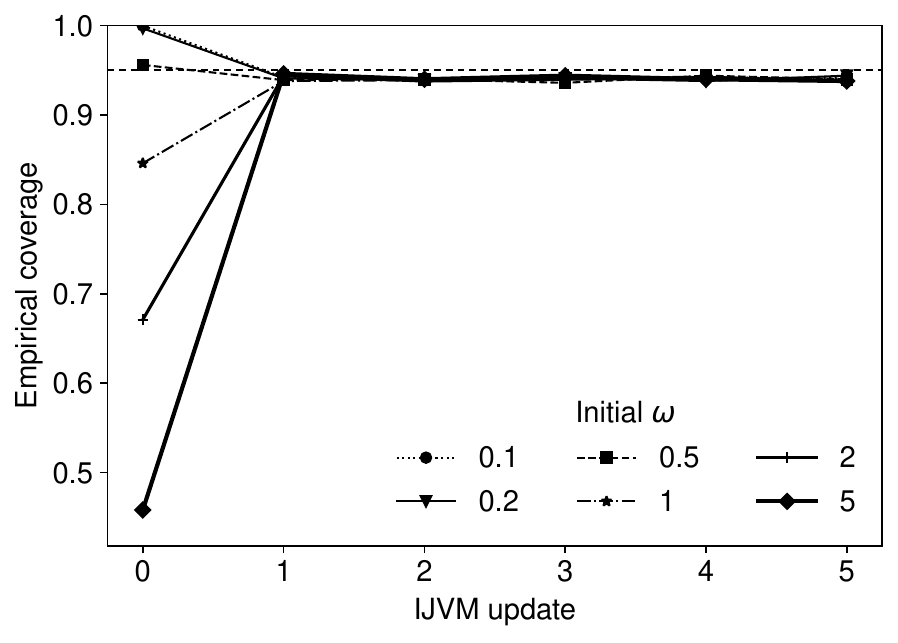}} \subfloat[$n=800$]{
\centering{}\includegraphics[width=0.4\columnwidth]{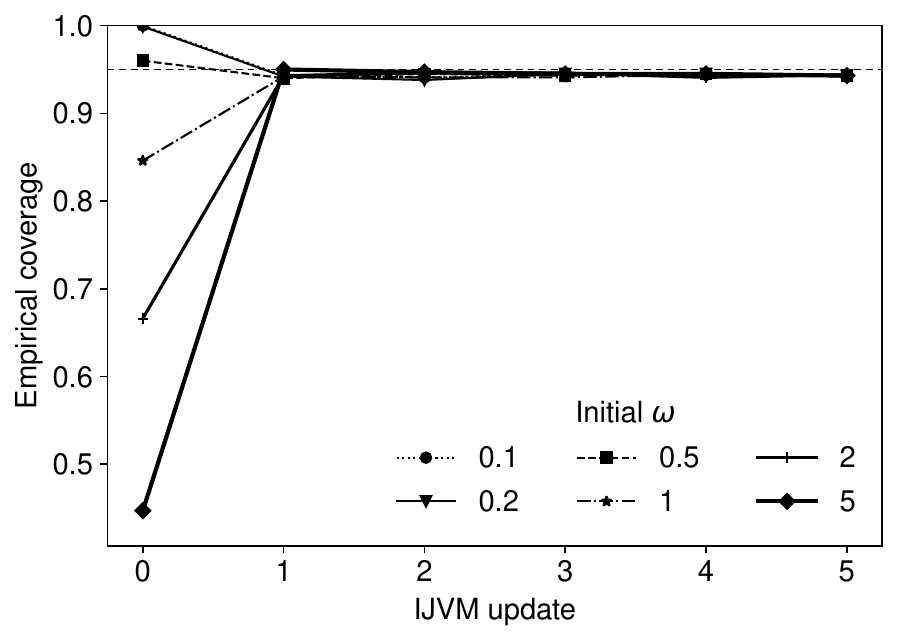}}
\end{figure}

To assess sensitivity to the pilot learning rate, we considered $\omega^{(0)}\in\{0.1,0.2,0.5,1,2,5\}$,
applying up to five IJVM updates to the same $1{,}000$ datasets for
each DGP and sample size. Figures~\ref{fig:ablation-pilot-dgp-1}--\ref{fig:ablation-pilot-dgp-4}
show that a single IJVM update substantially reduced sensitivity to
the initial learning rate across all four designs. Subsequent updates
produced little change in the two quadratic-loss regression designs
and SVM. In QuantReg, however, particularly at smaller sample sizes,
one or two further updates reduced the remaining sensitivity to the
pilot rate.

Figures~\ref{sm:fig:tau-dgp1}--\ref{sm:fig:tau-dgp4} report the
median absolute log variance ratio $d_{t}$. This diagnostic generally
decreased most sharply after the first update, with smaller improvements
thereafter. These results suggest that one-step IJVM provides an effective
initial calibration, while limited additional refinement can be worthwhile
in some finite-sample settings. The appropriate number of updates
depends on the task and the remaining variance-matching discrepancy.

\section{Discussion}

\label{sec:discussion}

This paper proposed infinitesimal-jackknife variance matching (IJVM)
to calibrate the learning rate of a generalized posterior for a prespecified
scalar functional. IJVM matches the posterior variance of the functional
to an infinitesimal-jackknife estimate of the sampling variance of
its posterior mean, using posterior covariances with observation-level
losses. The calibration calculation does not require explicit estimation
of the loss Hessian or the covariance matrix of loss gradients. The
basic procedure uses one pilot posterior fit for calibration and a
second fit for inference, thereby avoiding the repeated bootstrap
posterior fits required by generalized posterior calibration (GPC).

The theoretical results establish uniform convergence of the exact-posterior
update map to the functional-specific first-order calibration target
and consistency of the computed one-step rate under suitable Monte
Carlo conditions. The principal distinction from functional-specific
WSM is the information required to estimate this target, not the target
itself. WSM-g attained comparable coverage at lower computational
cost in the quadratic-loss regression designs, where curvature and
score-covariance estimates are easy to obtain. In contrast, IJVM requires
neither estimate as an input to calibration, which allows the same
algorithm to be applied to regular non-smooth convex losses, including
check loss and hinge loss. Here the population-risk curvature can
be nonzero even though observation-level Hessians vanish almost everywhere,
making the original WSM plug-in formula unsuitable without additional
curvature estimation. These examples therefore illustrate a broader
range of direct applications, rather than a claim of uniform superiority
over WSM or its possible non-smooth extensions.

In the simulations, IJVM achieved coverage generally close to the
nominal 95\% level in quadratic-loss regression and hinge-loss classification,
although finite-sample undercoverage remained in quantile regression.
A single update substantially reduced sensitivity to the pilot learning
rate, while one or two further updates sometimes provided additional
finite-sample improvements. These gains were task-dependent, however,
and increasing the number of updates did not necessarily improve coverage.
Thus, one-step IJVM offers an economical starting point, with further
refinement available when warranted by finite-sample diagnostics.
Compared with GPC, IJVM required substantially fewer posterior fits,
although wall-clock advantages depended on the implementation.

Several limitations remain. A rate calibrated for one scalar functional
need not calibrate other functionals or joint credible regions. The
theoretical results require regular local behavior and do not cover
nonregular losses with nonquadratic population risk or non-Gaussian
limiting posteriors. Moreover, variance matching alone does not guarantee
finite-sample coverage, and Monte Carlo consistency for dependent
MCMC draws requires additional conditions. Extensions beyond i.i.d.~clusters,
nonregular losses, and simultaneous calibration of multiple functionals
are left for future work.

\section*{Acknowledgement}

This work was supported by JSPS KAKENHI Grant Number 25K21168.

\appendix

\part*{Appendix}

\numberwithin{equation}{section}
\numberwithin{table}{section}
\numberwithin{figure}{section}

\section{Proofs}

\label{sm:proofs}

\subsection{Proof of Theorem~\ref{thm:convex-verification}}

\label{sm:proof-convex-verification}

We give the details because they establish posterior moment control
rather than just convergence in distribution. Throughout this proof,
$\ell_{i}(\theta)=\ell(\theta;Y_{i})$ and $\psi_{i}=\psi(Y_{i})$.
Differentiability at $\theta^{\dagger}$ and the Lipschitz envelope
imply $\|\psi_{i}\|\leq L(Y_{i})$ almost surely. Dominated differentiation
of $R$ at its minimizer gives $E_{P}\psi(Y)=0$, and the law of large
numbers yields \eqref{eq:ass-1}.

\paragraph{Local empirical quadraticity.}

For $t\in\mathbb{R}^{d}$ let 
\[
r_{i}(t)=\ell_{i}(\theta^{\dagger}+t)-\ell_{i}(\theta^{\dagger})-\psi^{\top}_{i}t,\qquad e_{i}(\varepsilon)=\sup_{0<\|t\|\leq\varepsilon}\frac{|r_{i}(t)|}{\|t\|}.
\]
Take a measurable version of the supremum, which can be obtained by
taking it over a countable dense set, since the loss is continuous.
By almost-sure differentiability, $e_{i}(\varepsilon)\to0$ almost
surely as $\varepsilon\downarrow0$; and $e_{i}(\varepsilon)\leq2L(Y_{i})$.
Dominated convergence therefore gives 
\begin{equation}
E_{P}\{e_{i}(\varepsilon)^{2}\}\longrightarrow0\quad\text{as }\varepsilon\downarrow0.\label{eq:proof-eps-modulus}
\end{equation}
Define $Z_{n}=n^{-1/2}\sum^{n}_{i=1}\psi_{i}$ and 
\[
H_{n}(h)=\sum^{n}_{i=1}\left\{ \ell_{i}(\theta^{\dagger}+h/\sqrt{n})-\ell_{i}(\theta^{\dagger})\right\} .
\]
For each fixed $h$, the population quadratic expansion gives 
\[
nE_{P}\{r_{i}(h/\sqrt{n})\}=\tfrac{1}{2}h^{\top}Jh+o(1),
\]
whereas independence and \eqref{eq:proof-eps-modulus} imply 
\[
\operatorname{Var}_{P}\left\{ \sum^{n}_{i=1}r_{i}(h/\sqrt{n})\right\} \leq nE_{P}\{r_{i}(h/\sqrt{n})^{2}\}\leq\|h\|^{2}E_{P}\{e_{i}(\|h\|/\sqrt{n})^{2}\}\longrightarrow0.
\]
Consequently, for fixed $h$, $H_{n}(h)=h^{\top}Z_{n}+\frac{1}{2}h^{\top}Jh+o_{P}(1)$.
Since $H_{n}(h)-h^{\top}Z_{n}$ is convex in $h$ and the limiting
quadratic is continuous, the convexity lemma for stochastic convex
functions upgrades this to uniform convergence on every compact set:
\begin{equation}
\sup_{\|h\|\leq M}\left|H_{n}(h)-h^{\top}Z_{n}-\tfrac{1}{2}h^{\top}Jh\right|=o_{P}(1),\qquad M<\infty.\label{eq:proof-convex-laq}
\end{equation}
This is the standard convexity argument used in non-smooth $M$-estimation
\citep{Pollard1991,HjortPollard2011}.

\paragraph{Empirical minimizer.}

By the multivariate central limit theorem, $Z_{n}=O_{P}(1)$. Completing
the square in \eqref{eq:proof-convex-laq}, together with positive
definiteness of $J$ and convexity, shows that every measurable global
minimizer satisfies 
\[
\sqrt{n}(\widehat{\theta}_{n}-\theta^{\dagger})+J^{-1}Z_{n}=o_{P}(1).
\]
For completeness, on events where $\|Z_{n}\|$ is bounded, the quadratic
approximation is strictly positive relative to its minimum on the
boundary of a sufficiently large fixed ball. Convexity excludes a
global minimizer outside this ball. On the ball the approximation
is uniform, and strict convexity of the limiting quadratic forces
the minimizer within $o_{P}(1)$ of $-J^{-1}Z_{n}$. This proves \eqref{eq:convex-mest}
and, by the central limit theorem, 
\begin{equation}
\sqrt{n}\{g(\widehat{\theta}_{n})-g^{\dagger}\}\rightsquigarrow N(0,B_{g})\label{eq:proof-convex-clt}
\end{equation}
whenever $g$ is differentiable at $\theta^{\dagger}$.

\paragraph{Uniform posterior normality and moment bounds.}

Write $v_{n}=\sqrt{n}(\widehat{\theta}_{n}-\theta^{\dagger})$ and
\[
K_{n}(x)=\sum^{n}_{i=1}\left\{ \ell_{i}(\widehat{\theta}_{n}+x/\sqrt{n})-\ell_{i}(\widehat{\theta}_{n})\right\} =H_{n}(v_{n}+x)-H_{n}(v_{n}).
\]
Equation \eqref{eq:proof-convex-laq} and $v_{n}=-J^{-1}Z_{n}+o_{P}(1)$
yield 
\begin{equation}
\sup_{\|x\|\leq M}|K_{n}(x)-\tfrac{1}{2}x^{\top}Jx|=o_{P}(1),\qquad M<\infty.\label{eq:proof-centered-laq}
\end{equation}
By construction $K_{n}(0)=0$, $K_{n}$ is convex, and $K_{n}(x)\geq0$.
The posterior density of $X_{n}$ is proportional to 
\[
x\longmapsto e^{-\omega K_{n}(x)}p(\widehat{\theta}_{n}+x/\sqrt{n}).
\]
For every fixed ball, \eqref{eq:proof-centered-laq} and continuity
and positivity of $p$ at $\theta^{\dagger}$ give a uniform-in-$\omega$
Gaussian approximation of this unnormalized density.

We next establish uniform control of the posterior tails, which is
needed to upgrade the local Gaussian approximation to total-variation
convergence and convergence of moments. Let $\lambda_{\min}(J)>0$
and fix $0<c<\lambda_{\min}(J)/2$. For each fixed $M\geq1$, equation
\eqref{eq:proof-centered-laq} implies 
\[
P\left\{ \inf_{\|x\|=M}K_{n}(x)\geq cM^{2}\right\} \longrightarrow1.
\]
On this event, convexity and $K_{n}(0)=0$ imply, for every $\|x\|>M$,
\begin{equation}
K_{n}(x)\geq cM\|x\|.\label{eq:proof-tail-convex}
\end{equation}

Write 
\[
Z_{n,\omega}=\int_{\mathbb{R}^{d}}e^{-\omega K_{n}(x)}p(\widehat{\theta}_{n}+x/\sqrt{n})\,dx
\]
for the normalizing integral of the rescaled posterior density. Since
$\widehat{\theta}_{n}\to_{P}\theta^{\dagger}$, $p$ is continuous
and strictly positive at $\theta^{\dagger}$, and \eqref{eq:proof-centered-laq}
holds, there exists a deterministic $c_{0}>0$ such that 
\[
P\left\{ \inf_{\omega\in\Omega}Z_{n,\omega}\geq c_{0}\right\} \longrightarrow1.
\]
Indeed, the integral over $\{\|x\|\leq1\}$ alone provides such a
lower bound, uniformly over $\omega\in\Omega$.

Let $C_{p}=\sup_{\theta}p(\theta)<\infty$ and $a=\underline{\omega}c>0$.
For any fixed $q\geq0$, the preceding bounds imply that, on an event
whose probability tends to one for each fixed $M$, 
\begin{align}
 & \sup_{\omega\in\Omega}E_{\pi_{\omega}}\left[\|X_{n}\|^{q}\mathbf{1}\{\|X_{n}\|>M\}\right]\nonumber \\
 & \quad\leq\frac{C_{p}}{c_{0}}\int_{\|x\|>M}\|x\|^{q}e^{-aM\|x\|}\,dx.\label{eq:proof-tail-moment}
\end{align}
For $M\geq1$ and $\|x\|>M$, 
\[
e^{-aM\|x\|}\leq e^{-aM^{2}/2}e^{-a\|x\|/2}.
\]
Consequently, there exists a finite constant $C_{q}$, independent
of $n$, $M$, and $\omega$, such that 
\[
\frac{C_{p}}{c_{0}}\int_{\|x\|>M}\|x\|^{q}e^{-aM\|x\|}\,dx\leq C_{q}e^{-aM^{2}/2}.
\]
In particular, for every $\varepsilon>0$, 
\begin{equation}
\lim_{M\to\infty}\limsup_{n\to\infty}P\left\{ \sup_{\omega\in\Omega}E_{\pi_{\omega}}\left[\|X_{n}\|^{q}\mathbf{1}\{\|X_{n}\|>M\}\right]>\varepsilon\right\} =0.\label{eq:proof-uniform-tail-ui}
\end{equation}
Thus posterior moments of every fixed polynomial order are uniformly
integrable in probability over $\Omega$.

On every fixed compact set, the unnormalized posterior densities converge
uniformly over $\omega\in\Omega$ to 
\[
p(\theta^{\dagger})\exp\{-\omega x^{\top}Jx/2\}.
\]
Combining this compact-set convergence with \eqref{eq:proof-uniform-tail-ui}
for $q=0$ and the uniform Gaussian tail bound shows that the posterior
normalizing integrals converge uniformly to their Gaussian counterparts.
It follows that 
\[
\sup_{\omega\in\Omega}\left\Vert \mathcal{L}_{\pi_{\omega}}(X_{n})-N_{\omega}\right\Vert _{\mathrm{TV}}=o_{P}(1).
\]

Furthermore, applying \eqref{eq:proof-uniform-tail-ui} with sufficiently
large fixed $q$, together with the corresponding uniform Gaussian
moment bounds, upgrades this total-variation convergence to convergence
of fixed-order polynomial moments. In particular, 
\[
\sup_{\omega\in\Omega}\left\Vert \operatorname{Var}_{\pi_{\omega}}(X_{n})-\omega^{-1}J^{-1}\right\Vert =o_{P}(1),
\]
and, for every fixed $q>0$, 
\[
\sup_{\omega\in\Omega}E_{\pi_{\omega}}\|X_{n}\|^{q}=O_{P}(1).
\]
This establishes \eqref{eq:convex-bvm} and the required uniform posterior
moment bounds. 

\paragraph{Averaged loss linearization.}

Let $\Delta_{ni}(x)=\sqrt{n}\{\ell_{i}(\widehat{\theta}_{n}+x/\sqrt{n})-\ell_{i}(\widehat{\theta}_{n})\}-\psi^{\top}_{i}x$.
Since $v_{n}=O_{P}(1)$, on $\|v_{n}\|\leq K$ and for $\|x\|\leq M$,
\begin{align*}
\Delta_{ni}(x) & =\sqrt{n}\left\{ r_{i}((v_{n}+x)/\sqrt{n})-r_{i}(v_{n}/\sqrt{n})\right\} ,\\
|\Delta_{ni}(x)| & \leq(2K+M)e_{i}((K+M)/\sqrt{n}).
\end{align*}
For each fixed $K,M$, Markov's inequality and \eqref{eq:proof-eps-modulus}
give 
\[
\frac{1}{n}\sum^{n}_{i=1}e_{i}((K+M)/\sqrt{n})^{2}=o_{P}(1).
\]
The corresponding averaged posterior squared error on $\{\|X_{n}\|\leq M\}$
is therefore $o_{P}(1)$ uniformly in $\omega$. On the complement,
global Lipschitz continuity and $\|\psi_{i}\|\leq L(Y_{i})$ imply
$|\Delta_{ni}(X_{n})|\leq2L(Y_{i})\|X_{n}\|$. Hence 
\begin{align*}
 & \sup_{\omega\in\Omega}\frac{1}{n}\sum^{n}_{i=1}E_{\pi_{\omega}}\left[\Delta_{ni}(X_{n})^{2}\mathbf{1}\{\|X_{n}\|>M\}\right]\\
 & \quad\leq4M^{-\delta}\left\{ \frac{1}{n}\sum^{n}_{i=1}L(Y_{i})^{2}\right\} \sup_{\omega\in\Omega}E_{\pi_{\omega}}\|X_{n}\|^{2+\delta}=O_{P}(M^{-\delta}).
\end{align*}
First choose $K$ large enough that $P(\|v_{n}\|>K)$ is small, then
let $n\to\infty$ for each fixed $M$, and finally let $M\to\infty$.
This proves the second part of \eqref{eq:ass-3}. For the functional,
write 
\[
r_{g}(t)=g(\theta^{\dagger}+t)-g^{\dagger}-\dot{g}^{\top}t.
\]
Differentiability at $\theta^{\dagger}$ implies that for every $\varepsilon>0$
there is $\eta>0$ such that $|r_{g}(t)|\leq\varepsilon\|t\|$ whenever
$\|t\|\leq\eta$. On $\{\|v_{n}\|\leq K\}$ and $\{\|\theta-\theta^{\dagger}\|\leq\eta\}$,
this gives 
\[
|G_{n}-\dot{g}^{\top}X_{n}|\leq\varepsilon(\|X_{n}\|+2\|v_{n}\|)
\]
for all sufficiently large $n$. The uniform posterior second-moment
bound and $v_{n}=O_{P}(1)$ control the squared expectation of this
term as $\varepsilon\downarrow0$. For the complementary region, the
growth condition \eqref{eq:func-growth-poly} and $\|v_{n}\|\leq K$
yield, for sufficiently large $n$, 
\[
|G_{n}-\dot{g}^{\top}X_{n}|^{2}\leq C_{K,\eta}n(1+\|X_{n}\|^{2m_{g}}),\qquad\|X_{n}\|>\eta\sqrt{n}/2,
\]
where $C_{K,\eta}$ is deterministic and independent of $n$ and $\omega$.
For any fixed $q>2m_{g}+2$, the posterior expectation of the right-hand
side, restricted to this region, is bounded by 
\[
O_{P}\bigl(n^{1-q/2}+n^{1-(q-2m_{g})/2}\bigr)=o_{P}(1)
\]
uniformly over $\omega\in\Omega$, using the established uniform posterior
$q$th-moment bound. Letting $K$ increase and then $\varepsilon$
decrease proves the first part of \eqref{eq:ass-3}. The same local--tail
decomposition, using sufficiently high fixed posterior moments of
$X_{n}$, shows $\sup_{\omega\in\Omega}E_{\pi_{\omega}}|G_{n}|^{q}=O_{P}(1)$
for any fixed $q>0$. Differentiability also gives \eqref{eq:func-sampling-al}
because $v_{n}=O_{P}(1)$. Finally, the uniform total-variation approximation
for $X_{n}$ implies that the conditional distribution functions of
$\dot{g}^{\top}X_{n}$ converge uniformly in both their argument and
$\omega\in\Omega$ to those of $N(0,A_{g}/\omega)$. The first part
of \eqref{eq:ass-3} transfers this conclusion to $G_{n}$. To see
this explicitly, for any $\eta>0$, the difference between the distribution
functions of $G_{n}$ and $\dot{g}^{\top}X_{n}$ at any point is bounded
by $\eta^{-2}E_{\pi_{\omega}}(G_{n}-\dot{g}^{\top}X_{n})^{2}$ plus
the probability that $\dot{g}^{\top}X_{n}$ belongs to an interval
of length $2\eta$. Uniform total-variation convergence bounds the
latter by $O(\eta)+o_{P}(1)$, uniformly in the argument and $\omega$,
because the limiting normal densities are uniformly bounded on $\Omega$.
Letting $n\to\infty$ and then $\eta\downarrow0$ proves the stated
uniform distribution-function convergence. Since $A_{g}>0$ and $\Omega$
is compact and bounded away from zero, the limiting normal density
is uniformly positive in a neighborhood of its $\beta$-quantile for
each fixed $0<\beta<1$. Uniform inversion therefore yields \eqref{eq:quant-expansion}.

\paragraph{Posterior-mean equivalence.}

We next establish \eqref{eq:convex-postmean}. Recall that 
\[
X_{n}=\sqrt{n}(\theta-\widehat{\theta}_{n}),\qquad G_{n}=\sqrt{n}\{g(\theta)-g(\widehat{\theta}_{n})\}.
\]
By the first part of \eqref{eq:ass-3} and Cauchy--Schwarz, 
\begin{align*}
 & \sup_{\omega\in\Omega}\left|E_{\pi_{\omega}}G_{n}-\dot{g}^{\top}E_{\pi_{\omega}}X_{n}\right|\\
 & \quad\leq\left\{ \sup_{\omega\in\Omega}E_{\pi_{\omega}}(G_{n}-\dot{g}^{\top}X_{n})^{2}\right\} ^{1/2}=o_{P}(1).
\end{align*}

It remains to show that 
\[
\sup_{\omega\in\Omega}\|E_{\pi_{\omega}}X_{n}\|=o_{P}(1).
\]
Let 
\[
d_{n}=\sup_{\omega\in\Omega}\left\Vert \mathcal{L}_{\pi_{\omega}}(X_{n})-N_{\omega}\right\Vert _{\mathrm{TV}},
\]
where $N_{\omega}=N_{d}(0,\omega^{-1}J^{-1})$. By \eqref{eq:convex-bvm},
$d_{n}=o_{P}(1)$.

For any fixed $M>0$, truncate the expectations to $\{\|x\|\leq M\}$.
Since $N_{\omega}$ has mean zero, the total-variation bound and $\|x\|\mathbf{1}\{\|x\|>M\}\leq M^{-1}\|x\|^{2}$
give 
\begin{align*}
\sup_{\omega\in\Omega}\|E_{\pi_{\omega}}X_{n}\| & \leq2Md_{n}\\
 & \quad+\frac{1}{M}\sup_{\omega\in\Omega}E_{\pi_{\omega}}\|X_{n}\|^{2}\\
 & \quad+\frac{1}{M}\sup_{\omega\in\Omega}E_{N_{\omega}}\|X\|^{2}.
\end{align*}
The second term involves a quantity that is $O_{P}(1)$ by the established
posterior moment bound, while 
\[
\sup_{\omega\in\Omega}E_{N_{\omega}}\|X\|^{2}\leq\underline{\omega}^{-1}\operatorname{tr}(J^{-1})<\infty.
\]
First letting $n\to\infty$ for fixed $M$, and then $M\to\infty$,
proves 
\[
\sup_{\omega\in\Omega}\|E_{\pi_{\omega}}X_{n}\|=o_{P}(1).
\]

Combining the preceding results yields 
\begin{align*}
 & \sup_{\omega\in\Omega}\sqrt{n}\left|E_{\pi_{\omega}}\{g(\theta)\}-g(\widehat{\theta}_{n})\right|\\
 & \qquad=\sup_{\omega\in\Omega}|E_{\pi_{\omega}}G_{n}|=o_{P}(1),
\end{align*}
which proves \eqref{eq:convex-postmean}.

For any possibly random $\omega_{n}\in\Omega$, it follows that 
\begin{align*}
 & \sqrt{n}\left[E_{\pi_{\omega_{n}}}\{g(\theta)\}-g^{\dagger}\right]\\
 & \quad=\sqrt{n}\{g(\widehat{\theta}_{n})-g^{\dagger}\}+o_{P_{\mathrm{joint}}}(1).
\end{align*}
The estimator central limit theorem \eqref{eq:proof-convex-clt} and
Slutsky's theorem therefore give \eqref{eq:convex-postmean-clt}.
\qed

\subsection{Details for the non-smooth examples}

\label{sm:proof-nonsmooth-examples}

\paragraph{Quantile regression with check loss.}

Let $q(X)=x^{\top}\theta^{\dagger}$ and $u=Y-q(X)$. For $h\in\mathbb{R}^{d}$,
put $v=x^{\top}h$. Knight's identity \citep{Knight1998} gives 
\begin{align*}
\rho_{\tau}(u-v)-\rho_{\tau}(u)={} & -v\{\tau-\mathbf{1}(u<0)\}\\
 & +\int^{v}_{0}\{\mathbf{1}(u<s)-\mathbf{1}(u<0)\}\,ds.
\end{align*}
Under the conditional-quantile restriction $F_{Y\mid X}(q(X)\mid X)=\tau$
and the assumed conditional density, we have $P(u<0\mid X)=\tau$.
Taking conditional expectations therefore yields 
\begin{align*}
 & E\{\rho_{\tau}(u-v)-\rho_{\tau}(u)\mid X\}\\
 & \quad=\int^{v}_{0}\{F_{Y\mid X}(q(X)+s\mid X)-\tau\}\,ds\\
 & \quad=v^{2}\int^{1}_{0}(1-t)f_{Y\mid X}(q(X)+tv\mid X)\,dt.
\end{align*}
The last identity holds for both positive and negative $v$.

By conditional density continuity at $q(X)$ and the local domination
condition stated in Section~\ref{subsec:nonsmooth-examples}, dominated
convergence gives, as $h\to0$, 
\begin{align*}
 & R(\theta^{\dagger}+h)-R(\theta^{\dagger})\\
 & \quad=\frac{1}{2}E\left[f_{Y\mid X}(q(X)\mid X)(x^{\top}h)^{2}\right]+o(\|h\|^{2})\\
 & \quad=\frac{1}{2}h^{\top}Jh+o(\|h\|^{2}),
\end{align*}
where 
\[
J=E\left[f_{Y\mid X}(x^{\top}\theta^{\dagger}\mid X)xx^{\top}\right].
\]
Thus the population risk has the required local quadratic expansion,
although the observation-level check loss is non-smooth.

Moreover, the check loss is differentiable at $\theta^{\dagger}$
almost surely, with score 
\[
\psi(Y,X)=x\{\mathbf{1}(Y<x^{\top}\theta^{\dagger})-\tau\}.
\]
The conditional-quantile restriction implies 
\[
E\{\psi(Y,X)\mid X\}=0
\]
and 
\[
E\{\psi(Y,X)\psi(Y,X)^{\top}\mid X\}=\tau(1-\tau)xx^{\top}.
\]
Consequently, 
\[
V=E\{\psi(Y,X)\psi(Y,X)^{\top}\}=\tau(1-\tau)E(xx^{\top}).
\]

Finally, the check loss is finite, convex, nonnegative, and globally
Lipschitz in $\theta$, with envelope $L(Y,X)=\|x\|$. Under the stated
moment, positive-definiteness, prior, and empirical-minimizer conditions,
the convex-loss verification theorem therefore applies. In particular,
for $g(\theta)=e^{\top}_{j}\theta$, it establishes the first-order
calibration target \eqref{eq:quantile-target}. 

\paragraph{Support vector machine classifier with hinge loss.}

Write $T=Z-1$. Since $T\sim N(0,1)$ and $P(Y=1\mid T=t)=F_{3}(-t)$,
the joint law of $(T,Y)$ is invariant under $(T,Y)\mapsto(-T,-Y)$.
Reparameterize the linear predictor as $\theta_{1}+\theta_{2}Z=\alpha+\beta T$,
where $\alpha=\theta_{1}+\theta_{2}$ and $\beta=\theta_{2}$. For
each fixed $\beta$, the population hinge risk is an even and convex
function of $\alpha$, and hence is minimized at $\alpha=0$. It therefore
suffices to minimize the risk along $\theta=(c,-c)^{\top}$.

Put $W=YT$. Its density is 
\[
f_{W}(w)=2\phi(w)F_{3}(-w),\qquad w\in\mathbb{R},
\]
where $\phi$ denotes the standard normal density. In particular,
$f_{W}$ is continuous and strictly positive on $\mathbb{R}$. Along
the symmetry line, the population risk is 
\[
r(c):=R(c,-c)=2E(1+cW)_{+}.
\]
Since $E|W|<\infty$ and $W$ has a continuous density, differentiation
under the expectation gives, for $c>0$, 
\[
r'(c)=2E\{W\mathbf{1}(W>-1/c)\}=2\int^{\infty}_{-1/c}wf_{W}(w)\,dw.
\]
Moreover, 
\[
r''(c)=\frac{2}{c^{3}}f_{W}(-1/c)>0,\qquad c>0.
\]
Thus $r'$ is strictly increasing on $(0,\infty)$.

At $c=0$, differentiability gives $r'(0)=2EW<0$. Indeed, for every
$w>0$, 
\[
f_{W}(w)-f_{W}(-w)=2\phi(w)\{F_{3}(-w)-F_{3}(w)\}<0,
\]
which implies $EW<0$. On the other hand, dominated convergence yields
\[
\lim_{c\to\infty}r'(c)=2E\{W\mathbf{1}(W>0)\}>0.
\]
By continuity and strict monotonicity of $r'$, there exists exactly
one $c^{\dagger}>0$ satisfying 
\[
E\{W\mathbf{1}(1+c^{\dagger}W>0)\}=0.
\]
Since $r$ is convex on $\mathbb{R}$, this $c^{\dagger}$ is its
unique global minimizer. Consequently, $\theta^{\dagger}=c^{\dagger}(1,-1)^{\top}$
is a population minimizer. Numerical one-dimensional integration gives
$c^{\dagger}=1.260499\ldots$.

The curvature calculation also applies to the general hinge-loss setting
in Section~\ref{subsec:nonsmooth-examples}, not only the simulation
design above. For $\theta_{2}\ne0$, the population score is 
\[
\nabla R(\theta)=-2\sum_{y\in\{-1,1\}}y\int_{\{z:\,y(\theta_{1}+\theta_{2}z)<1\}}x(z)\,p_{y}(z)f_{Z}(z)\,dz.
\]
The moment assumption permits formation of this score. Continuity
of $p_{y}f_{Z}$ at the two population margin points, without differentiability
of these densities, implies by taking differences of the integrals
over the moving boundaries that $\nabla R$ is differentiable at $\theta^{\dagger}$,
with derivative 
\[
J=\frac{2}{|\theta^{\dagger}_{2}|}\sum_{y\in\{-1,1\}}p_{y}(z_{y})f_{Z}(z_{y})x(z_{y})x(z_{y})^{\top},\qquad z_{y}=\frac{y-\theta^{\dagger}_{1}}{\theta^{\dagger}_{2}}.
\]
Differentiability of the gradient at $\theta^{\dagger}$ gives the
required local quadratic expansion of the population risk. Since $z_{1}\ne z_{-1}$,
the vectors $x(z_{1})$ and $x(z_{-1})$ are linearly independent.
Positivity of the two margin weights therefore gives $J\succ0$. If
$\theta^{\dagger}$ is a population minimizer, convexity and strictly
positive local curvature also imply that it is the unique minimizer.
For the simulation design, the minimizer shown above is $\theta^{\dagger}=(c^{\dagger},-c^{\dagger})^{\top}$.

The score is $\psi(Y,Z)=-2Yx\mathbf{1}(Yx^{\top}\theta^{\dagger}<1)$.
The optimality condition gives $E_{P}\psi=0$; since $Y^{2}=1$, $V$
is exactly \eqref{eq:hinge-v}. Both labels have strictly positive
conditional probability on open intervals of $Z$, which ensures that
$V$ is positive definite. For the simulation design, the envelope
$2\|x\|$ has moments of every order. Empirical hinge-risk minimization
is a feasible finite-dimensional linear program and has a minimizer;
the specified Laplace prior is bounded, proper, and positive and continuous
at $\theta^{\dagger}$. Thus all conditions of Theorem~\ref{thm:convex-verification}
are met in that design. The general hinge-loss example additionally
requires the stated moment, prior, and empirical-minimizer conditions.
For $g(\theta)=\theta_{2}$, evaluating the one-dimensional integrals
in \eqref{eq:hinge-j}--\eqref{eq:hinge-v} for the simulation design
gives \eqref{eq:hinge-target}. \qed

\subsection{Proof of Theorem~\ref{thm:unif-limit-of-update-map}}

\label{sm:subsec:proof-thm:unif-limit}

Write $r_{g,n}=G_{n}-\dot{g}^{\top}X_{n}$ and $r_{ni}=D_{ni}-\psi(Y_{i})^{\top}X_{n}$.
All covariances in this proof are under $\pi_{\omega}$ conditional
on the data. First, 
\[
n\omega U_{n}(\omega)=\omega\operatorname{Var}_{\pi_{\omega}}(G_{n}).
\]
The uniform second-moment bound in \eqref{eq:ass-2}, the first condition
in \eqref{eq:ass-3}, and Cauchy--Schwarz give 
\[
\sup_{\omega\in\Omega}\left|\operatorname{Var}_{\pi_{\omega}}(G_{n})-\dot{g}^{\top}\operatorname{Var}_{\pi_{\omega}}(X_{n})\dot{g}\right|=o_{P}(1).
\]
The variance approximation in \eqref{eq:ass-2} and $\overline{\omega}<\infty$
prove the first limit in \eqref{eq:limits-ab}.

For the IJ term, removing constants from posterior covariances gives
the exact identity 
\[
n\xi_{i,\omega}=\operatorname{Cov}_{\pi_{\omega}}(G_{n},D_{ni}).
\]
Expand $G_{n}=\dot{g}^{\top}X_{n}+r_{g,n}$ and $D_{ni}=\psi(Y_{i})^{\top}X_{n}+r_{ni}$.
Cauchy--Schwarz then yields 
\begin{equation}
\sup_{\omega\in\Omega}\frac{1}{n}\sum^{n}_{i=1}\left|n\omega\xi_{i,\omega}-\omega\dot{g}^{\top}\operatorname{Var}_{\pi_{\omega}}(X_{n})\psi(Y_{i})\right|^{2}=o_{P}(1).\label{eq:proof-cov-linear}
\end{equation}
Indeed, the averaged squared covariance between $r_{g,n}$ and $\psi(Y_{i})^{\top}X_{n}$
is bounded by a constant multiple of 
\[
\sup_{\omega\in\Omega}E_{\pi_{\omega}}r^{2}_{g,n}\;\sup_{\omega\in\Omega}E_{\pi_{\omega}}\|X_{n}\|^{2}\;\frac{1}{n}\sum^{n}_{i=1}\|\psi(Y_{i})\|^{2}=o_{P}(1).
\]
The other two remainder covariances are controlled by the averaged
second condition in \eqref{eq:ass-3} and by the product of the two
remainder second-moment bounds. By \eqref{eq:ass-1}--\eqref{eq:ass-2},
this implies 
\begin{equation}
\sup_{\omega\in\Omega}\frac{1}{n}\sum^{n}_{i=1}\left\{ n\omega\xi_{i,\omega}-t_{i}\right\} ^{2}=o_{P}(1),\qquad t_{i}=\dot{g}^{\top}J^{-1}\psi(Y_{i}).\label{eq:proof-cov-limit}
\end{equation}
This is \eqref{eq:ij-score-limit}. Set $a_{i,\omega}=n\omega\xi_{i,\omega}$
and $\overline{a}_{\omega}=n^{-1}\sum_{i}a_{i,\omega}$. Exactly,
\[
nW_{n}(\omega)=\frac{1}{n}\sum^{n}_{i=1}(a_{i,\omega}-\overline{a}_{\omega})^{2}.
\]
The centering operator is an orthogonal projection in the empirical
Euclidean norm; consequently \eqref{eq:proof-cov-limit} and Cauchy--Schwarz
show 
\[
\sup_{\omega\in\Omega}\left|nW_{n}(\omega)-\frac{1}{n}\sum^{n}_{i=1}(t_{i}-\overline{t})^{2}\right|=o_{P}(1),\qquad\overline{t}=\frac{1}{n}\sum^{n}_{i=1}t_{i}.
\]
Here $\overline{t}=o_{P}(1)$ by $E_{P}\psi(Y)=0$, while the empirical
second moment of $t_{i}$ tends to $B_{g}$ by \eqref{eq:ass-1}.
This establishes the second limit in \eqref{eq:limits-ab}. Finally,
\[
\mathcal{T}_{n}(\omega)=\frac{n\omega U_{n}(\omega)}{nW_{n}(\omega)}.
\]
Because $B_{g}>0$, the denominator is uniformly positive with probability
tending to one, and the ratio converges uniformly to $A_{g}/B_{g}$.
Projection onto $\Omega$ is continuous and fixes $\omega^{\circ}_{g}\in\Omega$,
proving the last assertion. \qed

\subsection{Proof of Proposition~\ref{prop:stab-of-onestep-calib}}

\label{sm:subsec:proof-prop:stab-of}

Let $a_{n}(\omega)=n\omega U_{n}(\omega)=A_{g}+e_{U}(\omega)$ and
$b_{n}(\omega)=nW_{n}(\omega)=B_{g}+e_{W}(\omega)$. By assumption
$b_{n}\geq B_{g}-\delta_{W}>0$, and 
\[
\mathcal{T}_{n}(\omega)-\omega^{\circ}_{g}=\frac{B_{g}e_{U}(\omega)-A_{g}e_{W}(\omega)}{B_{g}\{B_{g}+e_{W}(\omega)\}}.
\]
The error bounds give \eqref{eq:map-bound}. Projection is nonexpansive
and fixes $\omega^{\circ}_{g}$, so $|\omega^{(1)}_{n}-\omega^{\circ}_{g}|\leq\varepsilon_{n}$.
Each matching root is a fixed point of $\mathcal{T}_{n}$ and is within
$\varepsilon_{n}$ of $\omega^{\circ}_{g}$; the triangle inequality
proves \eqref{eq:omega1-bound}. For each $\omega\in\Omega$, 
\[
n|U_{n}(\omega)-W_{n}(\omega)|=\frac{nW_{n}(\omega)}{\omega}|\mathcal{T}_{n}(\omega)-\omega|.
\]
At $\omega=\omega^{(1)}_{n}$ the last factor is at most $2\varepsilon_{n}$,
proving \eqref{eq:onestep-calib-bound}. Under the final conditions,
the bounds imply $\mathcal{T}_{n}(\underline{\omega})-\underline{\omega}\geq0$
and $\mathcal{T}_{n}(\overline{\omega})-\overline{\omega}\leq0$.
Continuity gives a fixed point of $\mathcal{T}_{n}$ in $\Omega$,
and positivity of $W_{n}$ identifies it with a matching root. \qed

\subsection{Proof of Theorem~\ref{thm:exit-consistent-appr-root}}

\label{sm:subsec:proof:thm:exit-consistent}

The uniform limits in \eqref{eq:limits-ab} and $\underline{\omega}>0$
give 
\begin{equation}
\sup_{\omega\in\Omega}\left|n\{U_{n}(\omega)-W_{n}(\omega)\}-h(\omega)\right|=o_{P}(1),\qquad h(\omega)=\frac{A_{g}}{\omega}-B_{g}.\label{eq:proof-discrepancy}
\end{equation}
The function $h$ is continuous and has its unique zero at the interior
point $\omega^{\circ}_{g}$, with opposite signs on either side. Continuity
of $U_{n}-W_{n}$ and the intermediate-value theorem give existence
of a root with probability tending to one. For every $\epsilon>0$,
the compact set $\{\omega\in\Omega:|\omega-\omega^{\circ}_{g}|\geq\epsilon\}$,
when nonempty, has strictly positive minimum of $|h|$. Equation \eqref{eq:proof-discrepancy}
excludes exact and approximate roots from this set with probability
tending to one. This proves the claims. \qed

\subsection{Proof of Theorem~\ref{thm:computed}}

\label{sm:subsec:proof-thm:computed}

At any evaluated rate $\omega$, set 
\[
a_{i}=n\omega(\xi_{i,\omega}-\overline{\xi}_{\omega}),\quad e_{i}=n\omega(\widehat{\xi}_{i,\omega}-\xi_{i,\omega}),\quad\eta_{n}=\frac{1}{n}\sum^{n}_{i=1}e^{2}_{i}.
\]
Writing $\overline{e}=n^{-1}\sum_{i}e_{i}$, we have 
\[
nW_{n}(\omega)=\frac{1}{n}\sum_{i}a^{2}_{i},\qquad n\widehat{W}_{n}(\omega)=\frac{1}{n}\sum_{i}(a_{i}+e_{i}-\overline{e})^{2}.
\]
Cauchy--Schwarz and $n^{-1}\sum_{i}(e_{i}-\overline{e})^{2}\leq\eta_{n}$
yield 
\begin{equation}
n|\widehat{W}_{n}(\omega)-W_{n}(\omega)|\leq2\{nW_{n}(\omega)\}^{1/2}\eta^{1/2}_{n}+\eta_{n}.\label{eq:proof-mc-bound}
\end{equation}
Theorem~\ref{thm:unif-limit-of-update-map} implies $\sup_{\omega\in\Omega}nW_{n}(\omega)=O_{P}(1)$.
Thus \eqref{eq:mc-bound-xi}, together with \eqref{eq:mc-bound-u},
makes both scaled estimated variances asymptotically equivalent to
their exact counterparts at the possibly random pilot rate. The exact
IJ denominator is uniformly bounded away from zero on the $nW_{n}$
scale in probability. Taking the ratio and projecting therefore gives
$\widehat{\omega}^{(1)}\xrightarrow{P_{\mathrm{joint}}}\omega^{\circ}_{g}$.

For part (b), apply \eqref{eq:proof-mc-bound} at the frozen $\widetilde{\omega}_{n}$,
using the validation estimates. Equations \eqref{eq:val-bound-u}--\eqref{eq:verif-requir}
imply 
\begin{align*}
 & n|U_{n}(\widetilde{\omega}_{n})-W_{n}(\widetilde{\omega}_{n})|\\
 & \quad\leq n|U_{n}-\widehat{U}^{\mathrm{val}}_{n}|+n|\widehat{U}^{\mathrm{val}}_{n}-\widehat{W}^{\mathrm{val}}_{n}|+n|\widehat{W}^{\mathrm{val}}_{n}-W_{n}|=o_{P_{\mathrm{joint}}}(1),
\end{align*}
where the suppressed arguments are all $\widetilde{\omega}_{n}$.
The approximate-root argument applied to \eqref{eq:proof-discrepancy}
proves consistency. \qed

\subsection{Proof of Corollary~\ref{cor:mc-iid}}

\label{sm:proof-mc-iid}

Write $\widehat{\theta}_{n}$ for the empirical risk minimizer and
define 
\[
X^{(s)}_{n}=\sqrt{n}(\theta^{(s)}-\widehat{\theta}_{n}),
\]
\[
G_{s}=\sqrt{n}\{g(\theta^{(s)})-g(\widehat{\theta}_{n})\},
\]
and 
\[
D_{is}=\sqrt{n}\{\ell_{i}(\theta^{(s)})-\ell_{i}(\widehat{\theta}_{n})\}.
\]
Let $L_{i}=L(Y_{i})$. The loss Lipschitz condition gives
\[
|D_{is}-D_{it}|\leq L_{i}\|X^{(s)}_{n}-X^{(t)}_{n}\|.
\]
The functional need not be globally Lipschitz. Instead, the functional
moment bound established in the proof of Theorem~\ref{thm:convex-verification}
implies 
\[
M_{g,n,4}=\sup_{\omega\in\Omega}E_{\pi_{\omega}}|G_{n}|^{4}=O_{P}(1),\qquad Q_{n}=\sup_{\omega\in\Omega}E_{\pi_{\omega}}\|X_{n}\|^{4}=O_{P}(1).
\]
Sample variances and covariances are unchanged by additive constants.
Consequently, $n\widehat{U}_{n}(\omega_{n})$ and $n\widehat{\xi}_{i,\omega_{n}}$
can be written as second-order U-statistics with respective kernels
\[
H_{U}(s,t)=\tfrac{1}{2}(G_{s}-G_{t})^{2},\qquad H_{i}(s,t)=\tfrac{1}{2}(G_{s}-G_{t})(D_{is}-D_{it}).
\]
Their conditional expectations, given $\mathcal{F}_{n}$, are $nU_{n}(\omega_{n})$
and $n\xi_{i,\omega_{n}}$, respectively. For two conditionally independent
draws, elementary fourth-moment inequalities and Cauchy--Schwarz
yield a universal constant $C$ such that
\[
E\{H_{U}(s,t)^{2}\mid\mathcal{F}_{n}\}\leq CM_{g,n,4},
\]
\[
E\{H_{i}(s,t)^{2}\mid\mathcal{F}_{n}\}\leq CL^{2}_{i}(M_{g,n,4}Q_{n})^{1/2}.
\]
The standard second-order U-statistic variance bound now gives
\[
E\left[\{n(\widehat{U}_{n}-U_{n})\}^{2}\mid\mathcal{F}_{n}\right]\leq\frac{CM_{g,n,4}}{S_{n}},
\]
\[
E\left[\{n(\widehat{\xi}_{i,\omega_{n}}-\xi_{i,\omega_{n}})\}^{2}\mid\mathcal{F}_{n}\right]\leq\frac{CL^{2}_{i}(M_{g,n,4}Q_{n})^{1/2}}{S_{n}}.
\]
Since $\omega_{n}\leq\overline{\omega}$ and $E_{P}L(Y)^{2+\delta}<\infty$
implies $n^{-1}\sum^{n}_{i=1}L^{2}_{i}=O_{P}(1)$, we obtain
\[
E\left[\frac{1}{n}\sum^{n}_{i=1}\{n\omega_{n}(\widehat{\xi}_{i,\omega_{n}}-\xi_{i,\omega_{n}})\}^{2}\mid\mathcal{F}_{n}\right]\leq\frac{C\overline{\omega}^{2}}{S_{n}}(M_{g,n,4}Q_{n})^{1/2}\left(\frac{1}{n}\sum^{n}_{i=1}L^{2}_{i}\right).
\]
Conditional Markov's inequality and the stated $O_{P}(1)$ bounds
prove \eqref{eq:mc-iid-u} and \eqref{eq:mc-iid-xi}, with the same
rates as before.

For independent validation draws, condition on the data and the selection-stage
simulation, including the frozen candidate $\widetilde{\omega}_{n}$.
The same argument applies with $\omega_{n}=\widetilde{\omega}_{n}$
and $S_{n}$ replaced by the validation sample size. This proves the
final assertion. \qed 

\subsection{Proof of Corollary~\ref{cor:coverage}}

\label{sm:subsec:proof-cor:coverage}

Evaluate \eqref{eq:quant-expansion} at $\widehat{\omega}_{n}$. Consistency,
continuity of $\omega\mapsto\sqrt{A_{g}/\omega}$ on $\Omega$, and
$A_{g}/\omega^{\circ}_{g}=B_{g}$ give 
\[
\sqrt{n}\{q_{\beta,n}(\widehat{\omega}_{n})-g(\widehat{\theta}_{n})\}=z_{\beta}\sqrt{B_{g}}+o_{P_{\mathrm{joint}}}(1),\qquad\beta\in\{\beta_{L},\beta_{U}\}.
\]
The sampling central limit theorem and Slutsky's theorem imply limiting
coverage $\Phi(z_{\beta_{U}})-\Phi(z_{\beta_{L}})=1-\alpha$. Equation
\eqref{eq:cond-endpoint} makes the simulated endpoint errors negligible
at the root-$n$ scale, giving the final assertion. \qed 

\section{Simulation design and implementation}

\label{sm:sim-design}

Sections~\ref{subsec:sim-design} and~\ref{subsec:sim-protocol}
specify the data-generating processes, losses, priors, and simulation
protocol. Here we give the exact Gaussian linear-regression posterior,
population calibration targets, baseline calibration formulas, and
computational details.

\subsection{Regression calculations and population targets}

\label{sm:sim-dgp}

For LinReg-homo and LinReg-hetero, let $X$ be the $n\times4$ design
matrix and $y=(Y_{1},\ldots,Y_{n})^{\top}$. The Gaussian prior and
quadratic loss in Section~\ref{subsec:sim-design} imply the exact
generalized posterior 
\[
\theta\mid X,y,\omega\sim N_{4}\!\left(H^{-1}_{\omega}\omega X^{\top}y,\;H^{-1}_{\omega}\right),\qquad H_{\omega}=\omega X^{\top}X+10^{-2}I_{4}.
\]
This expression was used to generate independent draws for GPC's evaluations.
The target rate in LinReg-homo is $\omega^{\circ}_{g}=1$. For LinReg-hetero,
the population matrices for the quadratic loss are 
\[
J=E(X_{i}X^{\top}_{i})=\Sigma,\qquad V=E(\sigma^{2}_{i}X_{i}X^{\top}_{i}),
\]
and hence, for $e_{1}=(1,0,0,0)^{\top}$, 
\[
\omega^{\circ}_{g}=\frac{e^{\top}_{1}\Sigma^{-1}e_{1}}{e^{\top}_{1}\Sigma^{-1}V\Sigma^{-1}e_{1}}\approx0.270.
\]

\paragraph{QuantReg: exact population quantities.}

In QuantReg, $x=(1,X)^{\top}$, $X+2\sim\chi^{2}_{4}$, and $Y=x^{\top}\theta^{\dagger}+\varepsilon$
with $\varepsilon\sim N(0,4)$ independent of $X$. Thus $E(X)=2$,
$E(X^{2})=12$, and 
\[
M=E(xx^{\top})=\begin{pmatrix}1 & 2\\
2 & 12
\end{pmatrix},\qquad M^{-1}=\begin{pmatrix}3/2 & -1/4\\
-1/4 & 1/8
\end{pmatrix}.
\]
At the conditional median the normal density is $f_{Y\mid X}(x^{\top}\theta^{\dagger}\mid X)=1/(2\sqrt{2\pi})$.
The check-loss matrices in \eqref{eq:quantile-target} therefore reduce
to 
\[
J=\frac{M}{2\sqrt{2\pi}},\qquad V=\frac{M}{4}.
\]
For $g(\theta)=\theta_{2}$, straightforward matrix calculations give
\begin{equation}
A_{g}=\frac{\sqrt{2\pi}}{4},\qquad B_{g}=\frac{\pi}{4},\qquad\omega^{\circ}_{g}=\sqrt{\frac{2}{\pi}}\approx0.797885.\label{eq:quantreg-target}
\end{equation}
This scalar-functional oracle is not the joint-region oracle used
in the illustration of \citet{Martin2022}.

\paragraph{SVM: numerical population quantities.}

For the SVM design, write $T=Z-1$, so that $T\sim N(0,1)$ and $P(Y=1\mid T=t)=F_{3}(-t)$.
The symmetry of $(T,Y)$ under $(T,Y)\mapsto(-T,-Y)$ and convexity
of the population hinge risk imply that its minimizer has the form
$\theta^{\dagger}=(c^{\dagger},-c^{\dagger})^{\top}$. The constant
$c^{\dagger}>0$ uniquely solves 
\[
E\!\left[YT\,\mathbf{1}\{1+c^{\dagger}YT>0\}\right]=0.
\]
One-dimensional numerical integration yields $c^{\dagger}=1.260499\ldots$;
hence the slope truth is $g^{\dagger}=-1.260499\ldots$. The margin
locations are $z_{1}=1-1/c^{\dagger}$ and $z_{-1}=1+1/c^{\dagger}$.
Evaluating \eqref{eq:hinge-j}--\eqref{eq:hinge-v} gives 
\begin{equation}
A_{g}\approx2.270373,\qquad B_{g}\approx4.286019,\qquad\omega^{\circ}_{g}\approx0.529716.\label{eq:hinge-target}
\end{equation}
These numbers are population benchmarks, not empirical coverage or
timing results.

\subsection{Calibration and comparison algorithms}

\label{sm:sim-methods}

For the two quadratic-loss linear-regression designs, SafeBayes minimizes
the cumulative prequential posterior-expected loss
\[
\sum^{n}_{i=1}E_{\pi_{\omega}(\cdot\mid\mathcal{D}_{i-1})}\{\ell_{i}(\theta)\}
\]
over $\Omega$ \citep{Gruenwald2017}, where $\mathcal{D}_{0}=\varnothing$,
with the initial posterior equal to the prior. For the plug-in methods,
let $\widehat{\theta}$ be the ordinary least-squares estimate (the
empirical-risk minimizer), let $\widehat{\varepsilon}_{i}=Y_{i}-X^{\top}_{i}\widehat{\theta}$,
and define 
\[
\widehat{J}=n^{-1}\sum_{i}X_{i}X^{\top}_{i},\qquad\widehat{V}=n^{-1}\sum_{i}\widehat{\varepsilon}^{2}_{i}X_{i}X^{\top}_{i}.
\]

The four plug-in rates are 
\[
\begin{aligned}\widehat{\omega}_{\mathrm{FIM}} & =\frac{\operatorname{tr}(\widehat{J}\widehat{V}^{-1}\widehat{J})}{\operatorname{tr}(\widehat{J})}, & \widehat{\omega}_{\mathrm{WSM-I}} & =\frac{\operatorname{tr}(\widehat{J})}{\operatorname{tr}(\widehat{V})},\\
\widehat{\omega}_{\mathrm{WSM-M}} & =\frac{4}{\operatorname{tr}(\widehat{V}\widehat{J}^{-1})}, & \widehat{\omega}_{\mathrm{WSM-g}} & =\frac{e^{\top}_{1}\widehat{J}^{-1}e_{1}}{e^{\top}_{1}\widehat{J}^{-1}\widehat{V}\widehat{J}^{-1}e_{1}}.
\end{aligned}
\]
FIM matches the asymptotic Fisher-information number \citep{Lyddon2019}.
WSM-I and WSM-M use the identity and inverse-curvature weighting matrices
considered by \citet{Onizuka2026}, respectively. WSM-g is a functional-specific
rank-one specialization of the general WSM formula, included here
to compare methods with the same first-order target. For $g(\theta)=\theta_{1}$,
it uses $\widehat{Q}_{g}=\widehat{J}^{-1}e_{1}e^{\top}_{1}\widehat{J}^{-1}$.
All plug-in rates are projected onto $\Omega$, and each selected
rate is used in a final generalized-posterior fit. 

Each IJVM update uses the posterior variance and observation-level
loss covariances in \eqref{eq:comp-vars}, estimated from parameter
draws without numerical derivatives.

For GPC \citep{Syring2019}, we draw $B=500$ multinomial-bootstrap
count vectors $m^{(b)}=(m^{(b)}_{1},\ldots,m^{(b)}_{n})$ once per
dataset and retain them across learning-rate evaluations. At rate
$\omega$, let $I^{(b)}_{\omega}$ be the $95\%$ equal-tailed credible
interval for $g(\theta)$ under the weighted generalized posterior
$\pi_{\omega,m^{(b)}}(\theta)\propto\exp\{-\omega\sum_{i}m^{(b)}_{i}\ell_{i}(\theta)\}p(\theta)$.
The estimated bootstrap coverage is 
\[
\widehat{C}_{n}(\omega)=\frac{1}{B}\sum^{B}_{b=1}\mathbf{1}\{g(\widehat{\theta}_{n})\in I^{(b)}_{\omega}\},
\]
where $\widehat{\theta}_{n}$ minimizes the empirical risk on the
original dataset. The reference $g(\widehat{\theta}_{n})$ is fixed
across bootstrap samples and rate evaluations; it is computed by least
squares, check-loss minimization, or hinge-loss minimization, as appropriate.
We use \citeauthor{Kesten1958}'s (1958) stochastic search with starting
rate $0.5$, step size $l^{-0.51}$ at iteration $l$, tolerance $0.01$,
and at most $50$ coverage evaluations, selecting the evaluated rate
with the smallest $|\widehat{C}_{n}(\omega)-0.95|$. Each coverage
evaluation requires $B$ bootstrap posterior fits; the selected rate
also requires a final original-data posterior fit. We summarize the
number of GPC coverage evaluations over the $1{,}000$ statistical
replications of each completed design. 

\subsection{Posterior computation and performance measures}

\label{sm:sim-computation}

In QuantReg, let $v_{i}=1$ for an original-data fit and $v_{i}=m_{i}$
for a multinomial-bootstrap fit. With the fixed prior $\theta\sim N_{2}(0,10^{2}I_{2})$,
the target density is 
\[
\pi_{\omega,v}(\theta)\propto\exp\!\left\{ -\omega\sum_{i}v_{i}\rho_{1/2}(Y_{i}-x^{\top}_{i}\theta)\right\} \exp\!\left\{ -\frac{\theta^{\top}\theta}{2\cdot10^{2}}\right\} .
\]
We use the normal--exponential augmentation of \citet{Kozumi2011}.
The asymmetric-Laplace scale is $1/\omega$ for an unweighted fit;
bootstrap weights are absorbed into observation-specific rates, with
no additional scale parameter sampled. For each $i$ with $v_{i}>0$,
put $a_{i}=\omega v_{i}$ and introduce $z_{i}\sim\operatorname{Exp}(1)$,
so that 
\[
Y_{i}\mid\theta,z_{i},v_{i},\omega\sim N\!\left(x^{\top}_{i}\theta,\frac{8z_{i}}{a^{2}_{i}}\right).
\]
Integrating out $z_{i}$ gives a density proportional to $\exp\{-a_{i}\rho_{1/2}(Y_{i}-x^{\top}_{i}\theta)\}$.
Writing $d_{i}=a^{2}_{i}/(8z_{i})$, $Q=10^{-2}I_{2}+\sum_{i:v_{i}>0}d_{i}x_{i}x^{\top}_{i}$,
and $b=\sum_{i:v_{i}>0}d_{i}x_{i}Y_{i}$, the Gibbs full conditionals
are 
\begin{align*}
\theta\mid z,Y,X,v,\omega & \sim N_{2}(Q^{-1}b,Q^{-1}),\\
z^{-1}_{i}\mid\theta,Y_{i},x_{i},v_{i},\omega & \sim\operatorname{IG}\!\left(\frac{4}{a_{i}|Y_{i}-x^{\top}_{i}\theta|},2\right),\qquad v_{i}>0,
\end{align*}
where $\operatorname{IG}(\mu,a)$ is the inverse-Gaussian distribution
with mean $\mu$ and shape $a$. Observations with $v_{i}=0$ are
omitted. Thus the Gibbs sampler has exactly the required weighted
check-loss posterior as its marginal target. IJVM loss covariances
are computed from retained $\theta$ draws using the original losses
$\ell_{i}(\theta)$, without conditioning on $z$. 

For SVM, set $v_{i}=1$ for an original-data SVM fit and $v_{i}=m_{i}$
for a bootstrap fit with multinomial counts $m_{i}$. Introduce hinge-loss
augmentation scales $\phi_{i}$ for $v_{i}>0$ and prior augmentation
scales $\chi_{j}$. With $d_{i}=\omega v_{i}/\phi_{i}$, $c_{j}=(\nu^{2}\kappa^{2}_{j}\chi_{j})^{-1}$,
and 
\[
Q=\sum_{i:v_{i}>0}d_{i}x_{i}x^{\top}_{i}+\operatorname{diag}(c_{1},c_{2}),\qquad b=\sum_{i:v_{i}>0}d_{i}Y_{i}(1+\phi_{i})x_{i},
\]
the data-augmented Gibbs full conditionals are \citep{Polson2011}:
\begin{align*}
\theta\mid\phi,\chi,Y,X,v,\omega & \sim N_{2}(Q^{-1}b,Q^{-1}),\\
\phi^{-1}_{i}\mid\theta,Y_{i},x_{i},v_{i},\omega & \sim\operatorname{IG}\bigl(|1-Y_{i}x^{\top}_{i}\theta|^{-1},\omega v_{i}\bigr)\quad(v_{i}>0),\\
\chi^{-1}_{j}\mid\theta_{j} & \sim\operatorname{IG}\bigl(\nu\kappa_{j}/|\theta_{j}|,1\bigr),\quad j=1,2,
\end{align*}
where $\operatorname{IG}(\mu,a)$ is the inverse-Gaussian distribution
with mean $\mu$ and shape $a$. Observations with zero bootstrap
weight are omitted from the augmentation.

For the final equal-tailed interval $[L_{r},U_{r}]$ in replication
$r$, we calculate 
\[
\widehat{\mathrm{coverage}}=\frac{1}{N_{\mathrm{rep}}}\sum^{N_{\mathrm{rep}}}_{r=1}\mathbf{1}\{L_{r}\leq g^{\dagger}\leq U_{r}\},\qquad\widehat{\mathrm{length}}=\frac{1}{N_{\mathrm{rep}}}\sum^{N_{\mathrm{rep}}}_{r=1}(U_{r}-L_{r}).
\]
Wall-clock time covers learning-rate selection and the final posterior
fit, but excludes data generation and method-independent preprocessing.
Coverage and interval length are evaluated on all $1{,}000$ datasets
for these designs. For GPC, we report the minimum, median, and maximum
number of bootstrap-coverage evaluations across the $1{,}000$ statistical
replications. 

\section{Additional simulation results}

\label{sm:sec:add-sim-result}

\begin{figure}[H]
\caption{Length of $95\%$ credible intervals: LinReg-homo}

\label{sm:fig:length-dgp1}
\begin{centering}
\subfloat[$n=100$]{
\centering{}\includegraphics[width=0.4\columnwidth]{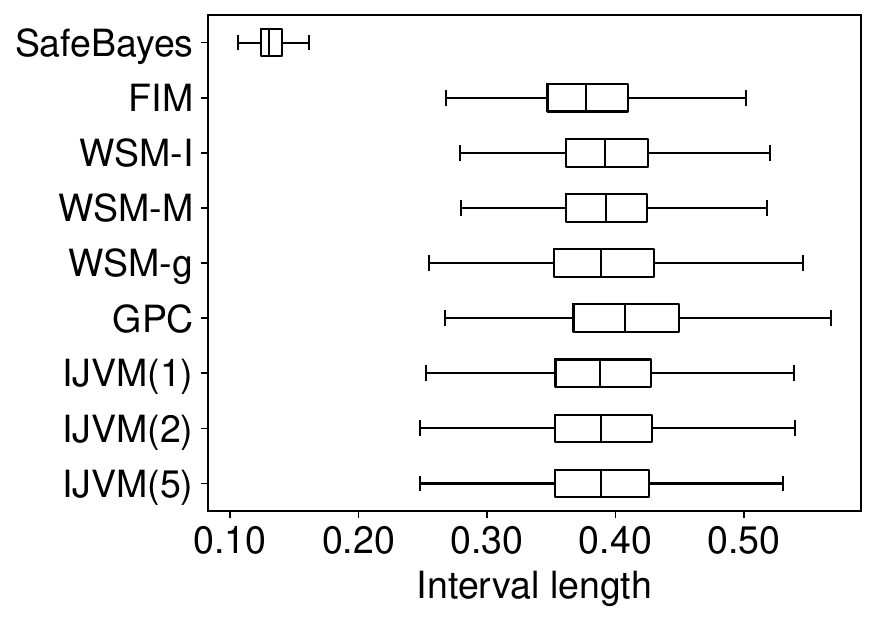}} \subfloat[$n=200$]{
\centering{}\includegraphics[width=0.4\columnwidth]{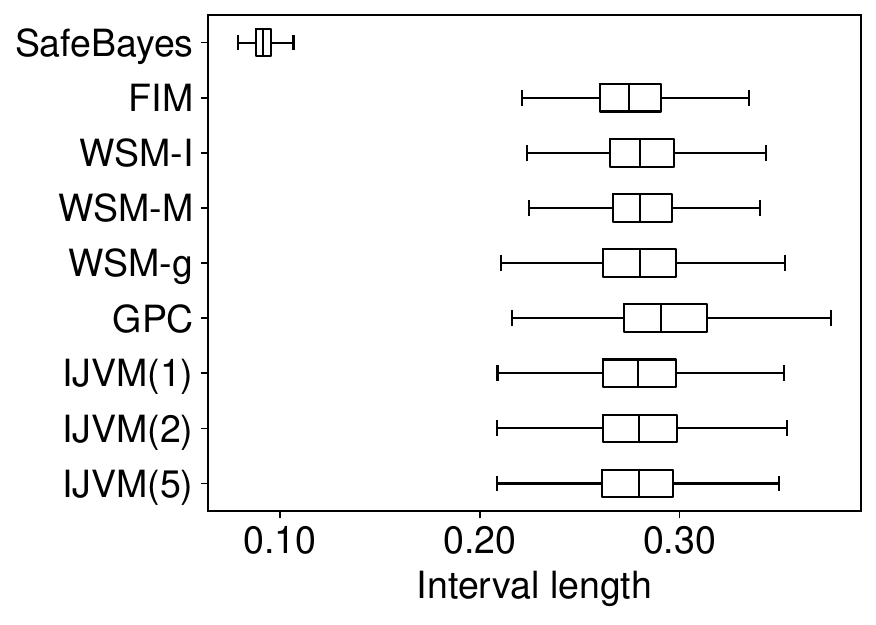}}
\par\end{centering}
\centering{}\subfloat[$n=400$]{
\centering{}\includegraphics[width=0.4\columnwidth]{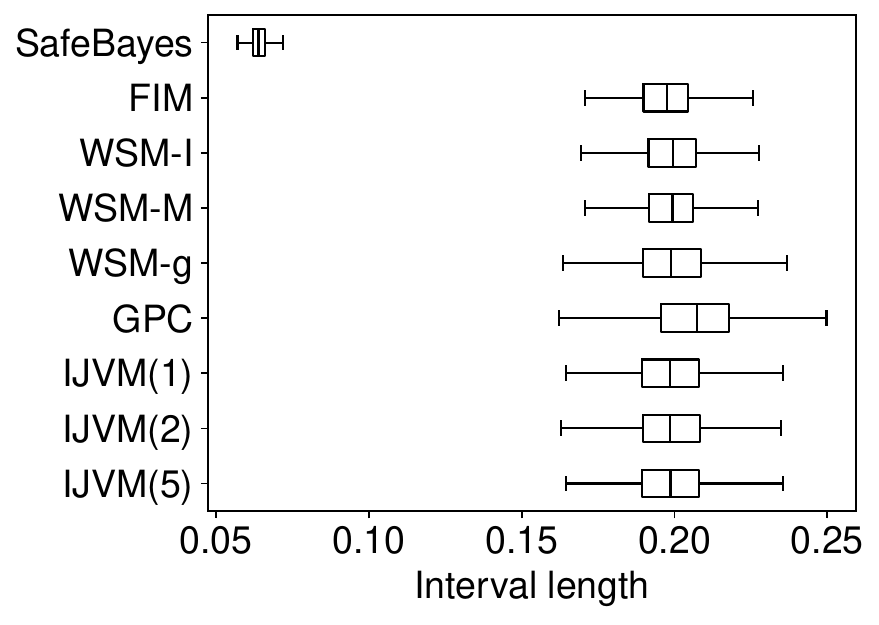}} \subfloat[$n=800$]{
\centering{}\includegraphics[width=0.4\columnwidth]{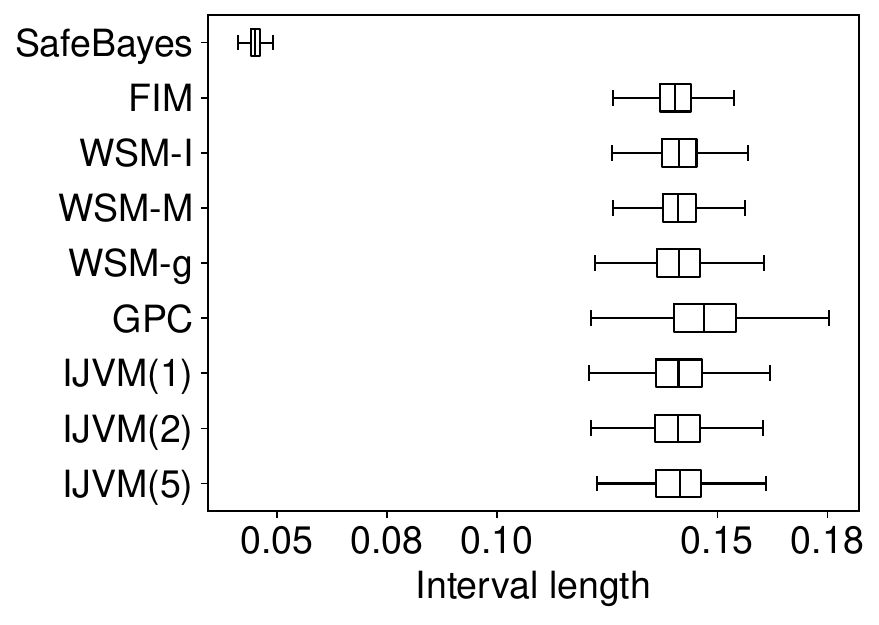}}
\end{figure}

\begin{figure}[H]

\caption{Length of $95\%$ credible intervals: LinReg-hetero}

\label{sm:fig:length-dgp2}
\begin{centering}
\subfloat[$n=100$]{
\centering{}\includegraphics[width=0.4\columnwidth]{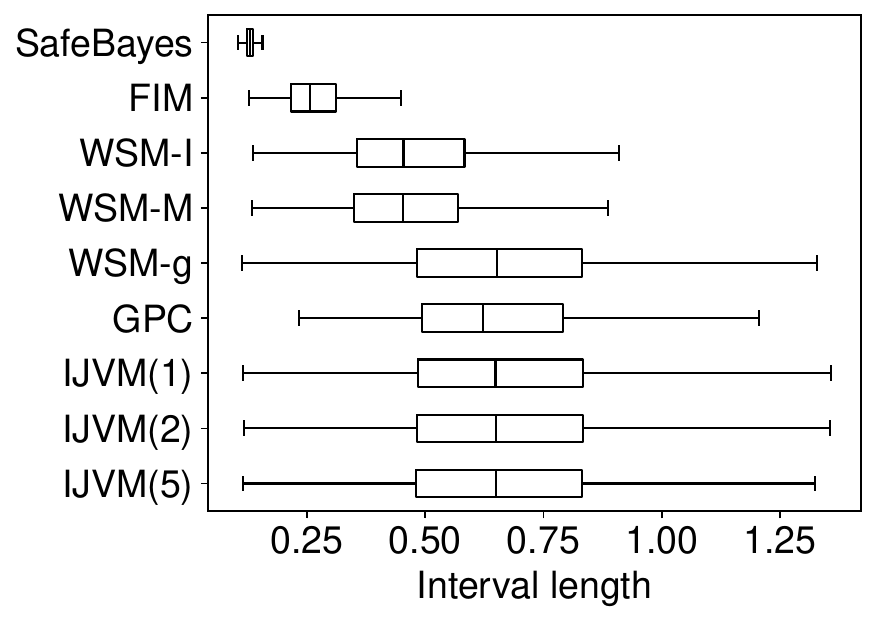}} \subfloat[$n=200$]{
\centering{}\includegraphics[width=0.4\columnwidth]{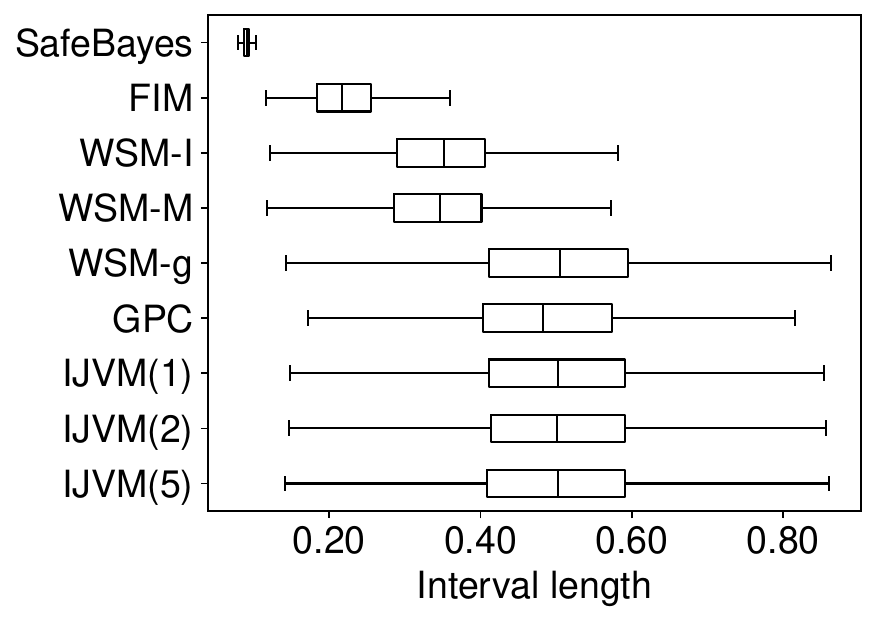}}
\par\end{centering}
\centering{}\subfloat[$n=400$]{
\centering{}\includegraphics[width=0.4\columnwidth]{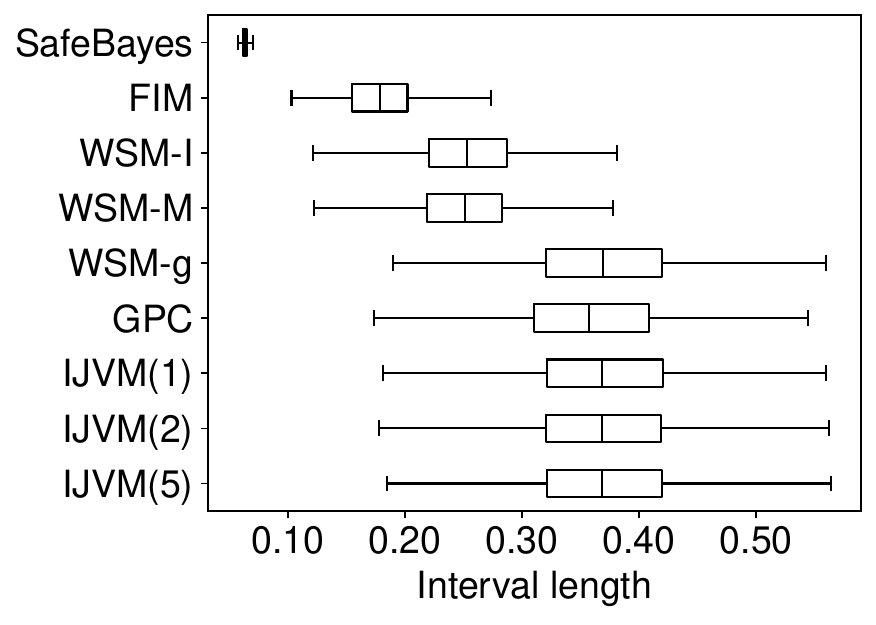}} \subfloat[$n=800$]{
\centering{}\includegraphics[width=0.4\columnwidth]{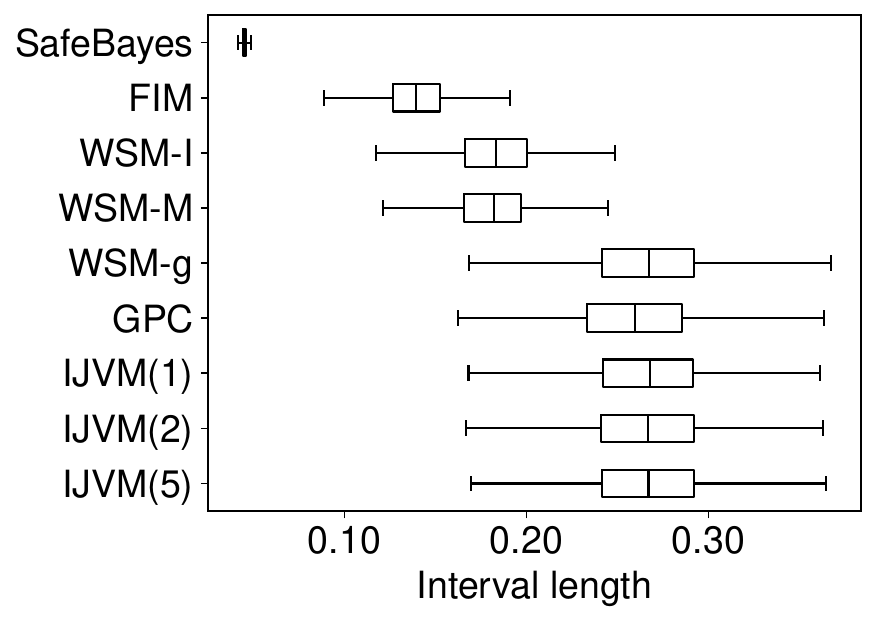}}
\end{figure}

\begin{figure}[H]
\caption{Length of $95\%$ credible intervals: QuantReg}

\label{sm:fig:length-dgp3}
\begin{centering}
\subfloat[$n=100$]{
\centering{}\includegraphics[width=0.4\columnwidth]{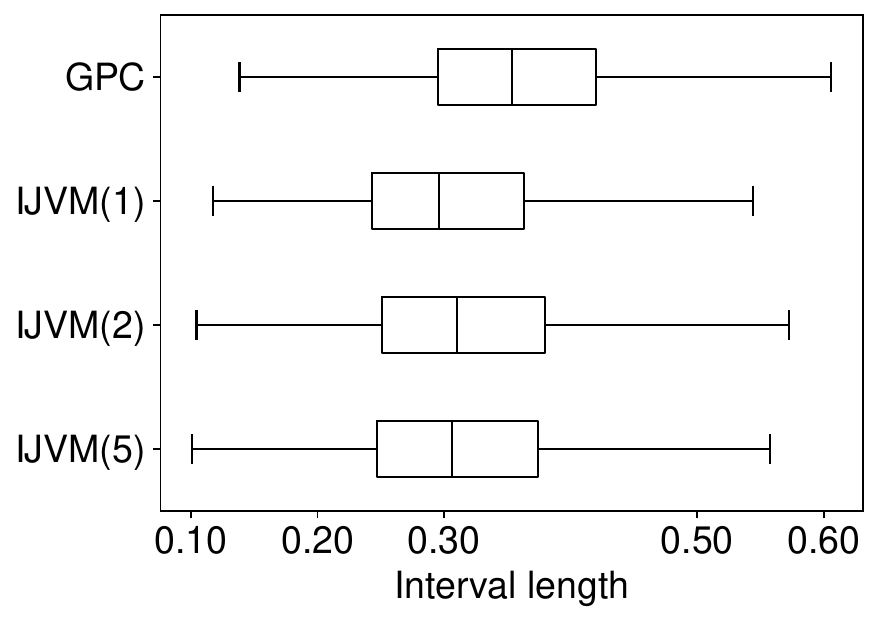}} \subfloat[$n=200$]{
\centering{}\includegraphics[width=0.4\columnwidth]{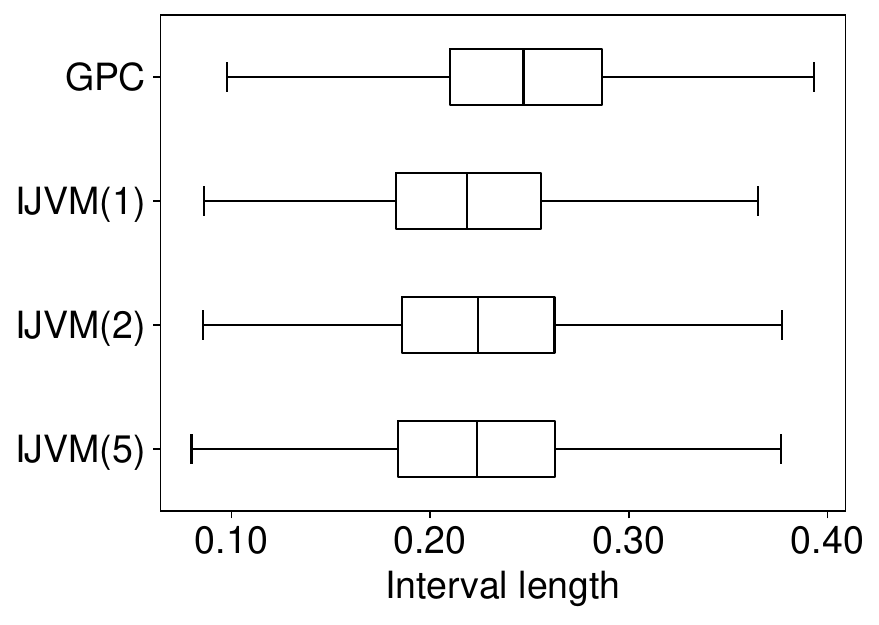}}
\par\end{centering}
\centering{}\subfloat[$n=400$]{
\centering{}\includegraphics[width=0.4\columnwidth]{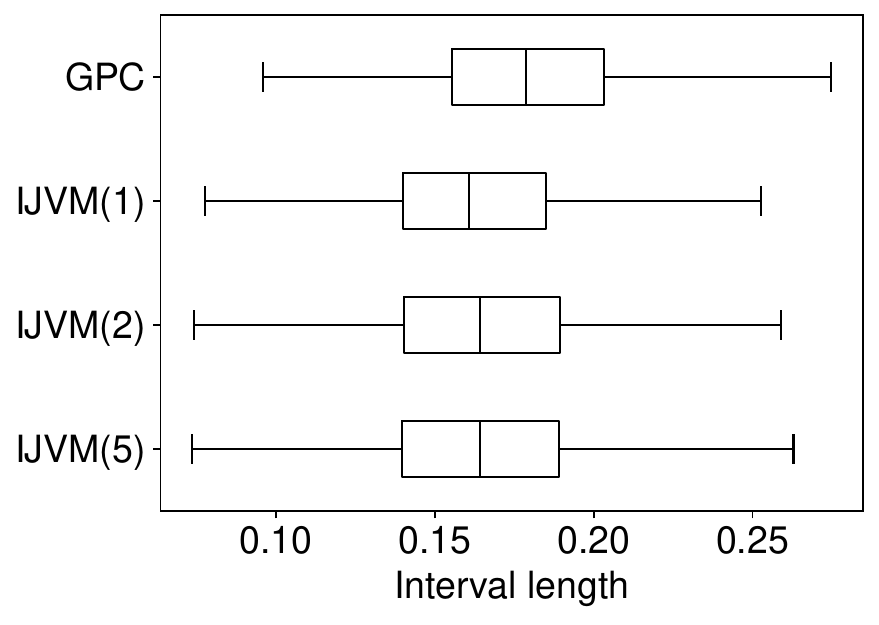}} \subfloat[$n=800$]{
\centering{}\includegraphics[width=0.4\columnwidth]{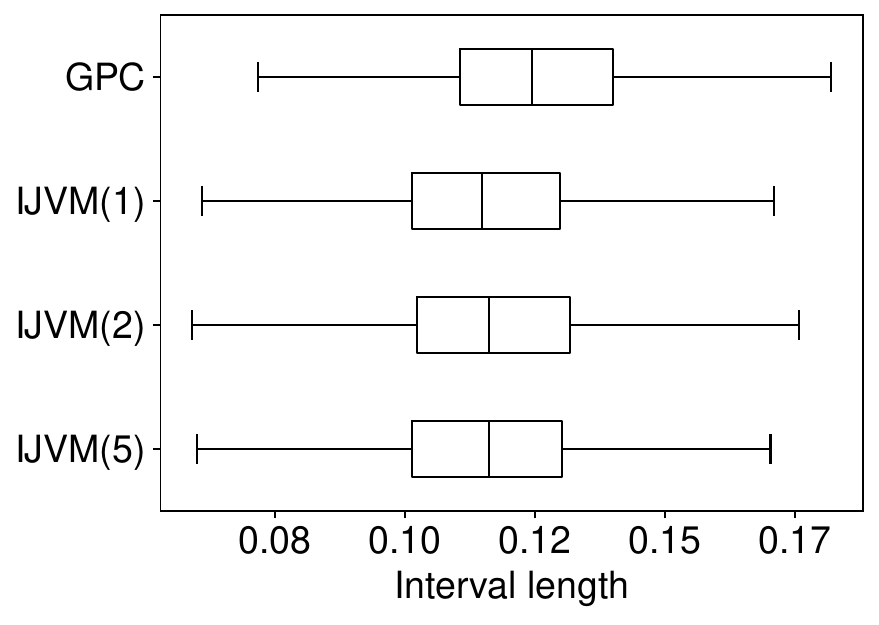}}
\end{figure}

\begin{figure}[H]
\caption{Length of $95\%$ credible intervals: SVM}

\label{sm:fig:length-dgp4}
\begin{centering}
\subfloat[$n=100$]{
\centering{}\includegraphics[width=0.4\columnwidth]{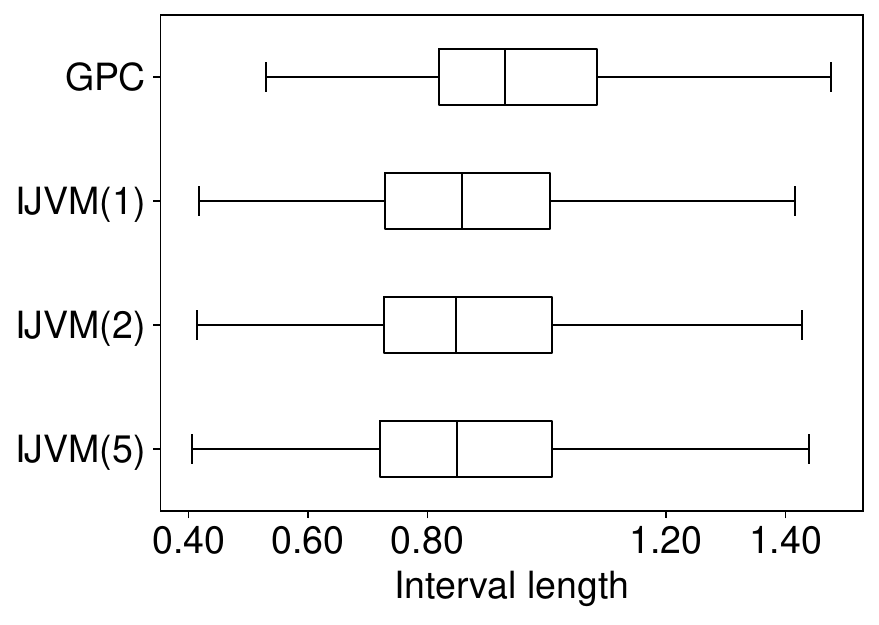}} \subfloat[$n=200$]{
\centering{}\includegraphics[width=0.4\columnwidth]{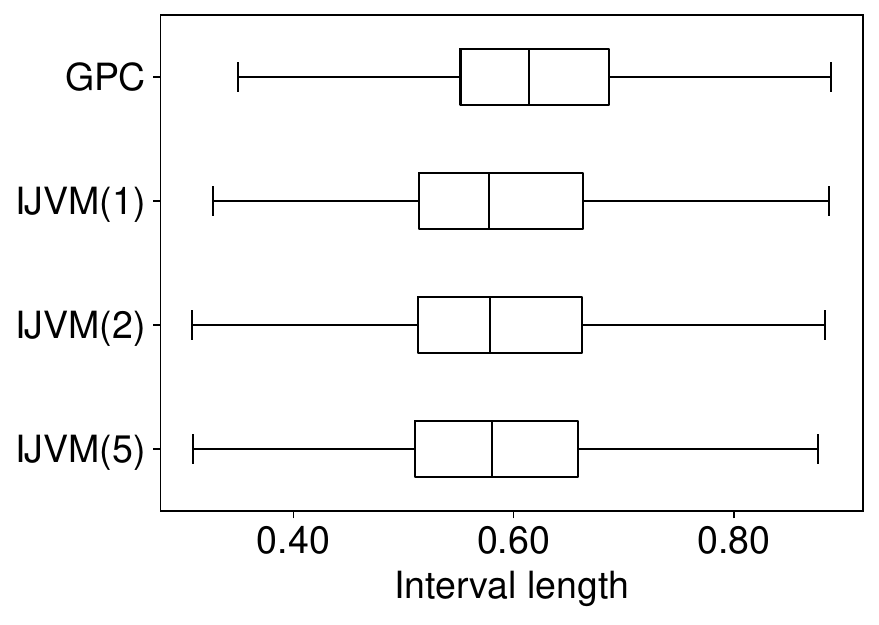}}
\par\end{centering}
\centering{}\subfloat[$n=400$]{
\centering{}\includegraphics[width=0.4\columnwidth]{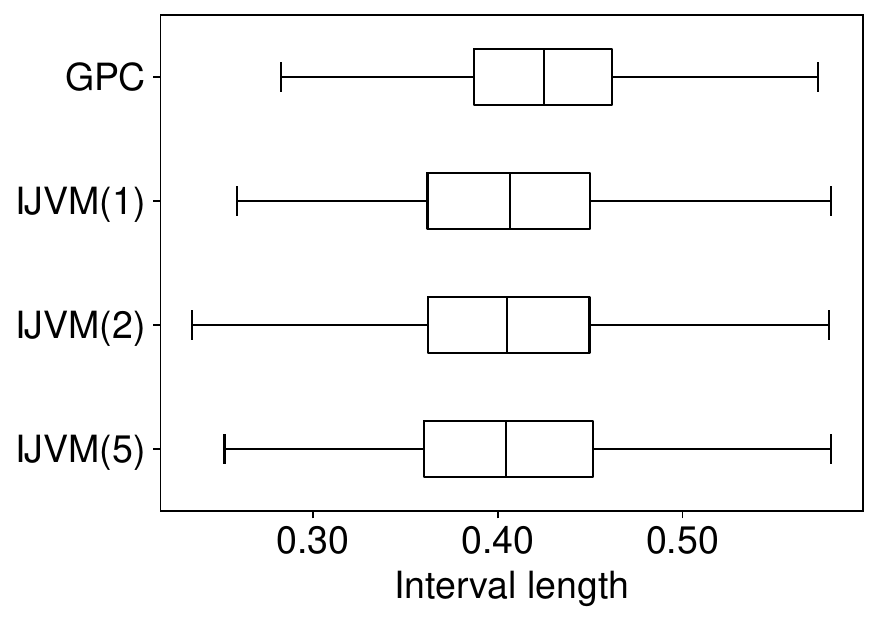}} \subfloat[$n=800$]{
\centering{}\includegraphics[width=0.4\columnwidth]{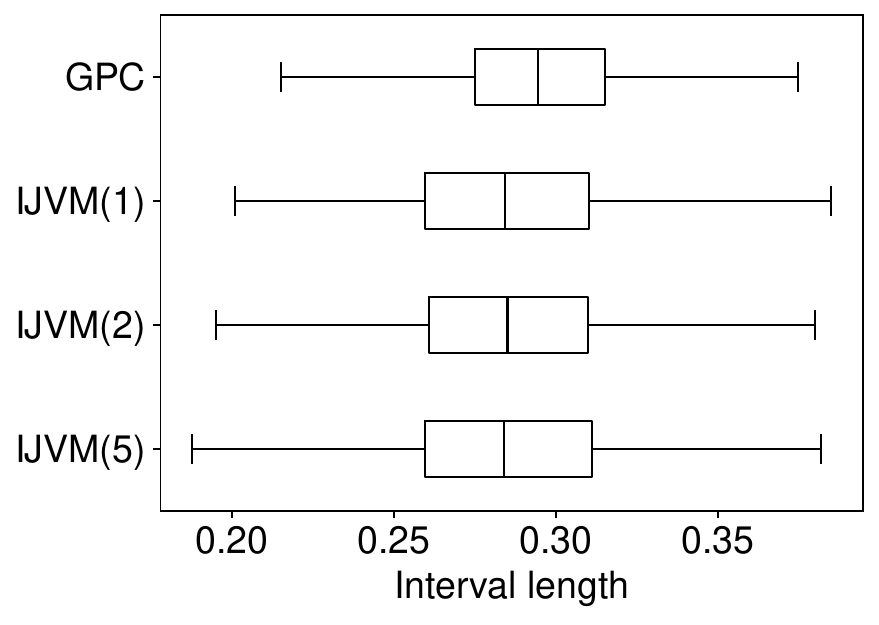}}
\end{figure}

\begin{figure}[H]
\caption{Total wall-clock time: LinReg-homo}

\label{sm:fig:time-dgp1}
\begin{centering}
\subfloat[$n=100$]{
\centering{}\includegraphics[width=0.4\columnwidth]{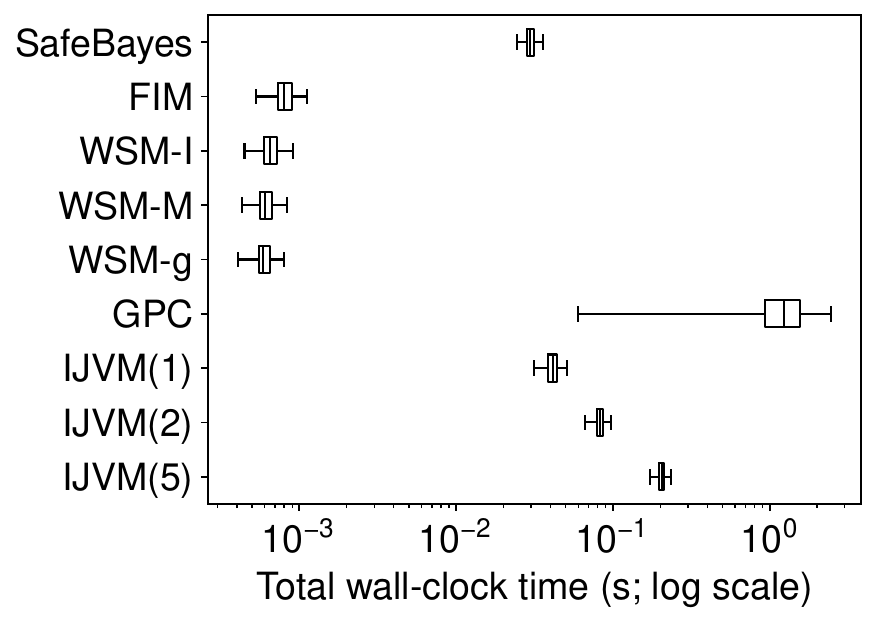}} \subfloat[$n=200$]{
\centering{}\includegraphics[width=0.4\columnwidth]{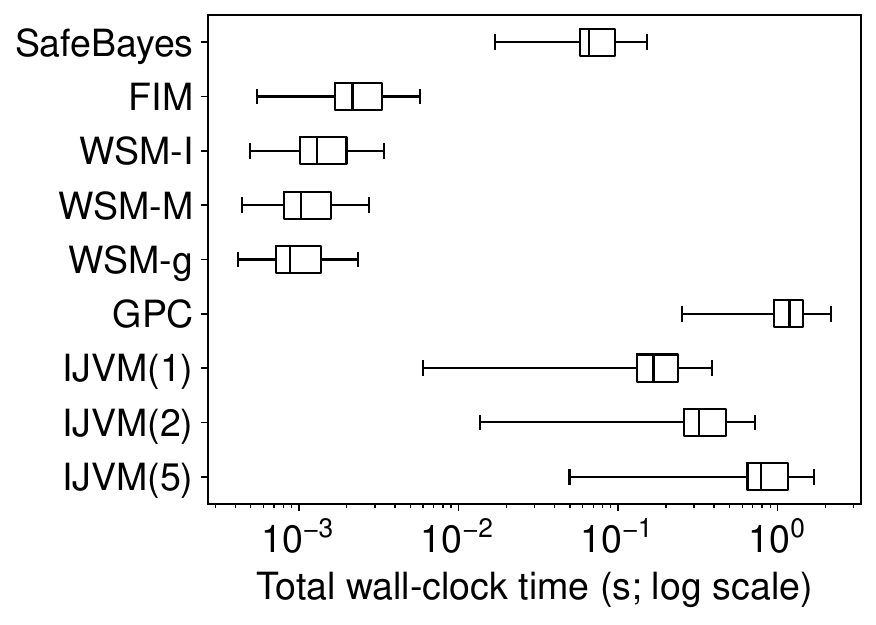}}
\par\end{centering}
\centering{}\subfloat[$n=400$]{
\centering{}\includegraphics[width=0.4\columnwidth]{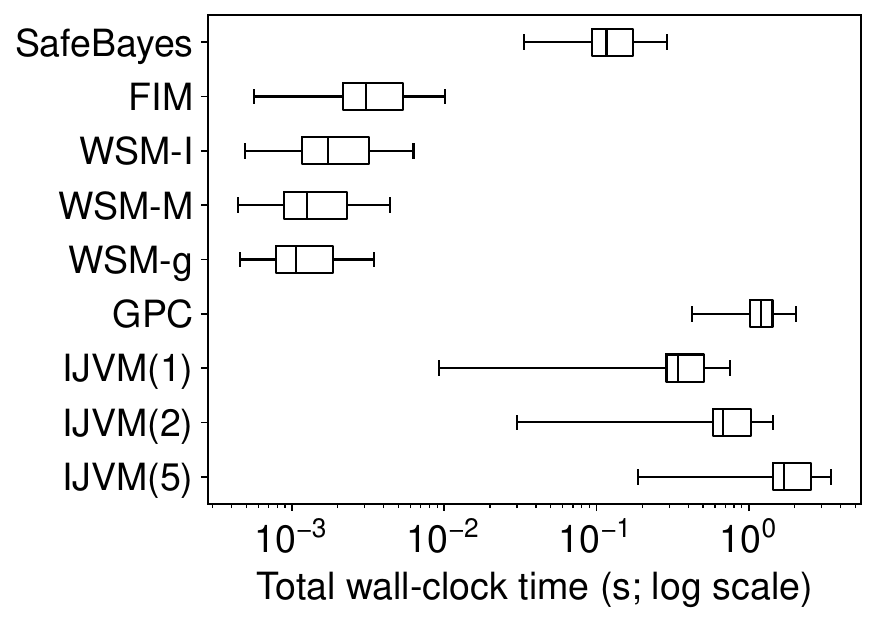}} \subfloat[$n=800$]{
\centering{}\includegraphics[width=0.4\columnwidth]{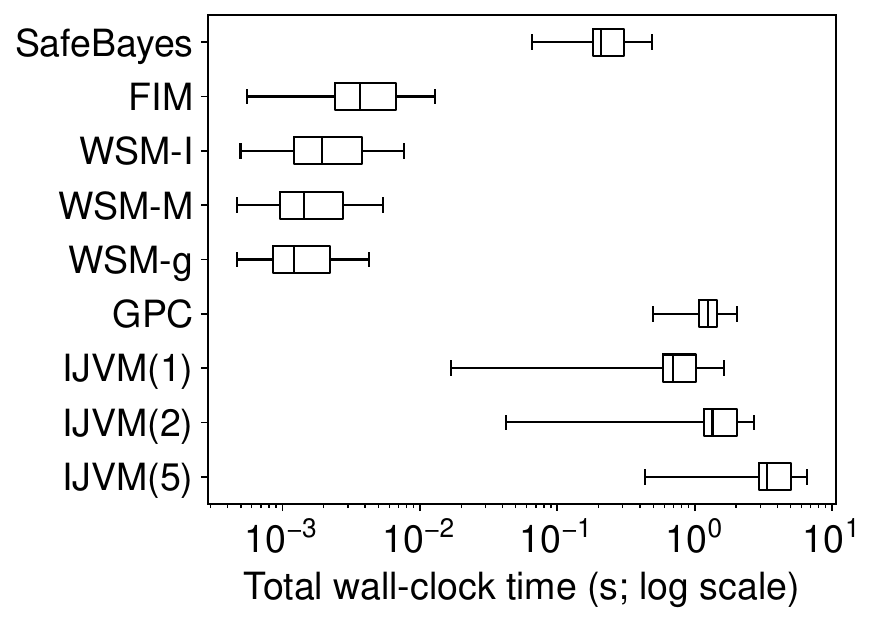}}
\end{figure}

\begin{figure}[H]

\caption{Total wall-clock time: LinReg-hetero}

\label{sm:fig-time-dgp2}
\begin{centering}
\subfloat[$n=100$]{
\centering{}\includegraphics[width=0.4\columnwidth]{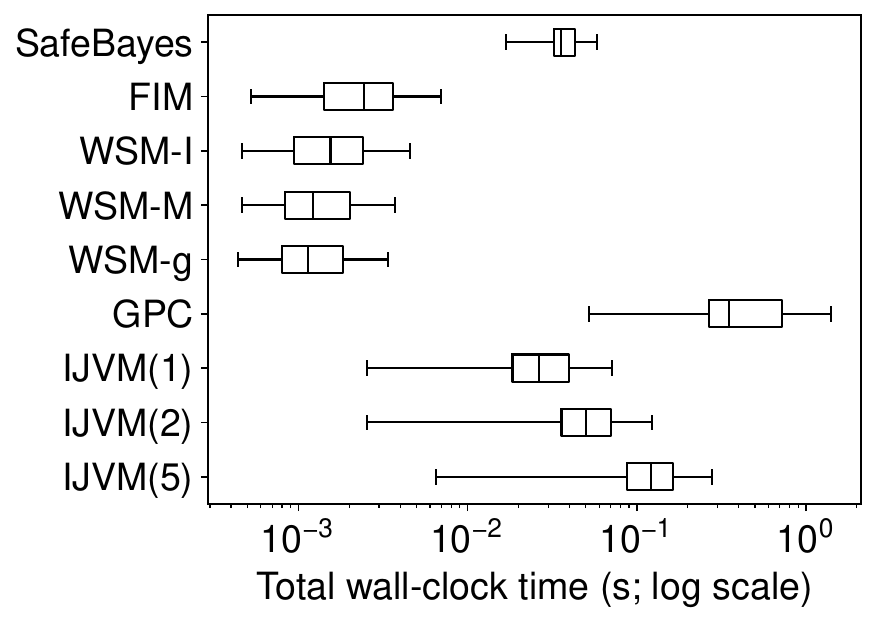}} \subfloat[$n=200$]{
\centering{}\includegraphics[width=0.4\columnwidth]{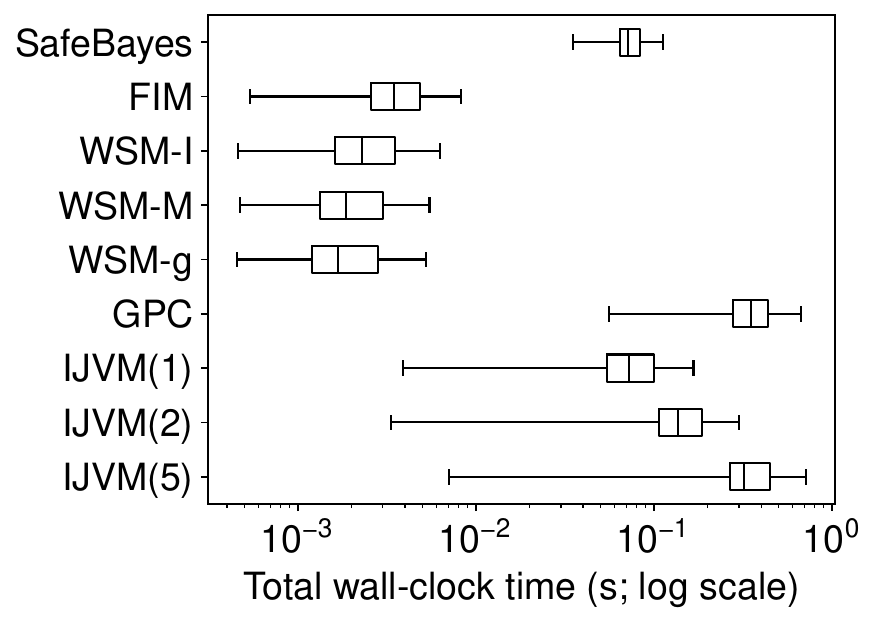}}
\par\end{centering}
\centering{}\subfloat[$n=400$]{
\centering{}\includegraphics[width=0.4\columnwidth]{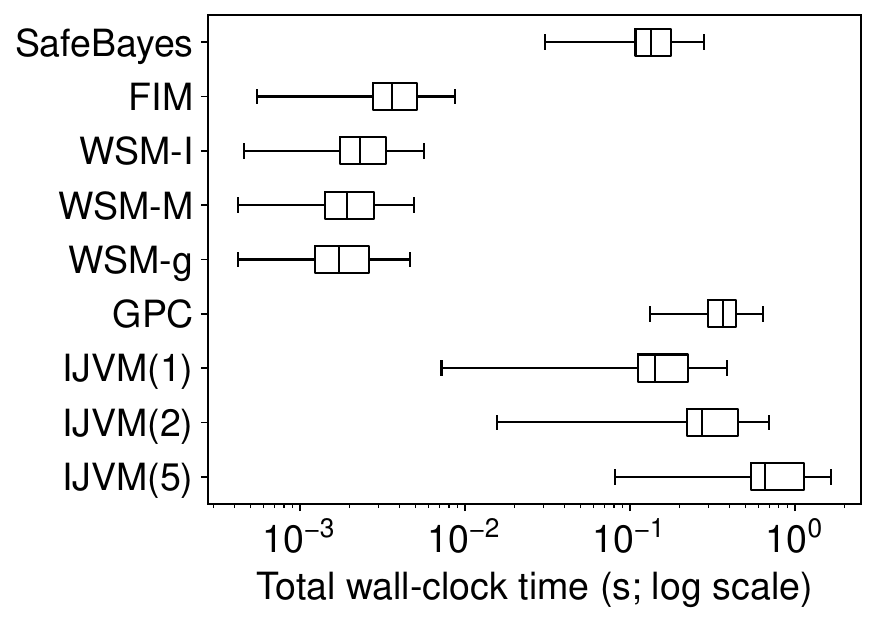}} \subfloat[$n=800$]{
\centering{}\includegraphics[width=0.4\columnwidth]{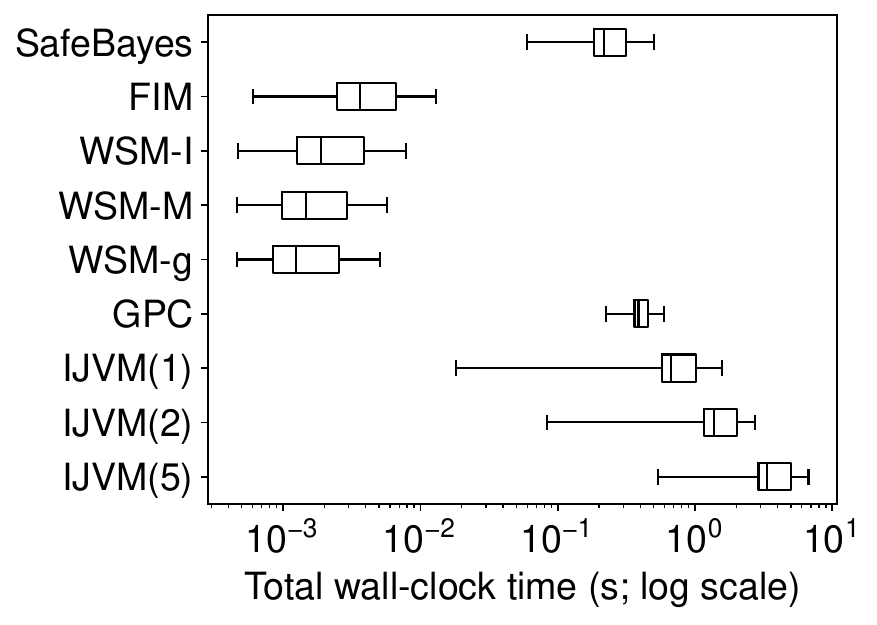}}
\end{figure}

\begin{figure}[H]
\caption{Total wall-clock time: QuantReg}

\label{sm:fig-time-dgp3}
\begin{centering}
\subfloat[$n=100$]{
\centering{}\includegraphics[width=0.4\columnwidth]{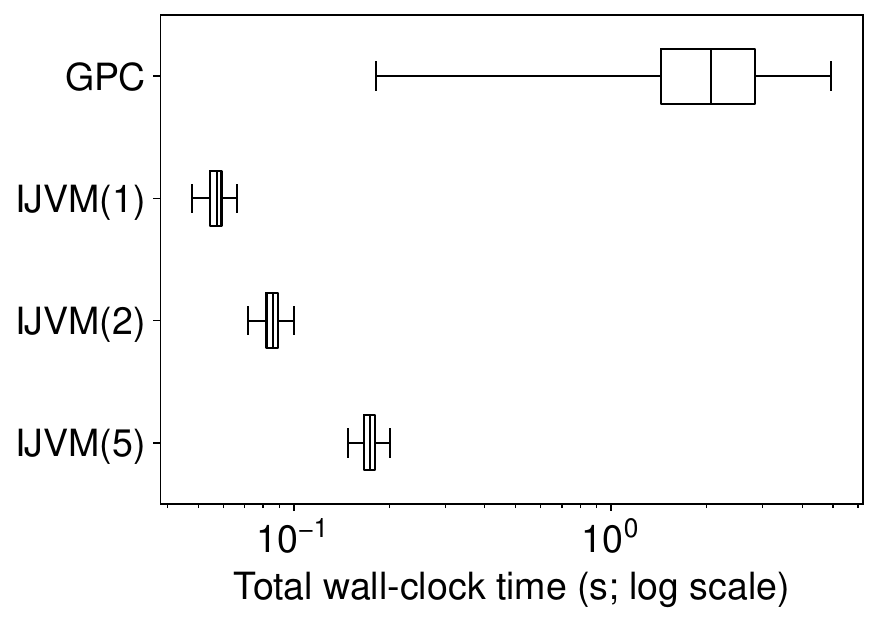}} \subfloat[$n=200$]{
\centering{}\includegraphics[width=0.4\columnwidth]{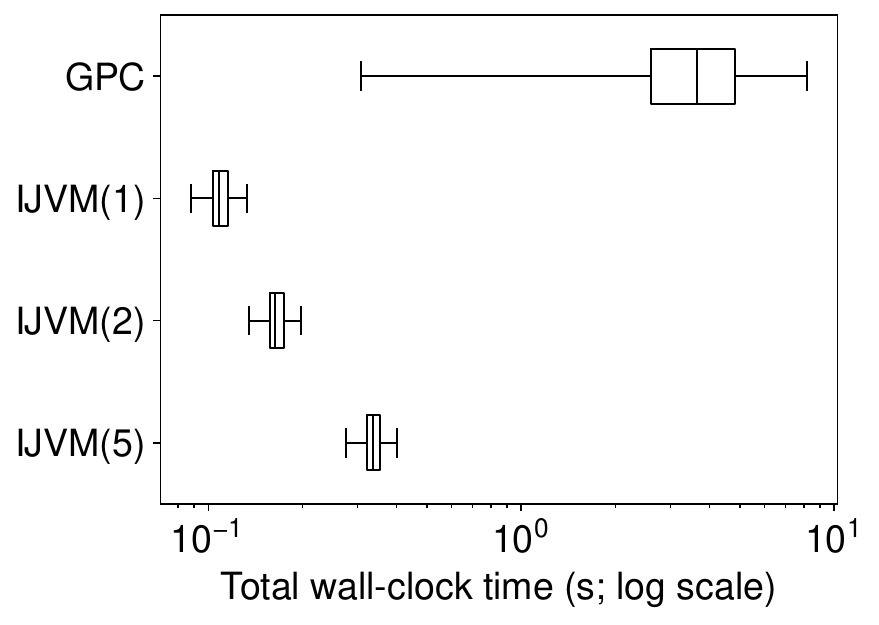}}
\par\end{centering}
\centering{}\subfloat[$n=400$]{
\centering{}\includegraphics[width=0.4\columnwidth]{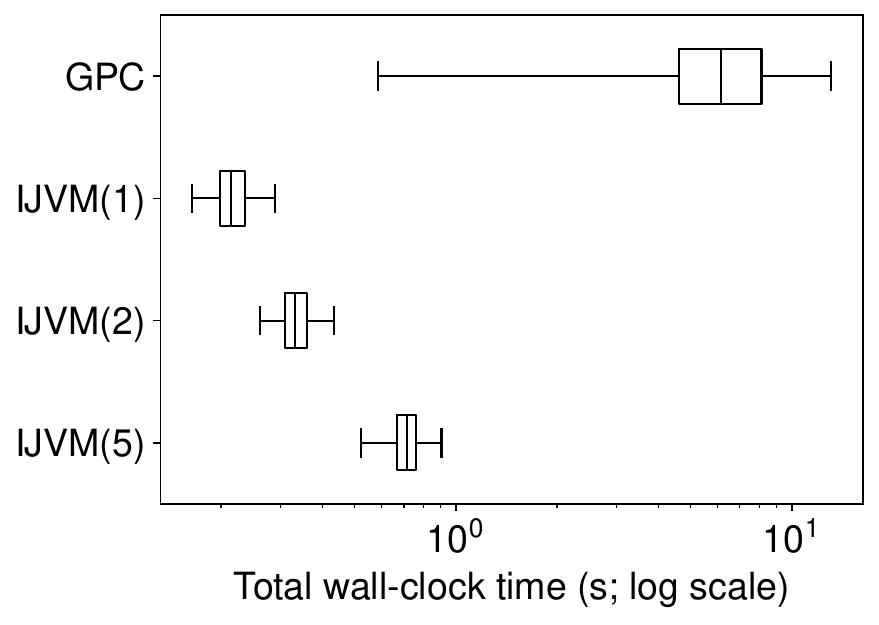}} \subfloat[$n=800$]{
\centering{}\includegraphics[width=0.4\columnwidth]{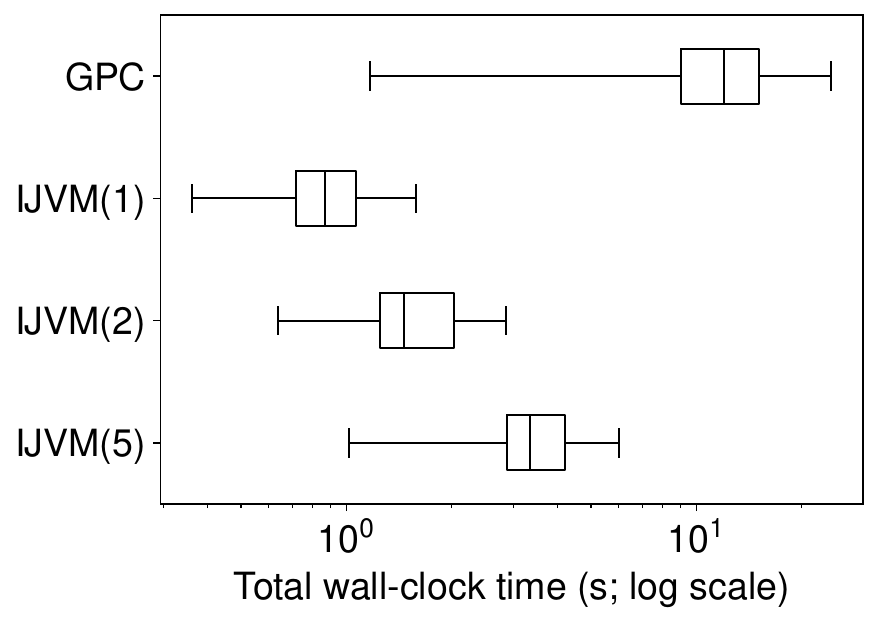}}
\end{figure}

\begin{figure}[H]
\caption{Total wall-clock time: SVM}

\label{sm:fig-time-dgp4}
\begin{centering}
\subfloat[$n=100$]{
\centering{}\includegraphics[width=0.4\columnwidth]{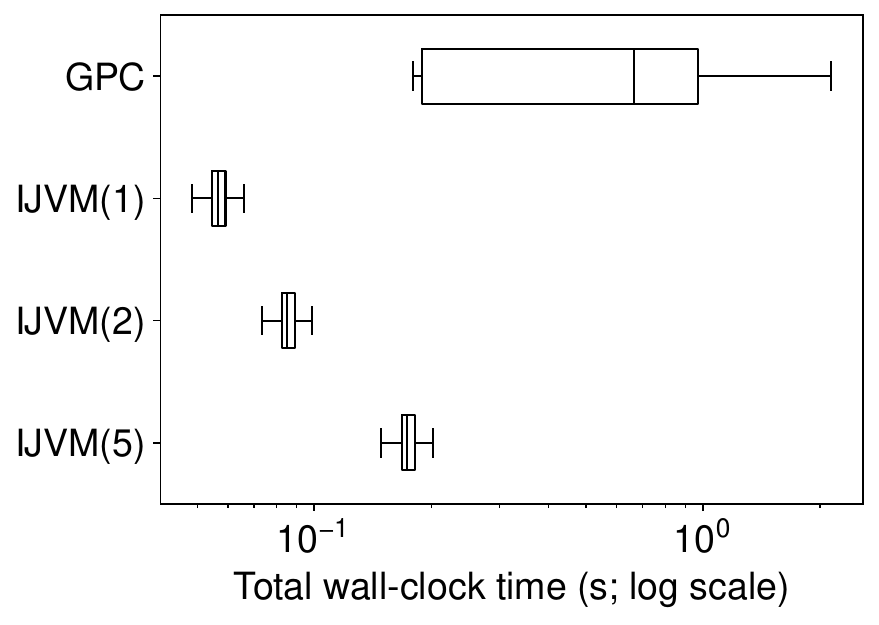}} \subfloat[$n=200$]{
\centering{}\includegraphics[width=0.4\columnwidth]{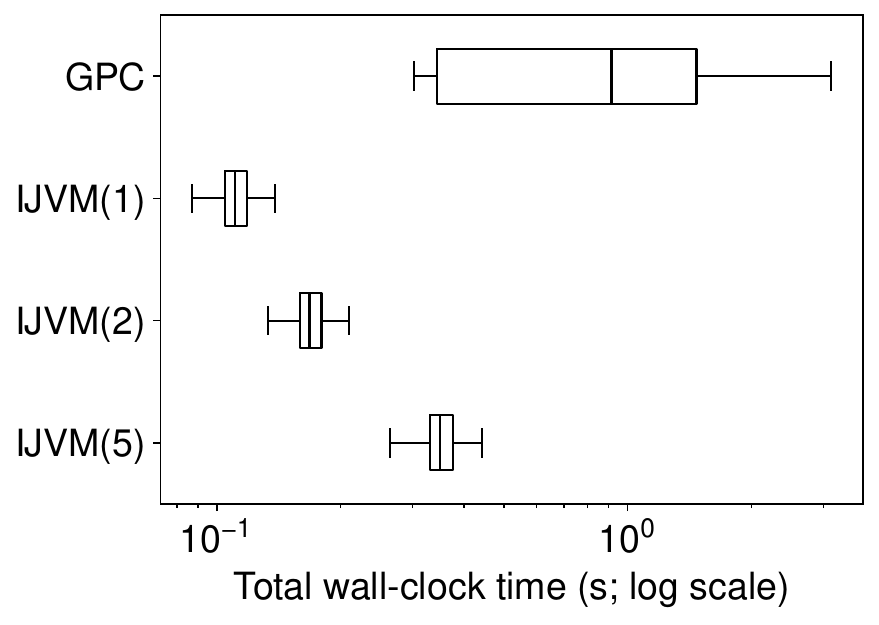}}
\par\end{centering}
\centering{}\subfloat[$n=400$]{
\centering{}\includegraphics[width=0.4\columnwidth]{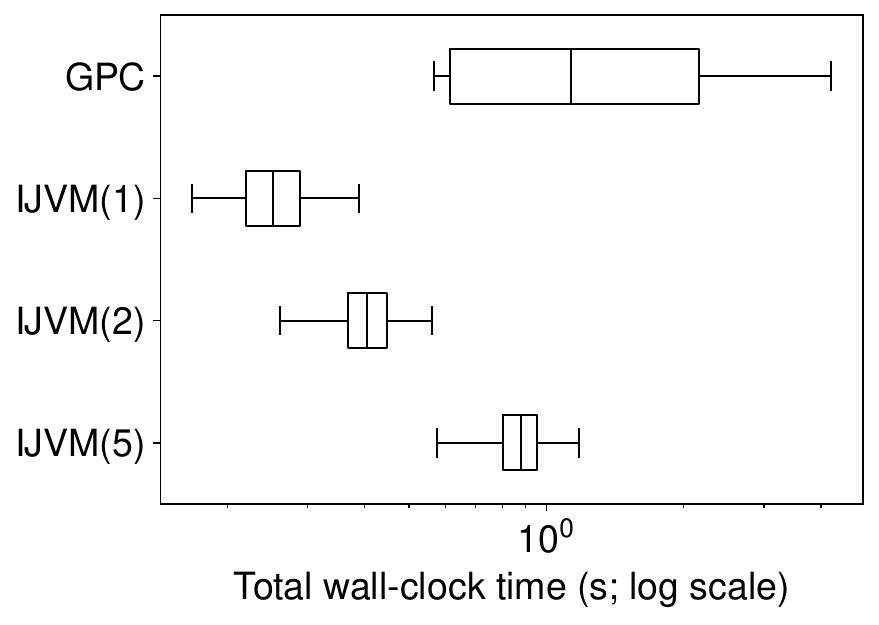}} \subfloat[$n=800$]{
\centering{}\includegraphics[width=0.4\columnwidth]{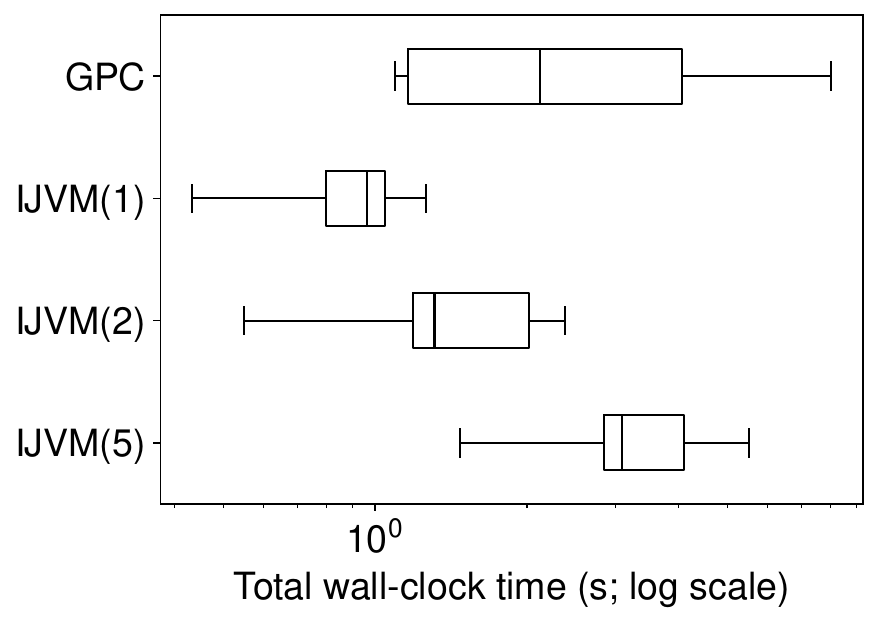}}
\end{figure}

\begin{table}[H]
\caption{Number of GPC bootstrap-coverage evaluations}

\label{sm:tab:gpc-niter}
\begin{centering}
\begin{tabular}{lcccc}
\hline 
DGP & \multicolumn{4}{c}{$n$}\tabularnewline
\cline{2-5}
 & 100 & 200 & 400 & 800\tabularnewline
\hline 
\multirow{2}{*}{LinReg-homo} & 20 & 18 & 18 & 17\tabularnewline
 & (1, 47) & (4, 49) & (5, 39) & (7, 33)\tabularnewline
\multirow{2}{*}{LinReg-hetero} & 5 & 5 & 5 & 5\tabularnewline
 & (1, 50) & (1, 50) & (1, 38) & (1, 13)\tabularnewline
\multirow{2}{*}{QuantReg} & 13 & 13 & 12 & 12\tabularnewline
 & (1, 50) & (1, 47) & (1, 38) & (1, 37)\tabularnewline
\multirow{2}{*}{SVM} & 4 & 3 & 2 & 2\tabularnewline
 & (1, 13) & (1, 12) & (1, 17) & (1, 13)\tabularnewline
\hline 
\end{tabular}
\par\end{centering}
\centering{}%
\begin{minipage}[t]{0.6\columnwidth}%
Note: Median values are reported, with the minimum and maximum values
shown in parentheses.%
\end{minipage}
\end{table}

\begin{figure}[H]
\caption{Median absolute log variance ratio: LinReg-homo}

\label{sm:fig:tau-dgp1}
\begin{centering}
\subfloat[$n=100$]{
\centering{}\includegraphics[width=0.4\columnwidth]{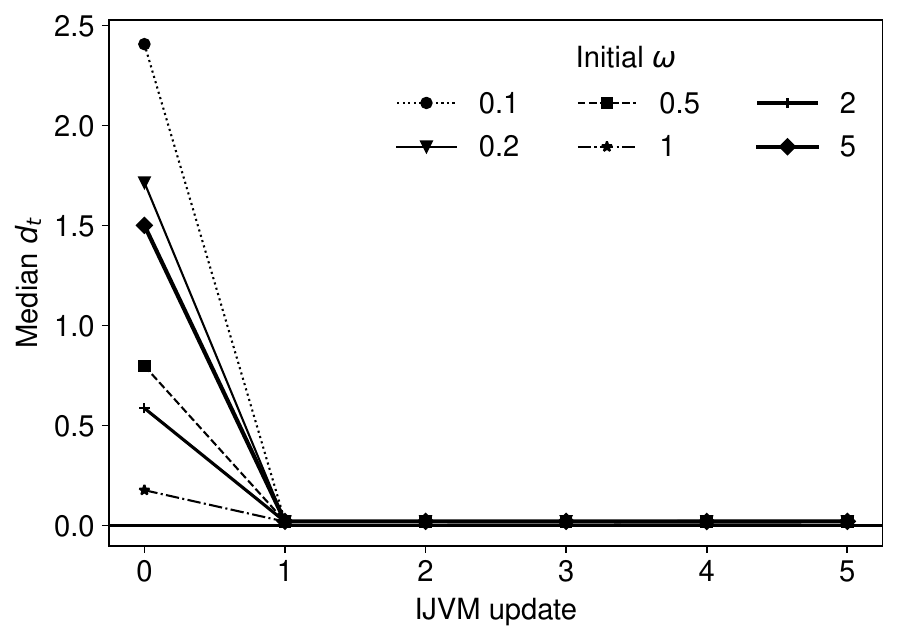}} \subfloat[$n=200$]{
\centering{}\includegraphics[width=0.4\columnwidth]{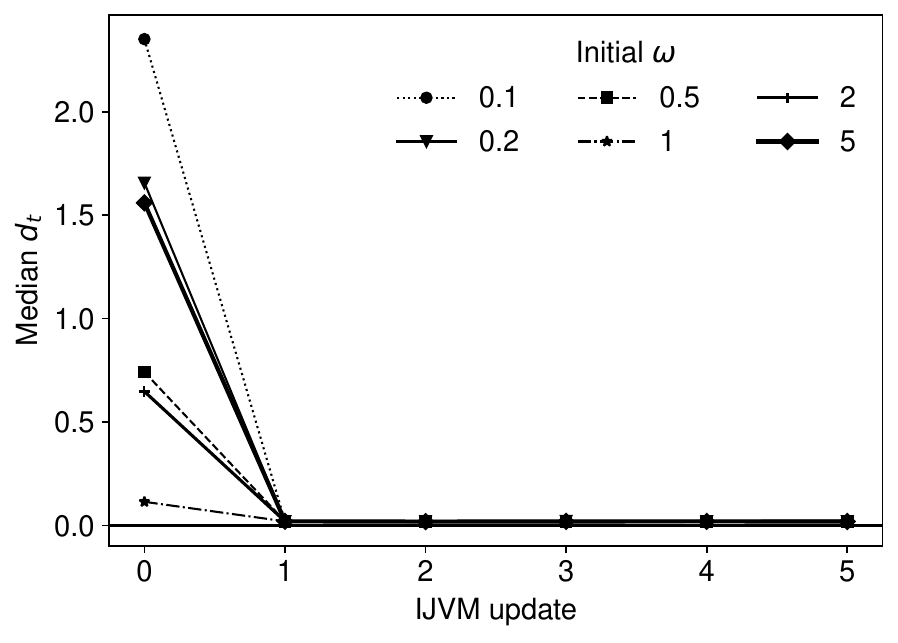}}
\par\end{centering}
\centering{}\subfloat[$n=400$]{
\centering{}\includegraphics[width=0.4\columnwidth]{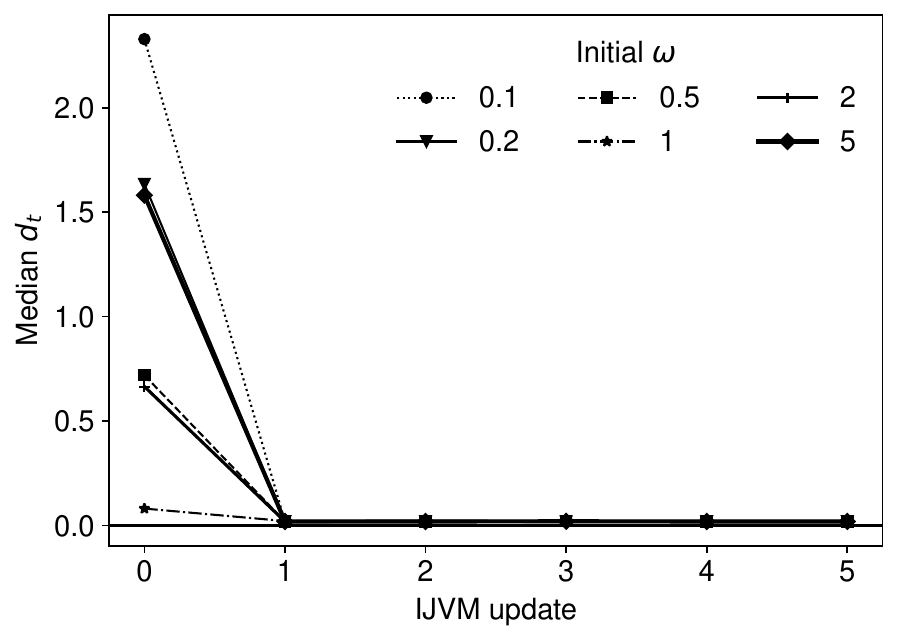}} \subfloat[$n=800$]{
\centering{}\includegraphics[width=0.4\columnwidth]{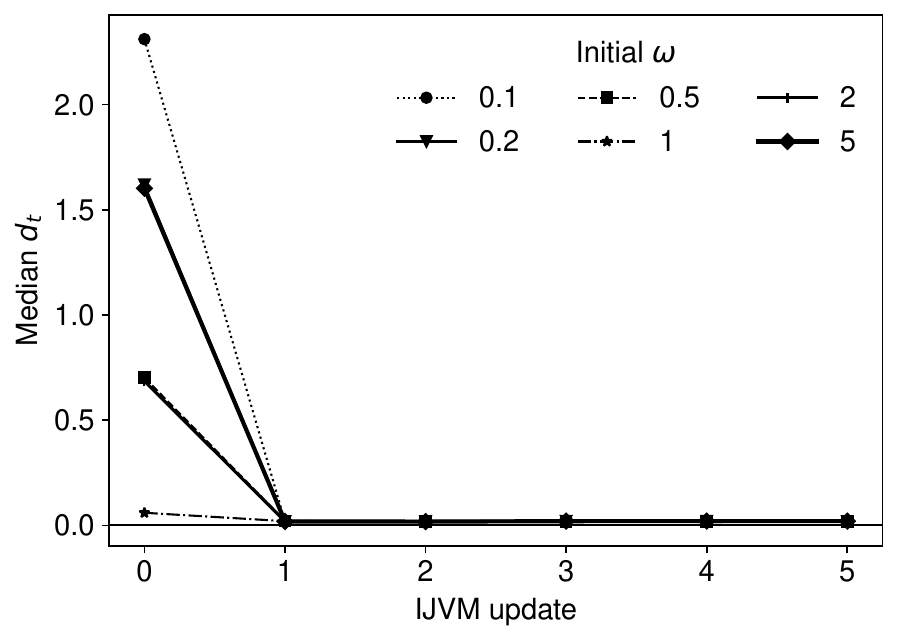}}
\end{figure}

\begin{figure}[H]
\caption{Median absolute log variance ratio: LinReg-hetero}

\label{sm:fig:tau-dgp2}
\begin{centering}
\subfloat[$n=100$]{
\centering{}\includegraphics[width=0.4\columnwidth]{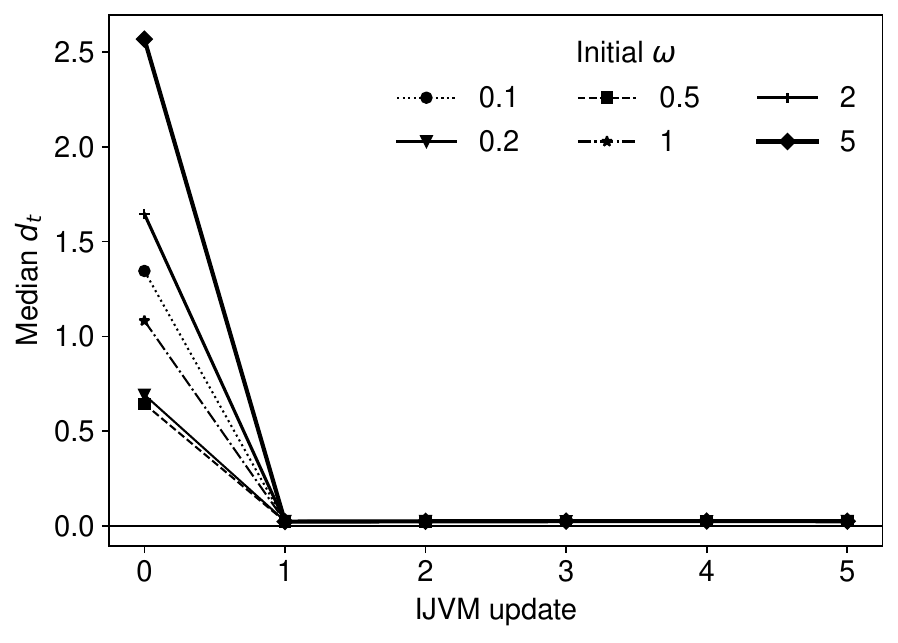}} \subfloat[$n=200$]{
\centering{}\includegraphics[width=0.4\columnwidth]{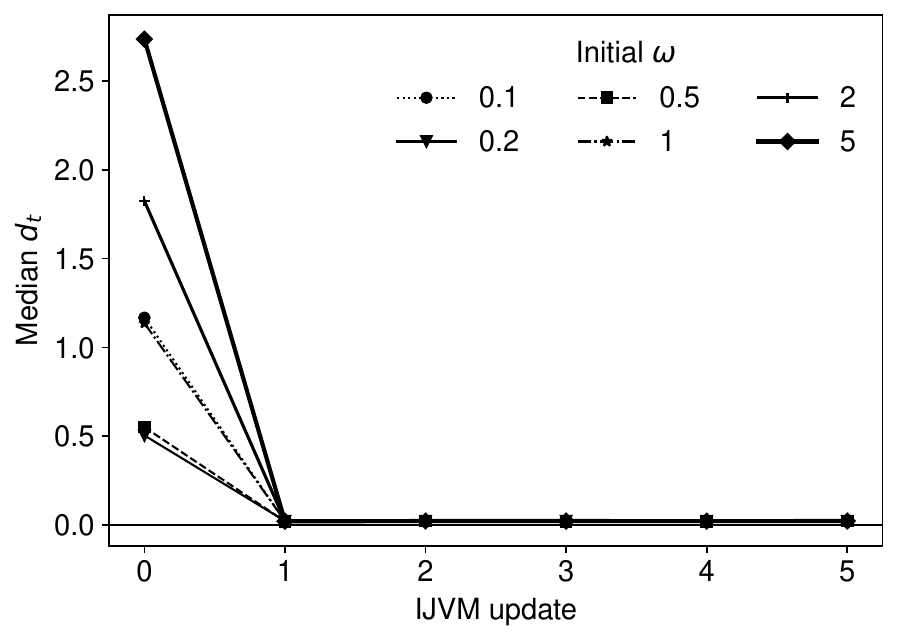}}
\par\end{centering}
\centering{}\subfloat[$n=400$]{
\centering{}\includegraphics[width=0.4\columnwidth]{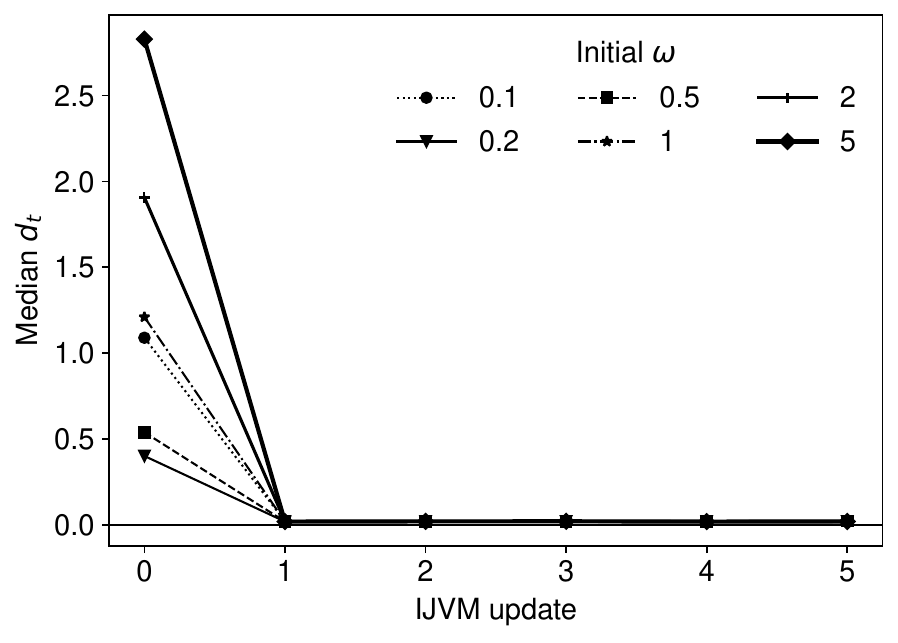}} \subfloat[$n=800$]{
\centering{}\includegraphics[width=0.4\columnwidth]{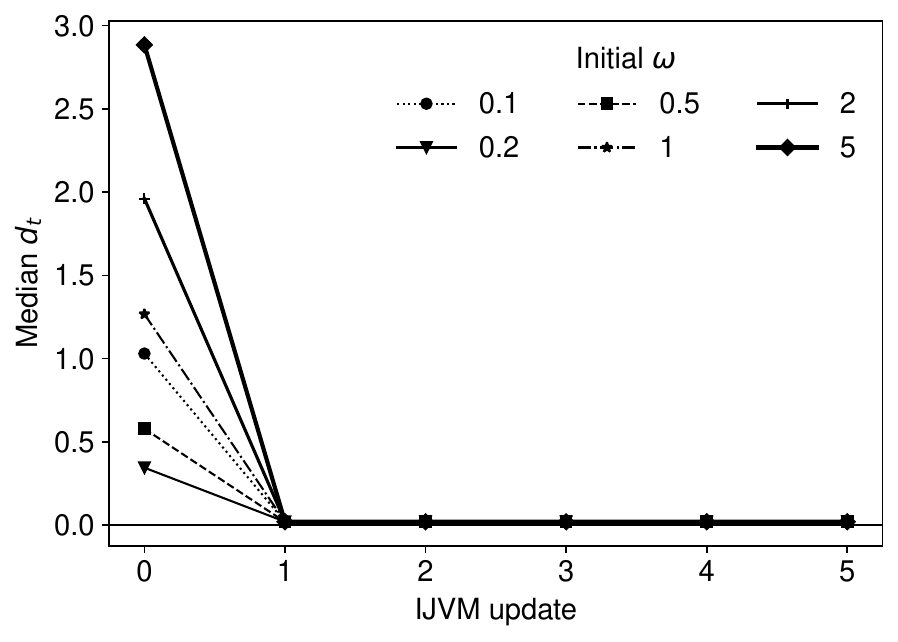}}
\end{figure}

\begin{figure}[H]

\caption{Median absolute log variance ratio: QuantReg}

\label{sm:fig:tau-dgp3}
\begin{centering}
\subfloat[$n=100$]{
\centering{}\includegraphics[width=0.4\columnwidth]{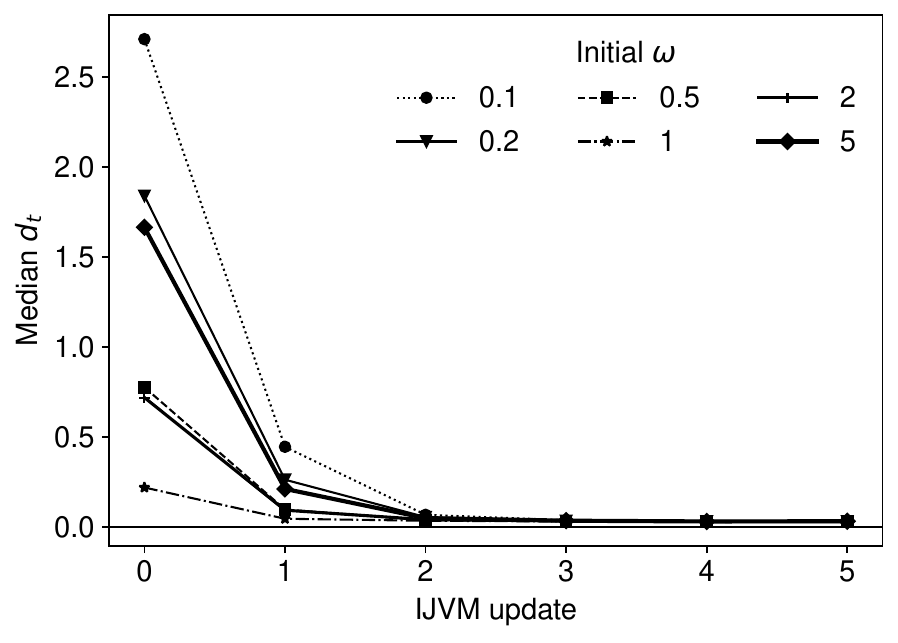}} \subfloat[$n=200$]{
\centering{}\includegraphics[width=0.4\columnwidth]{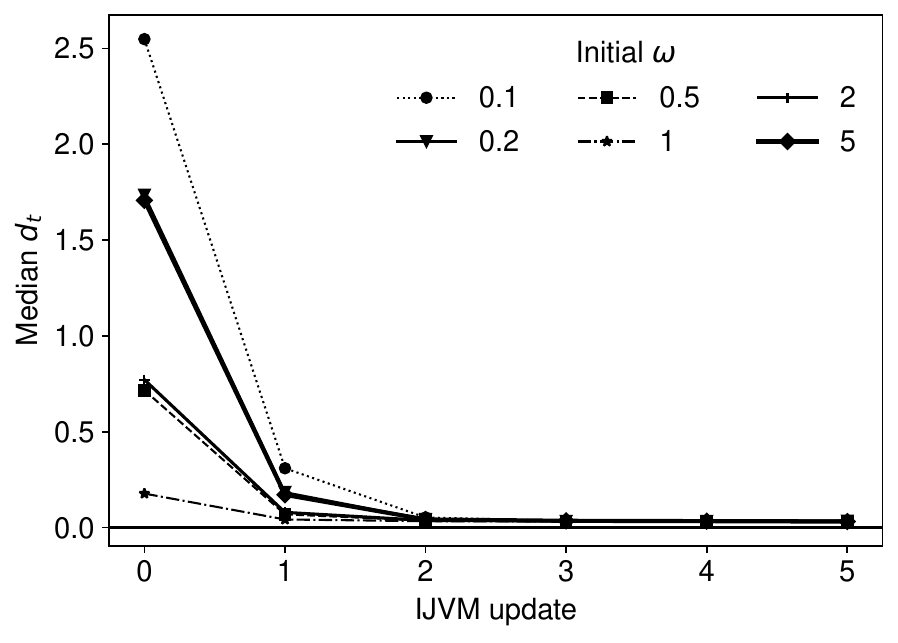}}
\par\end{centering}
\centering{}\subfloat[$n=400$]{
\centering{}\includegraphics[width=0.4\columnwidth]{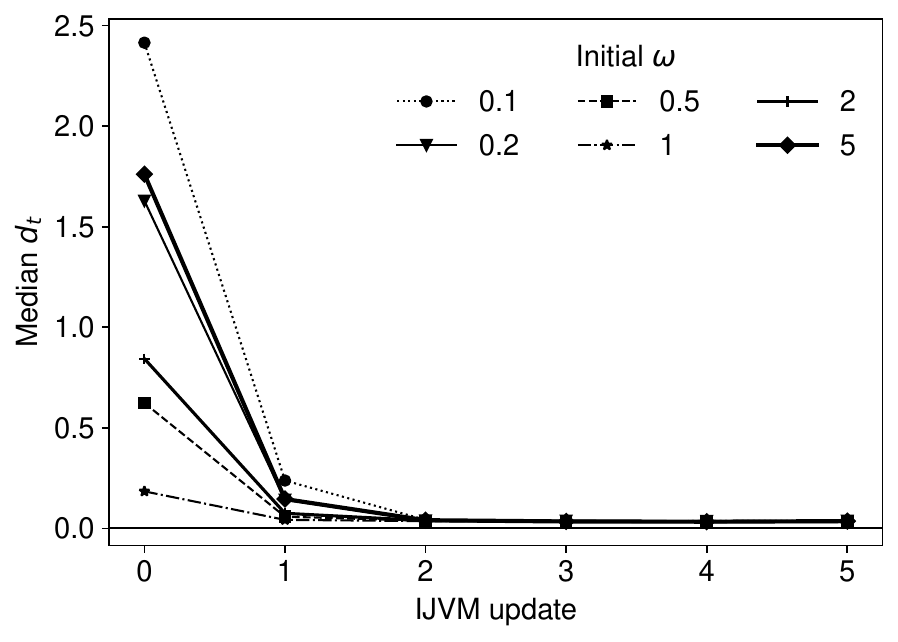}} \subfloat[$n=800$]{
\centering{}\includegraphics[width=0.4\columnwidth]{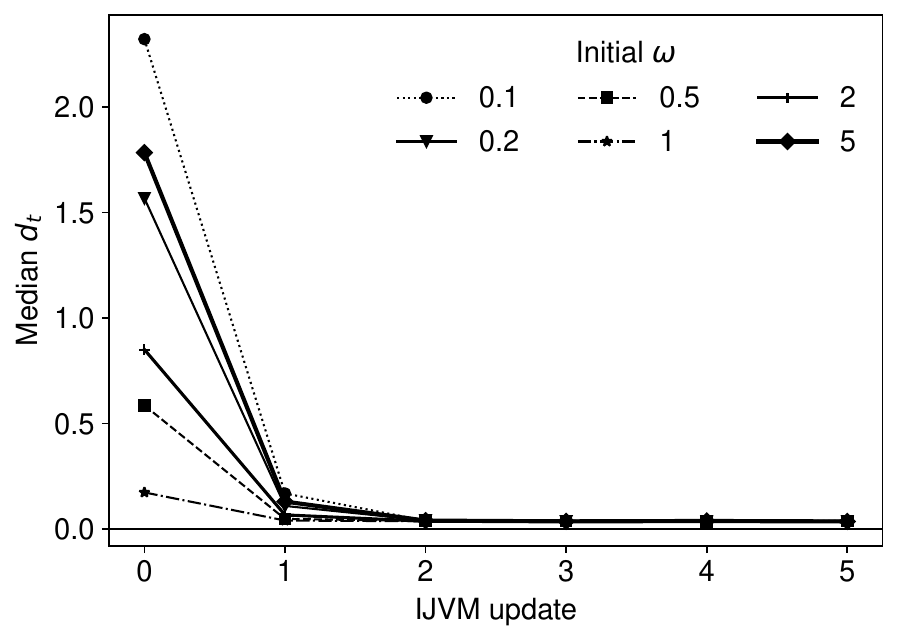}}
\end{figure}

\begin{figure}[H]
\caption{Median absolute log variance ratio: SVM}

\label{sm:fig:tau-dgp4}
\begin{centering}
\subfloat[$n=100$]{
\centering{}\includegraphics[width=0.4\columnwidth]{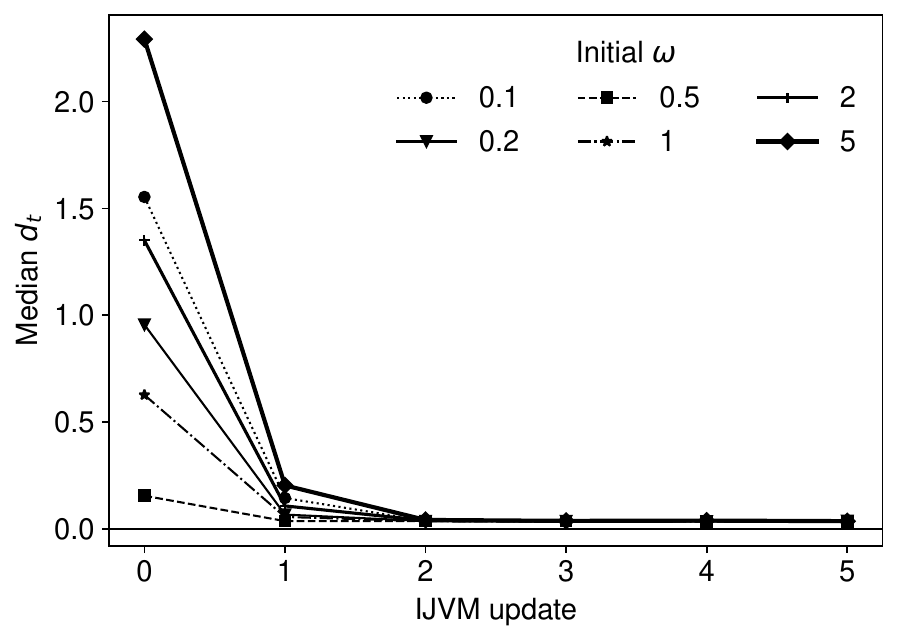}} \subfloat[$n=200$]{
\centering{}\includegraphics[width=0.4\columnwidth]{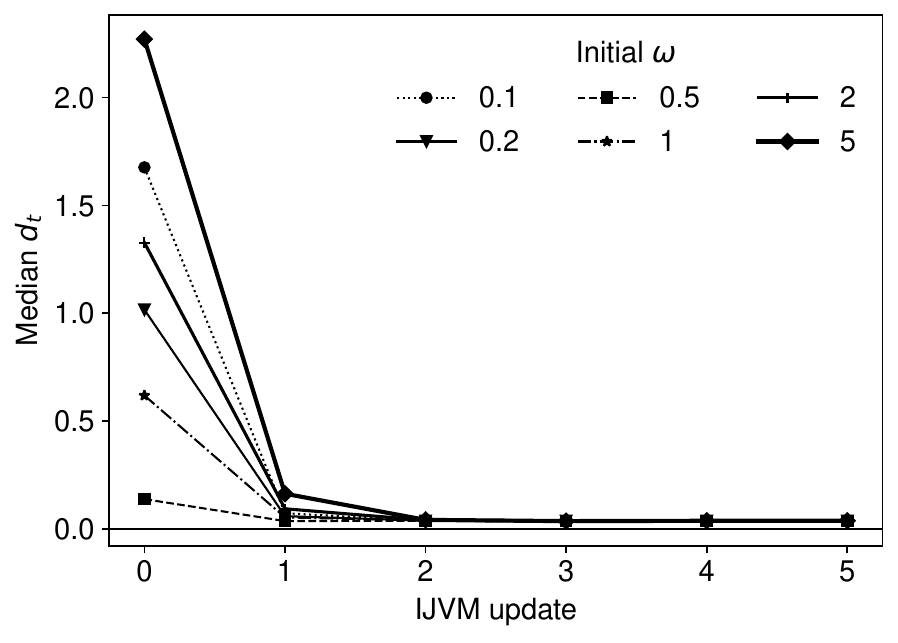}}
\par\end{centering}
\centering{}\subfloat[$n=400$]{
\centering{}\includegraphics[width=0.4\columnwidth]{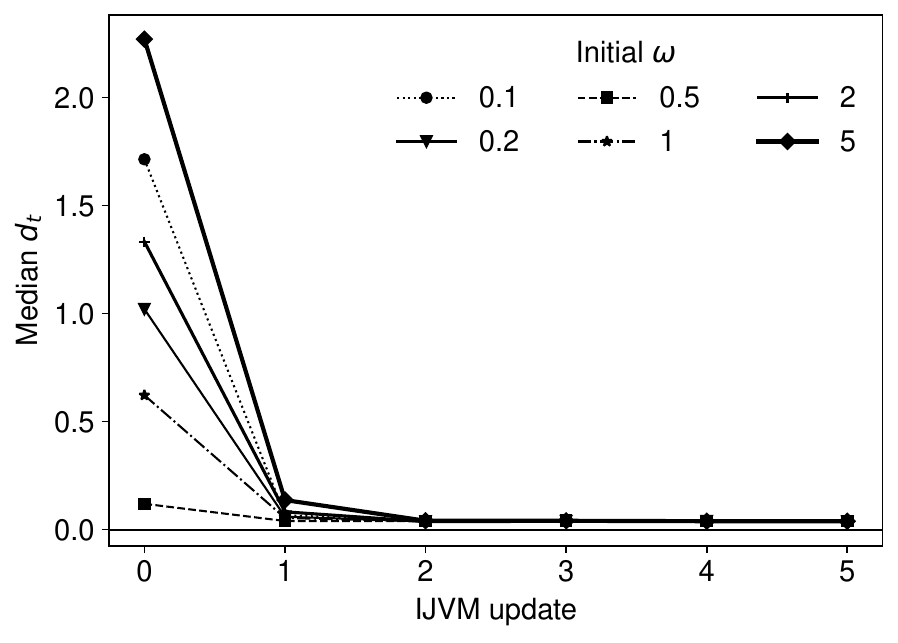}} \subfloat[$n=800$]{
\centering{}\includegraphics[width=0.4\columnwidth]{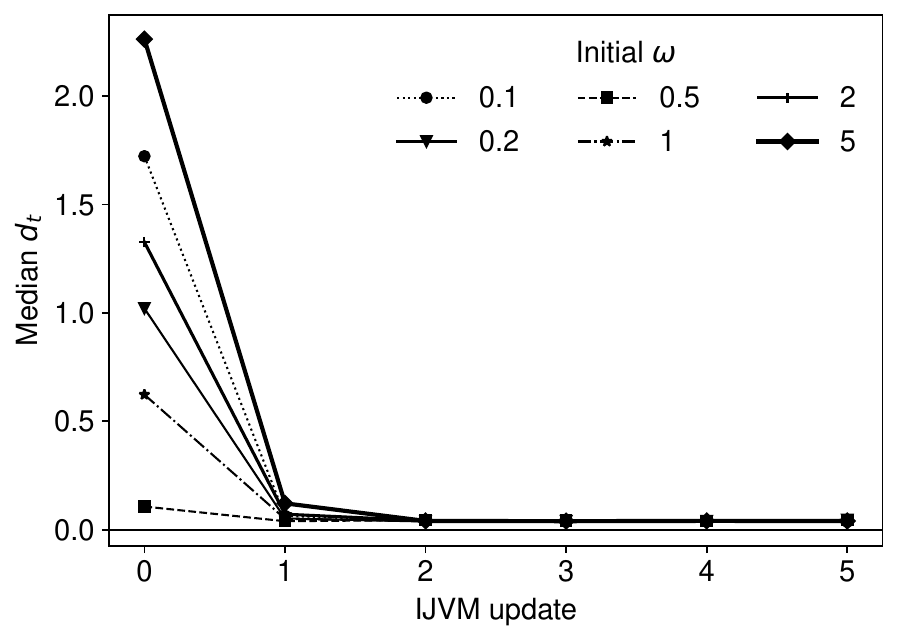}}
\end{figure}

\section{Independent-cluster sampling}

\label{app:cluster}

\subsection{Setup and first-order results}

\label{app:cluster-setup}

Let $\mathcal{Y}_{1},\ldots,\mathcal{Y}_{G}$ be i.i.d.~random clusters,
where $\mathcal{Y}_{c}=(N_{c},Y_{c1},\ldots,Y_{cN_{c}})$ and $N_{c}<\infty$
almost surely. Within-cluster dependence is unrestricted subject to
the assumptions below. Write $L_{c}(\theta)=\sum^{N_{c}}_{j=1}\ell(\theta;Y_{cj})$
and $R^{\mathrm{CL}}(\theta)=E\{L_{c}(\theta)\}$, and suppose that
this risk has a unique minimizer $\theta^{\mathrm{CL},\dagger}$.
Let $\widehat{\theta}_{G}$ minimize $G^{-1}\sum^{G}_{c=1}L_{c}(\theta)$.
Let $\dot{g}^{\mathrm{CL}}\in\mathbb{R}^{d}\setminus\{0\}$ denote
the cluster-level functional linearization vector, specified by \eqref{eq:cl-ass-lin}
below. When $g$ is differentiable at $\theta^{\mathrm{CL},\dagger}$,
it can be chosen as $\dot{g}^{\mathrm{CL}}=\nabla g(\theta^{\mathrm{CL},\dagger})$.
Let a cluster-level score $\psi_{c}=\psi^{\mathrm{CL}}(\mathcal{Y}_{c})$
with $E(\psi_{c})=0$. Let $J^{\mathrm{CL}}\succ0$ be defined by
\[
R^{\mathrm{CL}}(\theta^{\mathrm{CL},\dagger}+h)-R^{\mathrm{CL}}(\theta^{\mathrm{CL},\dagger})=\tfrac{1}{2}h^{\top}J^{\mathrm{CL}}h+o(\|h\|^{2}),\qquad V^{\mathrm{CL}}=E(\psi_{c}\psi^{\top}_{c})\succ0,
\]
and set 
\begin{equation}
A^{\mathrm{CL}}_{g}=(\dot{g}^{\mathrm{CL}})^{\top}(J^{\mathrm{CL}})^{-1}\dot{g}^{\mathrm{CL}},\quad B^{\mathrm{CL}}_{g}=(\dot{g}^{\mathrm{CL}})^{\top}(J^{\mathrm{CL}})^{-1}V^{\mathrm{CL}}(J^{\mathrm{CL}})^{-1}\dot{g}^{\mathrm{CL}},\quad\omega^{\mathrm{CL},\circ}_{g}=\frac{A^{\mathrm{CL}}_{g}}{B^{\mathrm{CL}}_{g}}.\label{eq:cl-ab}
\end{equation}
For smooth observation-level losses, $\psi_{c}=\sum^{N_{c}}_{j=1}\nabla\ell(\theta^{\mathrm{CL},\dagger};Y_{cj})$.
In particular, $V^{\mathrm{CL}}$ includes all within-cluster score
cross-covariances, without assuming independence of the observations
within a cluster.

The following conditions are the cluster-level versions of \eqref{eq:ass-1}--\eqref{eq:ass-3}.
First, assume that $\sqrt{G}(\widehat{\theta}_{G}-\theta^{\mathrm{CL},\dagger})=O_{P}(1)$
and 
\begin{equation}
\frac{1}{G}\sum^{G}_{c=1}\|\psi_{c}\|^{2}=O_{P}(1),\qquad\frac{1}{G}\sum^{G}_{c=1}\psi_{c}\psi^{\top}_{c}\xrightarrow{P}V^{\mathrm{CL}}.\label{eq:cl-ass-score}
\end{equation}
Put $X_{G}=\sqrt{G}(\theta-\widehat{\theta}_{G})$, $Q_{G}=\sqrt{G}\{g(\theta)-g(\widehat{\theta}_{G})\}$,
and $D_{Gc}=\sqrt{G}\{L_{c}(\theta)-L_{c}(\widehat{\theta}_{G})\}$.
Assume 
\begin{equation}
\sup_{\omega\in\Omega}E_{\pi_{\omega}}\|X_{G}\|^{2}=O_{P}(1),\qquad\sup_{\omega\in\Omega}\left\Vert \operatorname{Var}_{\pi_{\omega}}(X_{G})-\omega^{-1}(J^{\mathrm{CL}})^{-1}\right\Vert =o_{P}(1),\label{eq:cl-ass-post}
\end{equation}
and the two averaged posterior linearizations 
\begin{align}
\sup_{\omega\in\Omega}E_{\pi_{\omega}}(Q_{G}-(\dot{g}^{\mathrm{CL}})^{\top}X_{G})^{2} & =o_{P}(1),\nonumber \\
\sup_{\omega\in\Omega}\frac{1}{G}\sum^{G}_{c=1}E_{\pi_{\omega}}(D_{Gc}-\psi^{\top}_{c}X_{G})^{2} & =o_{P}(1).\label{eq:cl-ass-lin}
\end{align}
All posterior moments are assumed to exist and be measurable. We use
$U_{G}(\omega)=\operatorname{Var}_{\pi_{\omega}}\{g(\theta)\}$ and
the cluster IJ variance in \eqref{eq:cluster-ij}.
\begin{prop}[Independent-cluster IJVM]
\label{prop:cluster-ijvm} Under \eqref{eq:cl-ass-score}--\eqref{eq:cl-ass-lin},
\begin{equation}
\sup_{\omega\in\Omega}\bigl|G\omega U_{G}(\omega)-A^{\mathrm{CL}}_{g}\bigr|=o_{P}(1),\qquad\sup_{\omega\in\Omega}\bigl|GW^{\mathrm{CL}}_{G}(\omega)-B^{\mathrm{CL}}_{g}\bigr|=o_{P}(1).\label{eq:cl-limits}
\end{equation}
Consequently, $W^{\mathrm{CL}}_{G}$ is positive on $\Omega$ with
probability tending to one. On that event, defining $\mathcal{T}^{\mathrm{CL}}_{G}(\omega)=\omega U_{G}(\omega)/W^{\mathrm{CL}}_{G}(\omega)$,
\begin{equation}
\sup_{\omega\in\Omega}\bigl|\mathcal{T}^{\mathrm{CL}}_{G}(\omega)-\omega^{\mathrm{CL},\circ}_{g}\bigr|=o_{P}(1).\label{eq:cl-map}
\end{equation}
If $\omega^{\mathrm{CL},\circ}_{g}\in\Omega$, one projected update
from any fixed or data-dependent pilot in $\Omega$ is consistent.
If the target belongs to $\operatorname{int}(\Omega)$ and $U_{G}-W^{\mathrm{CL}}_{G}$
is continuous on $\Omega$ with probability tending to one, an exact
matching root exists with probability tending to one, all roots are
consistent, and any $\widetilde{\omega}_{G}\in\Omega$ satisfying
$G|U_{G}(\widetilde{\omega}_{G})-W^{\mathrm{CL}}_{G}(\widetilde{\omega}_{G})|=o_{P}(1)$
is consistent. 
\end{prop}

The Monte Carlo and coverage conclusions do not require smooth losses.
At a pilot rate $\omega^{(0)}\in\Omega$, possibly selected from earlier
simulation, suppose that fresh draws give estimators satisfying 
\begin{equation}
G\bigl|\widehat{U}_{G}(\omega^{(0)})-U_{G}(\omega^{(0)})\bigr|=o_{P_{\mathrm{joint}}}(1),\qquad\frac{1}{G}\sum^{G}_{c=1}\left[G\omega^{(0)}\{\widehat{\xi}^{\mathrm{CL}}_{c,\omega^{(0)}}-\xi^{\mathrm{CL}}_{c,\omega^{(0)}}\}\right]^{2}=o_{P_{\mathrm{joint}}}(1).\label{eq:cl-mc}
\end{equation}
If $\omega^{\mathrm{CL},\circ}_{g}\in\Omega$, the computed update
\eqref{eq:cluster-update} converges to $\omega^{\mathrm{CL},\circ}_{g}$
in joint probability. If, in addition, 
\begin{equation}
\sqrt{G}\{g(\widehat{\theta}_{G})-g(\theta^{\mathrm{CL},\dagger})\}\rightsquigarrow N(0,B^{\mathrm{CL}}_{g})\label{eq:cl-clt}
\end{equation}
and for $\beta\in\{\beta_{L},\beta_{U}\}$, 
\begin{equation}
\sup_{\omega\in\Omega}\left|\sqrt{G}\{q_{\beta,G}(\omega)-g(\widehat{\theta}_{G})\}-z_{\beta}\sqrt{A^{\mathrm{CL}}_{g}/\omega}\right|=o_{P}(1),\label{eq:cl-quantiles}
\end{equation}
then credible intervals at any consistent selected rate have coverage
converging to $\beta_{U}-\beta_{L}=1-\alpha$. The same conclusion
holds for simulated endpoints with errors $o_{P_{\mathrm{joint}}}(G^{-1/2})$.
The continuity, Monte Carlo, and coverage conditions are invoked only
for the corresponding conclusions, as in Theorems~\ref{thm:exit-consistent-appr-root}--\ref{thm:computed}
and Corollary~\ref{cor:coverage}.

\subsection{Proofs and sufficient conditions}

\label{app:cluster-proof}

\paragraph{Cluster reweighting identity.}

For a nonnegative cluster-weight vector $m$, consider $\pi_{\omega,m}(\theta)\propto\exp\{-\omega\sum^{G}_{c=1}m_{c}L_{c}(\theta)\}p(\theta)$.
Differentiation under the posterior integral, when justified, gives
\[
\left.\frac{\partial}{\partial m_{c}}E_{\pi_{\omega,m}}\{g(\theta)\}\right|_{m=\mathbf{1}}=-\omega\xi^{\mathrm{CL}}_{c,\omega}.
\]
For $M\sim\operatorname{Multinomial}(G;1/G,\ldots,1/G)$, $\operatorname{Cov}(M_{c},M_{d})=\mathbf{1}\{c=d\}-G^{-1}$.
Hence the conditional variance of the first-order perturbation of
the posterior mean is exactly 
\[
\operatorname{Var}_{M}\left\{ -\omega\sum^{G}_{c=1}(M_{c}-1)\xi^{\mathrm{CL}}_{c,\omega}\right\} =\omega^{2}\sum^{G}_{c=1}(\xi^{\mathrm{CL}}_{c,\omega}-\bar{\xi}^{\mathrm{CL}}_{\omega})^{2}=W^{\mathrm{CL}}_{G}(\omega).
\]
As with \eqref{eq:ij-var}, this is an exact identity for the linear
approximation, not for the variance of the nonlinear bootstrap posterior
mean.

\paragraph{Proof of Proposition~\ref{prop:cluster-ijvm}.}

For the posterior variance, $GU_{G}(\omega)=\operatorname{Var}_{\pi_{\omega}}(Q_{G})$.
The uniform moment and functional linearization conditions give 
\[
\sup_{\omega\in\Omega}\left|\operatorname{Var}_{\pi_{\omega}}(Q_{G})-(\dot{g}^{\mathrm{CL}})^{\top}\operatorname{Var}_{\pi_{\omega}}(X_{G})\dot{g}^{\mathrm{CL}}\right|=o_{P}(1).
\]
Together with \eqref{eq:cl-ass-post}, this proves the first limit
in \eqref{eq:cl-limits}.

For the cluster IJ term, the identity $G\xi^{\mathrm{CL}}_{c,\omega}=\operatorname{Cov}_{\pi_{\omega}}(Q_{G},D_{Gc})$
is exact. Write $r_{g,G}=Q_{G}-(\dot{g}^{\mathrm{CL}})^{\top}X_{G}$
and $r_{Gc}=D_{Gc}-\psi^{\top}_{c}X_{G}$. By Cauchy--Schwarz, \eqref{eq:cl-ass-score}--\eqref{eq:cl-ass-lin}
imply 

\[
\sup_{\omega\in\Omega}\frac{1}{G}\sum^{G}_{c=1}\left|G\omega\xi^{\mathrm{CL}}_{c,\omega}-\omega(\dot{g}^{\mathrm{CL}})^{\top}\operatorname{Var}_{\pi_{\omega}}(X_{G})\psi_{c}\right|^{2}=o_{P}(1).
\]
Indeed, the term involving $\operatorname{Cov}(r_{g,G},\psi^{\top}_{c}X_{G})$
is bounded in average squared norm by a constant multiple of $\sup_{\omega}E_{\pi_{\omega}}r^{2}_{g,G}$
times $\sup_{\omega}E_{\pi_{\omega}}\|X_{G}\|^{2}$ times $G^{-1}\sum_{c}\|\psi_{c}\|^{2}$.
The terms containing $r_{Gc}$ are controlled by $G^{-1}\sum_{c}E_{\pi_{\omega}}r^{2}_{Gc}$
and the same moment bounds. Thus \eqref{eq:cl-ass-post} yields 
\begin{equation}
\sup_{\omega\in\Omega}\frac{1}{G}\sum^{G}_{c=1}\left\{ G\omega\xi^{\mathrm{CL}}_{c,\omega}-t_{c}\right\} ^{2}=o_{P}(1),\qquad t_{c}=(\dot{g}^{\mathrm{CL}})^{\top}(J^{\mathrm{CL}})^{-1}\psi_{c}.\label{eq:cl-ij-score}
\end{equation}
Let $a_{c,\omega}=G\omega\xi^{\mathrm{CL}}_{c,\omega}$. Because centering
is a contraction for the empirical Euclidean norm, \eqref{eq:cl-ij-score}
implies 
\[
\sup_{\omega\in\Omega}\left|GW^{\mathrm{CL}}_{G}(\omega)-\frac{1}{G}\sum^{G}_{c=1}(t_{c}-\bar{t})^{2}\right|=o_{P}(1),\qquad\bar{t}=G^{-1}\sum^{G}_{c=1}t_{c}.
\]
The independent-cluster law of large numbers gives $\bar{t}=o_{P}(1)$
and $G^{-1}\sum_{c}t^{2}_{c}\xrightarrow{P}B^{\mathrm{CL}}_{g}$.
This proves the second limit and hence the uniform map limit. Consistency
of the projected one-step update follows by continuity of projection.
Finally, $G\{U_{G}(\omega)-W^{\mathrm{CL}}_{G}(\omega)\}$ converges
uniformly to $A^{\mathrm{CL}}_{g}/\omega-B^{\mathrm{CL}}_{g}$, which
has a unique zero at $\omega^{\mathrm{CL},\circ}_{g}$. Continuity
and the opposite signs at the endpoints establish root existence;
the same uniform convergence establishes consistency of all exact
or approximate roots. \qed

\paragraph{Monte Carlo accuracy and coverage.}

Let $\widehat{a}_{c}=G\omega\widehat{\xi}^{\mathrm{CL}}_{c,\omega}$
and $a_{c}=G\omega\xi^{\mathrm{CL}}_{c,\omega}$. The second condition
in \eqref{eq:cl-mc} implies $G^{-1}\sum_{c}(\widehat{a}_{c}-a_{c})^{2}=o_{P_{\mathrm{joint}}}(1)$.
Since $G^{-1}\sum_{c}(a_{c}-\bar{a})^{2}=O_{P}(1)$ uniformly in $\omega$,
Cauchy--Schwarz shows that $G|\widehat{W}^{\mathrm{CL}}_{G}-W^{\mathrm{CL}}_{G}|=o_{P_{\mathrm{joint}}}(1)$.
Together with the first condition in \eqref{eq:cl-mc} and $B^{\mathrm{CL}}_{g}>0$,
this proves consistency of the computed one-step ratio. For coverage,
\eqref{eq:cl-quantiles} evaluated at a consistent rate gives 
\[
\sqrt{G}\{q_{\beta,G}(\widehat{\omega}_{G})-g(\widehat{\theta}_{G})\}=z_{\beta}\sqrt{B^{\mathrm{CL}}_{g}}+o_{P_{\mathrm{joint}}}(1).
\]
Equation \eqref{eq:cl-clt} and Slutsky's theorem give limiting coverage
$\Phi(z_{\beta_{U}})-\Phi(z_{\beta_{L}})=1-\alpha$. Endpoint errors
of order $o_{P_{\mathrm{joint}}}(G^{-1/2})$ are asymptotically negligible.
\qed

\paragraph{Convex cluster losses.}

A convenient sufficient condition is that $L_{c}(\cdot)$ satisfies
the conditions of Theorem~\ref{thm:convex-verification} when the
cluster $\mathcal{Y}_{c}$ is treated as one observation. Specifically,
suppose $\Theta=\mathbb{R}^{d}$, $L_{c}(\cdot)$ is finite and convex,
and 
\[
|L_{c}(\theta)-L_{c}(\theta')|\leq\Lambda(\mathcal{Y}_{c})\|\theta-\theta'\|,\qquad E\{\Lambda(\mathcal{Y}_{c})^{2+\delta}\}<\infty
\]
for some $\delta>0$. Suppose further that the cluster loss is almost
surely differentiable at $\theta^{\mathrm{CL},\dagger}$, the cluster
population risk is locally quadratic with $J^{\mathrm{CL}}\succ0$,
the resulting cluster-score covariance $V^{\mathrm{CL}}$ is positive
definite, and a measurable empirical minimizer exists with probability
tending to one. Let the fixed prior be bounded, continuous and positive
at the minimizer, and suppose all posteriors are proper. Applying
Theorem~\ref{thm:convex-verification} to the i.i.d.~cluster losses
gives \eqref{eq:cl-ass-score}--\eqref{eq:cl-ass-lin}, \eqref{eq:cl-clt},
and \eqref{eq:cl-quantiles} for a $g$ differentiable at $\theta^{\mathrm{CL},\dagger}$
with $\dot{g}^{\mathrm{CL}}=\nabla g(\theta^{\mathrm{CL},\dagger})\ne0$
and satisfying the polynomial-growth condition \eqref{eq:func-growth-poly}.
It also gives uniform posterior moments of every fixed order.

Under these convex-cluster conditions, let $S_{G}\to\infty$ conditionally
independent posterior draws be generated after the rate to be evaluated
has been selected. The same second-order U-statistic argument as in
the proof of Corollary~\ref{cor:mc-iid}, with $G$ replacing $n$
and $\Lambda(\mathcal{Y}_{c})$ replacing $L(Y_{i})$, yields 
\[
G(\widehat{U}_{G}-U_{G})=O_{P_{\mathrm{joint}}}(S^{-1/2}_{G}),\qquad\frac{1}{G}\sum^{G}_{c=1}\left[G\omega\{\widehat{\xi}^{\mathrm{CL}}_{c,\omega}-\xi^{\mathrm{CL}}_{c,\omega}\}\right]^{2}=O_{P_{\mathrm{joint}}}(S^{-1}_{G}).
\]
To verify the latter bound, the covariance estimator is a second-order
U-statistic in independent posterior draws. Its squared kernel has
conditional expectation bounded by a constant multiple of $\Lambda(\mathcal{Y}_{c})^{2}$
times the square root of the product of the fourth posterior moments
of $Q_{G}$ and $X_{G}$. These fourth moments are uniformly $O_{P}(1)$
by the polynomial-growth argument in the proof of Theorem~\ref{thm:convex-verification},
and $G^{-1}\sum_{c}\Lambda(\mathcal{Y}_{c})^{2}=O_{P}(1)$. The bounds
prove \eqref{eq:cl-mc}. For dependent MCMC draws, additional simulation-error
control is needed. With unbounded cluster sizes, an observation-level
moment condition alone need not imply the required moment condition
for the aggregate envelope $\Lambda(\mathcal{Y}_{c})$; it must be
checked separately. This argument concerns regular convex losses,
not nonregular losses with nonquadratic local population risk. The
squared-error example below is not globally Lipschitz; its regularity
can instead be checked directly from the Gaussian design.

\subsection{Simulation study}

\label{app:cluster-simulation}

We use the independent-cluster regression design of \citet{Tanaka2026efficient},
a clustered counterpart of LinReg-homo in Section~\ref{subsec:sim-design}.
Each cluster contains $m=5$ observations. Let $\Sigma_{jk}=0.2^{|j-k|}$,
and independently generate $Z_{c},Z_{c1},\ldots,Z_{cm}\sim N_{4}(0,\Sigma)$
and $U_{c},E_{c1},\ldots,E_{cm}\sim N(0,1)$, with the two families
independent. Set 
\begin{align*}
X_{cj} & =\sqrt{\lambda_{X}}Z_{c}+\sqrt{1-\lambda_{X}}Z_{cj},\\
\varepsilon_{cj} & =\sqrt{\lambda_{\varepsilon}}U_{c}+\sqrt{1-\lambda_{\varepsilon}}E_{cj},\\
Y_{cj} & =X^{\top}_{cj}\theta_{0}+\varepsilon_{cj},\qquad\theta_{0}=(1,1,2,-1)^{\top},
\end{align*}
where $\lambda_{X}=\lambda_{\varepsilon}=0.5$. We use $L_{c}(\theta)=\frac{1}{2}\sum^{m}_{j=1}(Y_{cj}-X^{\top}_{cj}\theta)^{2}$,
$\theta\sim N_{4}(0,10^{2}I_{4})$, and the scalar target $g(\theta)=\theta_{1}$,
with $g^{\dagger}=1$. As shown by the cluster score calculation,
\[
J^{\mathrm{CL}}=m\Sigma,\qquad V^{\mathrm{CL}}=m\{1+(m-1)\lambda_{X}\lambda_{\varepsilon}\}\Sigma,
\]
thus the oracle rate is
\[
\omega^{\mathrm{CL},\circ}_{g}=\{1+(m-1)\lambda_{X}\lambda_{\varepsilon}\}^{-1}=0.5.
\]

For $G\in\{20,40,80,160\}$, with $n_{G}=mG$ total observations,
we generated $1{,}000$ independent datasets and evaluated nominal
$95\%$ equal-tailed intervals. IJVM used $5{,}000$ independent draws
from the exact Gaussian pilot posterior at $\omega^{(0)}=1$, with
the rate projected onto $\Omega=[0.01,10]$; the final intervals used
exact Gaussian posterior quantiles. We compare cluster-level IJVM
with observation-level IJVM (Obs. IJVM), which incorrectly treats
all lower-level observations as independent, and with GPC based on
cluster bootstrap resampling (Cluster GPC).

\begin{figure}[H]

\caption{Length of $95\%$ credible intervals}

\label{sm:fig:cl-length}
\begin{centering}
\subfloat[$n_{G}=100$]{
\begin{centering}
\includegraphics[width=0.4\columnwidth]{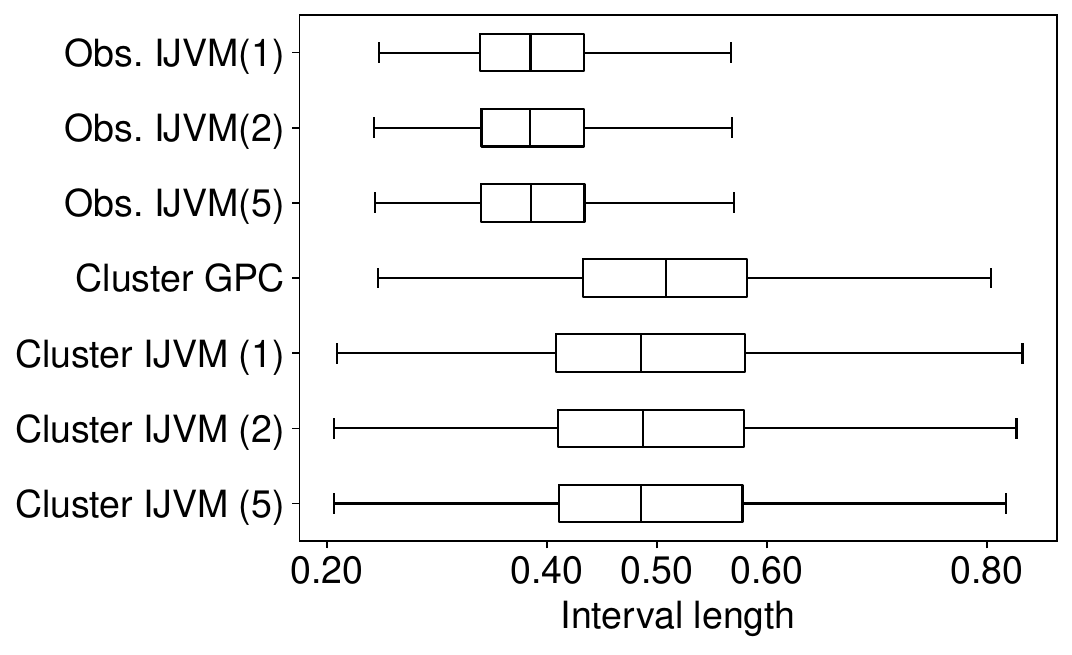}
\par\end{centering}
} \subfloat[$n_{G}=200$]{
\centering{}\includegraphics[width=0.4\columnwidth]{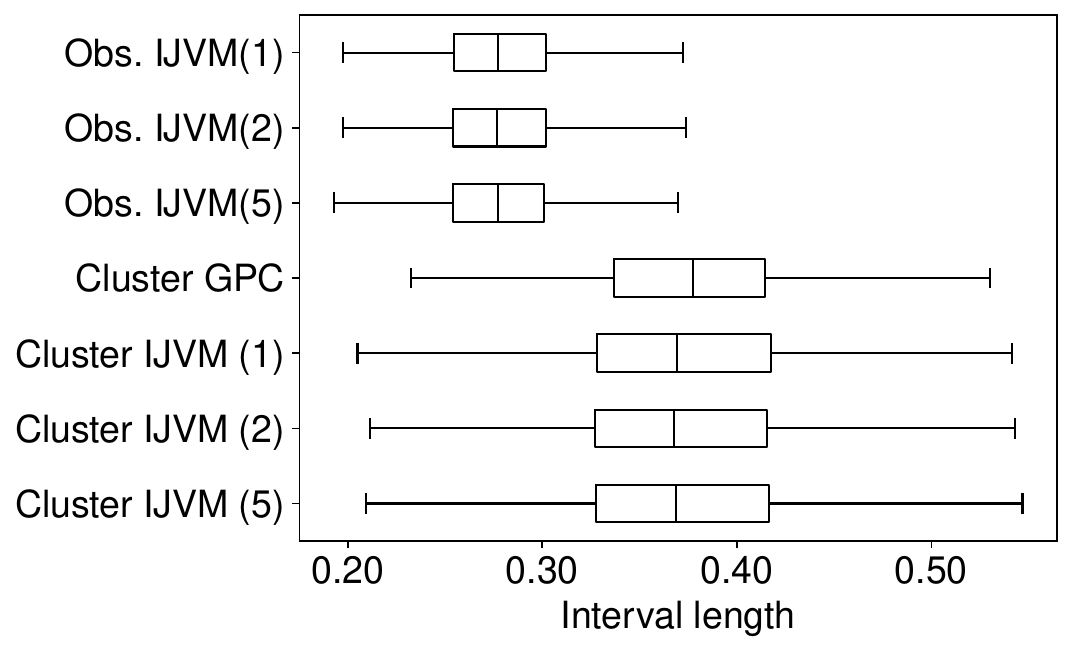}}
\par\end{centering}
\centering{}\subfloat[$n_{G}=400$]{
\centering{}\includegraphics[width=0.4\columnwidth]{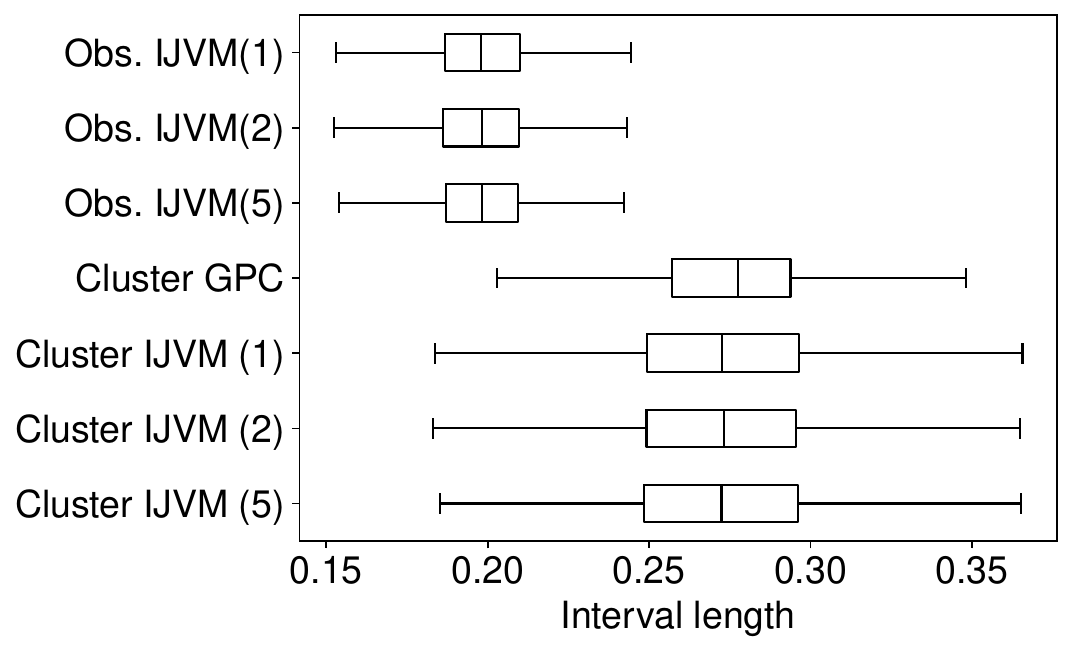}} \subfloat[$n_{G}=800$]{
\centering{}\includegraphics[width=0.4\columnwidth]{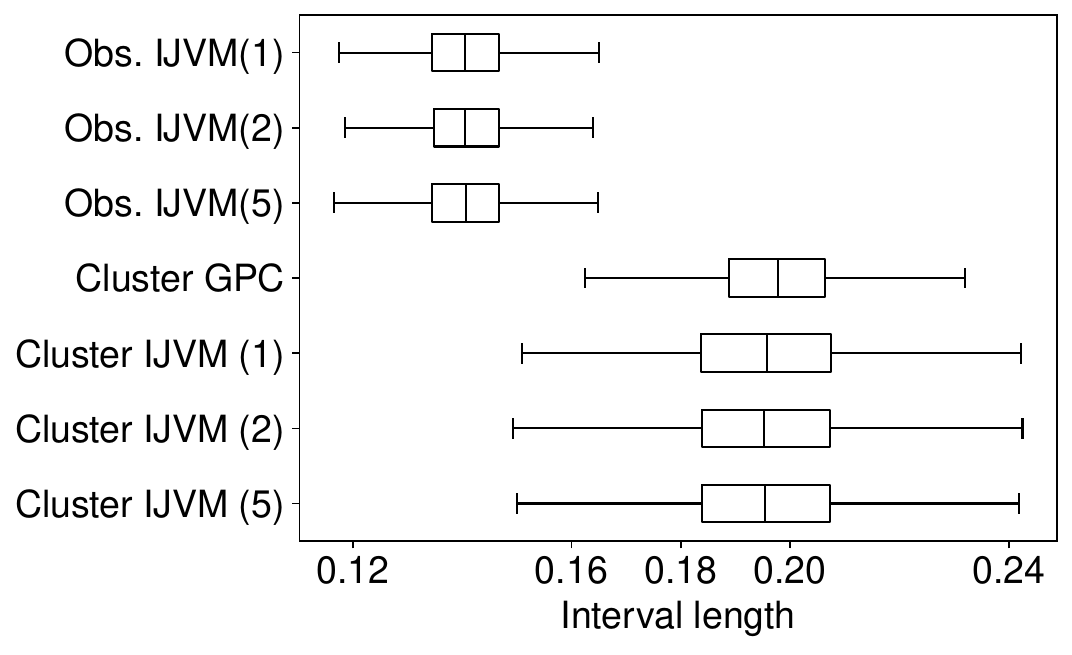}}
\end{figure}

Table~\ref{sm:tab:cl-coverage} reports the empirical coverage of
nominal $95\%$ credible intervals, while Figure~\ref{sm:fig:cl-length}
compares their lengths. Obs. IJVM substantially undercovers the target,
with coverage ranging from $0.808$ to $0.847$, reflecting its failure
to account for within-cluster dependence. In contrast, cluster-level
IJVM improves coverage considerably, approaching the nominal 95\%
level as the sample size increases. Its performance is broadly comparable
to cluster GPC, although some undercoverage remains at smaller sample
sizes. Figure~\ref{sm:fig:cl-length} shows that cluster-level IJVM
produces interval lengths similar to those of cluster GPC, whereas
obs. IJVM yields substantially shorter intervals. Additional IJVM
updates provide little systematic improvement in coverage or interval
length. These results support the effectiveness of cluster-level variance
matching for uncertainty quantification under within-cluster dependence.

\begin{table}[H]
\caption{Empirical coverage of nominal $95\%$ intervals}
\label{sm:tab:cl-coverage}
\centering{}%
\begin{tabular}{lr@{\extracolsep{0pt}.}lr@{\extracolsep{0pt}.}lr@{\extracolsep{0pt}.}lr@{\extracolsep{0pt}.}l}
\hline 
Method & \multicolumn{8}{c}{$n_{G}$}\tabularnewline
\cline{2-9}
 & \multicolumn{2}{c}{100} & \multicolumn{2}{c}{200} & \multicolumn{2}{c}{400} & \multicolumn{2}{c}{800}\tabularnewline
\hline 
Obs. IJVM(1) & 0&821 & 0&808 & 0&845 & 0&843\tabularnewline
Obs. IJVM(2) & 0&817 & 0&810 & 0&847 & 0&843\tabularnewline
Obs. IJVM(5) & 0&818 & 0&811 & 0&847 & 0&842\tabularnewline
Cluster GPC & 0&905 & 0&921 & 0&958 & 0&954\tabularnewline
Cluster IJVM (1) & 0&886 & 0&915 & 0&949 & 0&950\tabularnewline
Cluster IJVM (2) & 0&889 & 0&913 & 0&953 & 0&950\tabularnewline
Cluster IJVM (5) & 0&884 & 0&915 & 0&951 & 0&948\tabularnewline
\hline 
\end{tabular}
\end{table}

\bibliographystyle{apalike2}
\bibliography{reference}

\end{document}